\documentclass[11pt,a4paper]{article}
\usepackage{xy,amsmath,amssymb,amsthm,mathtools,xcolor,tikz,ifthen,cancel}
\usepackage{hyperref}
\mathtoolsset{showonlyrefs=true}

\numberwithin{equation}{section}

\newcommand{\qee}{\mbox{\hspace{0.2mm}}\hfill$\triangle$}
\let\oldcaption=\caption
\renewcommand{\caption}[1]{\oldcaption{\small #1}}
\newtheorem{thm}{Theorem}[section]

\theoremstyle{remark}
\newtheorem{rmkk}[thm]{Remark}
\newtheorem{exe}[thm]{Example}

\newcommand{\Q}{{\mathbb Q}}
\newcommand{\Z}{{\mathbb Z}}

\newcommand{\C}{{\mathbb C}}
\renewcommand{\P}{{\mathbb P}}
\newcommand{\M}{{\mathcal M}}
\newcommand{\vv}{\mathfrak{v}}
\newlength{\bibitemsep}
\newlength{\bibparskip}
\let\oldthebibliography\thebibliography
\renewcommand\thebibliography[1]{%
  \oldthebibliography{#1}%
  \setlength{\parskip}{\bibitemsep}%
  \setlength{\itemsep}{\bibparskip}%
}
\begin{document}
\begin{titlepage}
\begin{center}
{\large \bf  D3-BRANE SUPERGRAVITY SOLUTIONS FROM RICCI-FLAT METRICS   \\[3pt] ON CANONICAL
BUNDLES  OF     K\"AHLER-EINSTEIN SURFACES} \vskip 1cm {\sc Ugo
Bruzzo,$^{a,b,c,d}$ Pietro Fr\'e\,$^{e}$, Umar Shahzad$^{a,d}$ and
Mario Trigiante$^{f,g}$} \vskip 0.5cm
\smallskip
{\sl \small \frenchspacing
${}^a$   SISSA (Scuola Internazionale Superiore di Studi Avanzati), Via Bonomea 265, I-34136 Trieste, Italy \\ 
${}^b$Departamento de Mat\'ematica, Universidade   Federal da Para\'iba,
Jo\~ao Pessoa, PB, Brazil \\
${}^c$ INFN (Istituto Nazionale di Fisica Nucleare),  Sezione di
Trieste \\
${}^d$IGAP (Institute for Geometry and Physics), Trieste, Italy \\
${}^e$ Emeritus, Dipartimento di Fisica, Universit\`a di Torino, Via P. Giuria 1, I-10125 Torino, Italy \\
${}^f$ Dipartimento di Fisica, Politecnico di Torino,
C.so Duca degli Abruzzi 24, I-10129 Torino, Italy\\
${}^g$INFN,   Sezione di Torino\\
E-mail:  {\tt bruzzo@sissa.it, pietro.fre@unito.it, \\[-2pt]
ushahzad@sissa.it, mario.trigiante@polito.it} }
\end{center}
\begin{abstract}  \footnotesize
{D3 brane solutions of type IIB supergravity can be obtained by
means of a classical ansatz involving a  harmonic warp factor,
$H(\mathbf{y},\bar{\mathbf{y}})$ multiplying at power $-1/2$ the
first summand, \textit{i.e.} the Minkowski metric of the D3 brane
world-sheet and at power $1/2$ the second summand, \textit{i.e.} the
Ricci-flat metric  on a 6-dimensional transverse space
$\mathcal{M}_6$, whose complex coordinates $y$ are the arguments of
the warp factor}. Of particular interest is the case where
$\mathcal{M}_6=\operatorname{tot}[
K\left[\left(\mathcal{M}_B\right)\right]$ is the total space of the
canonical bundle over a complex K\"ahler 2-fold $\mathcal{M}_B$.
This situation emerges in many cases while considering the
resolution \`{a} la Kronheimer of singular manifolds of type
$\mathcal{M}_6=\mathbb{C}^3/\Gamma$, where
$\Gamma\subset\mathrm{SU(3)} $ is a discrete subgroup. When $\Gamma=
\mathbb{Z}_4$, the 2-fold $\mathcal{M}_B$ is the second Hirzebruch
surface endowed with a K\"ahler metric having $\mathrm{SU(2)\times
U(1)}$ isometry. {There is an entire class $\operatorname{Met}(\mathcal{FV})$ of
such cohomogeneity one K\"ahler metrics parameterized by a single
function $\mathcal{FK}(\mathfrak{v})$ that are best described in the
Abreu-Martelli-Sparks-Yau (AMSY) symplectic formalism.} We study in detail a two-parameter
subclass $\operatorname{Met}(\mathcal{FV})_{KE}\subset\operatorname{Met}(\mathcal{FV})$ of
K\"ahler-Einstein metrics of the aforementioned class, defined on
manifolds that  are homeomorphic to $S^2\times S^2$, but are
singular as complex manifolds. {Actually
$\operatorname{Met}(\mathcal{FV})_{KE}\subset \operatorname{Met}(\mathcal{FV})_{ext}\subset
\operatorname{Met}(\mathcal{FV})$ is a subset of a four parameter subclass
$\operatorname{Met}(\mathcal{FV})_{ext}$ of cohomogeneity one extremal K\"ahler
metrics originally introduced by Calabi in 1983 and translated by
Abreu into the AMSY action-angle formalism. $\operatorname{Met}(\mathcal{FV})_{ext}$
contains also a two-parameter subclass
$\operatorname{Met}(\mathcal{FV})_{ext\mathbb{F}_2}$ disjoint from
$\operatorname{Met}(\mathcal{FV})_{KE}$ of extremal smooth metrics on the second
Hirzebruch surface that we rederive using constraints on period
integrals of the Ricci $2$-form.}  The K\"ahler-Einstein nature of
the metrics in $\operatorname{Met}(\mathcal{FV})_{KE}$ allows the construction of the
Ricci-flat metric on their canonical bundle via the Calabi Ansatz,
which we recast in the AMSY formalism deriving some new elegant
formulae. {The metrics in $\operatorname{Met}(\mathcal{FV})_{KE}$ {are defined on the  base manifolds of  $U(1)$ fibrations supporting the}
family of Sasaki-Einstein  metrics $\mathrm{SE}met_5$ introduced by
Gauntlett et al.~in 2004, and already appeared in \cite{Gibbons-Pope}.  However, as we show in detail using Weyl
tensor polynomial invariants, the 6-dimensional Ricci-flat metric on
the \textit{metric cone} of $\M_5 \in\operatorname{Met}(SE)_5$ is
different from the Ricci-flat metric on
$\operatorname{tot}[ K\left[\left(\mathcal{M}_{KE}\right)\right]$
constructed via Calabi ansatz. This opens new research perspective.}
We also show the full integrability of the differential
system of geodesics equations on $\mathcal{M}_B$ thanks to a {certain}
conserved quantity    which is similar to
the Carter constant in the case of the Kerr metric.
\end{abstract}

\vfill
\noindent \parbox{6cm}{\hrulefill} \par\footnotesize
\noindent U.B.'s research is partly supported by Bolsa de Produtividade 313333/2020-3 of Brazilian CNPq,
by  PRIN ``Geometry of Algebraic Varieties'' and INdAM-GNSAGA.
He is a member of the VBAC group.\\[5pt]
\end{titlepage}
{\small \tableofcontents} \noindent {}
\newpage
\section{Introduction}
The paper  \cite{Bianchi:2021} reported, within the context of quiver
gauge theories, on
some advances   about  a special aspect of the
gauge/gravity correspondence,  i.e.,
\textit{the relevance of the generalized Kronheimer
construction}\footnote{{By ``generalized Kronheimer
construction'' we mean taking the K\"ahler quotient together with the construction of the metric that it naturally carries,
thus generalizing Kronheimer's construction of
the resolutions of the ADE singularities together with a hyperk\"ahler metric  on them \cite{kro1}.
In the case at hand of course the metric is not hyperk\"ahler.}}
\cite{Bruzzo:2017fwj,noietmarcovaldo} to the
resolution of $\mathbb{C}^3/\Gamma$ singularities. In particular, on the basis of the structure of exact D3-brane
supergravity solutions,  the issue of the construction of a
Ricci-flat metric on a smooth resolution $Y^\Gamma$ of
$\mathbb{C}^3/\Gamma$ was considered. The general framework
there and in this paper is  the   problem of
establishing holographic dual pairs, {made of}
\begin{enumerate} \itemsep=-2pt
  \item a gauge theory living on a D3-brane world volume;
  \item a classical D3-brane
solution of type IIB supergravity in $D=10$.
\end{enumerate}
Quiver gauge theories have been extensively
studied in the literature in this connection, see e.g.~\cite{Morrison:1998cs,Bianchi:2014qma,Feng:2007ur,Feng:2000mw,Feng:2001xr}.
Indeed the quiver diagram is a powerful
tool which encodes the data of a K\"ahler  quotient describing the
geometry of the six directions transverse to the brane.
The possibility of testing the holographic principle
\cite{Maldacena:1997re,Kallosh:1998ph,Ferrara:1998jm,Ferrara:1998ej,Ferrara:1997dh}
and resorting to the supergravity side of the correspondence in
order
 to perform, \textit{classically and in the bulk}, quantum
calculations that pertain to the boundary gauge theory, is tightly
connected with the quiver approach whenever the classical brane
solution has a conformal point corresponding to a limiting geometry
of the  type
{\begin{equation}\label{adsasak}
    M_{D} = \mathrm{AdS}_{p+2} \times  \mathrm{SE}^{D-p-2}.
\end{equation}
Here $\mathrm{AdS}_{p+2}$     denotes the  anti-de Sitter space of dimension  $p+2$, while $ \mathrm{SE}^{D-p-2}$
is    a Sasaki-Einstein manifold of dimension  $D-p-2$}
\cite{Fabbri:1999hw}.

A special class of quivers is that of
McKay quivers, which are associated with   the resolution of $\mathbb{C}^n/\Gamma$
quotient singularities by means of a Kronheimer-like  or GIT construction
\cite{kro1,kro2,mango}, where $\Gamma$ is a discrete subgroup in $\mathrm{SU}(n)$.
 The case $n=2$ corresponds to  ALE manifolds,  the discrete group $\Gamma$
being  given by
the ADE classification.\footnote{For a recent review of these matters
see chapter 8 of \cite{advancio}.}
\par
The case $n=3$ was studied from the mid 90s
\cite{itoriddo,62,Kin94,crawthesis,CrawIshii,SardoInfirri:1994is,SardoInfirri:1996ga,SardoInfirri:1996gb,degeratu}.
The 3-dimensional McKay correspondence provides a  group theoretical prediction of
the cohomology groups
$\mathrm{H}^\bullet (Y^\Gamma ,\Q)$ of a crepant
  resolution $Y^\Gamma $ of the quotient singularity
$\mathbb{C}^3/\Gamma$. Such  {complex} 3-folds carry a Ricci-flat K\"ahler
metric whose asymptotics at infinity depends on whether the locus in
$\C^3$ having nontrivial isotropy is compact (i.e., the origin) or
not. In the first case one has an ALE asymptotics, while  in the second
case the asymptotics depends on the direction: this is the Quasi-ALE
case
 (see e.g.~\cite{Joyce-QALE}, Thm.~3.3). As
it was stressed in \cite{Bianchi:2021}, {\it the Ricci-flat metric
 is not necessarily  the {metric} determined by the K\"ahler
quotient.}

In \cite{Bianchi:2021} the {relevant} conceptual landscape was
summarized as  follows.  The  finite group $\Gamma\subset \mathrm SU(3)$ singles out   a McKay quiver diagram which determines
\begin{enumerate} \itemsep=-2pt
  \item the gauge group $\mathcal{F}_\Gamma$;
  \item the matter field content $\Phi^I$ of the gauge theory;
  \item the  representation of  the gauge group factor
on the matter fields $\Phi^I$;
  \item the possible (mass)-deformations of the superpotential
  $\mathcal{W}(\phi)$.
\end{enumerate}
The Ricci-flat metric on $Y$ {in principle} can be inferred, by means of a
Monge-Amp\`{e}re equation, from the K\"ahler metric on the
exceptional compact divisor  in the
resolution of $\mathbb{C}^3/\Gamma$, which in  turn is determined by
the McKay quiver through the Kronheimer construction. In particular,
as discussed in \cite{Bianchi:2021} for the  case
$\mathbb{C}^3/\mathbb{Z}_4$,   although the Kronheimer metric on
$Y^\Gamma $ is not Ricci-flat, yet its restriction to the
exceptional divisor provides the appropriate starting point for an
iterative solution of a Monge-Amp\`{e}re equation which determines
the Ricci-flat metric.
\paragraph{The reversed view point of this paper.}
The explicit solution of the Monge-Amp\`{e}re equation for the case
of the smooth $Y^\Gamma $ resolution of the orbifold
$\mathbb{C}^3/\mathbb{Z}_4$, namely the summation of the series
implied by the iterative approach, proved to be quite unmanageable.  Only the case of a partial
resolution of the singularity,  which is the canonical bundle of the
(singular) weighted projective space $\mathbb{WP}[1,1,2]$, is quite  accessible, and   was
discussed in \cite{Bianchi:2021}  utilizing the powerful  AMSY  (Abreu-Martelli-Sparks-Yau) symplectic formalism of
\cite{abreu,Martelli:2005tp,Bykov:2017mgc}, whose simple results however
cannot always be explicitly  transferred back to the standard complex formalism
of K\"ahler geometry, as such a transcription involves the
inversion of higher transcendental functions.
 
On the other hand, Calabi's paper \cite{Calabi-Metriques} provides, in the
standard complex formalism, a recipe  for  constructing  a
K\"ahler metric $g_E$ on the total space of a holomorphic vector
bundle $E \to \M$, where $\M$ is a compact K\"ahler manifold,
satisfying the following conditions:
\begin{enumerate} \itemsep=-2pt
\item[C1:] the restriction of $g_{E}$
to the space tangent to the zero section of $E$ coincides with a
given  {K\"ahler-Einstein} (KE) metric $g_\M$ on
$\M$;
\item[C2:] the horizontal spaces given by the Chern connection of
the metric $g_E$ are the orthogonal complements of the tangent
spaces to the fibers of $E$ with respect to $g_E$;
\item [C3:] $g_E$ restricts on every fiber of $E$
to an hermitian metric.  \end{enumerate}
In the case of the singular variety $\mathbb{C}^3/\mathbb{Z}_4$, the
resolved variety is the total space of the canonical bundle of the
second Hirzebruch surface $\mathbb{F}_2$, and since  $\mathbb{F}_2$
admits no KE metric,  the Calabi Ansatz cannot be applied. On the other
hand, as we recall in  subsection \ref{pirilla}, the
resolution of singularities admits no infinitesimal deformation of
the complex structure, namely no $(2,1)$-forms, and this is an
obstacle for introducing fluxes in the supergravity D3-brane solution.
{In this paper, making use  of the AMSY formalism,   we
explore the possibility   of singling out
K\"ahler metrics, serving  as the starting point for the Calabi ansatz, within a
family, here called $\operatorname{Met}(\mathcal{FV})$, of 4D  
K\"ahler metrics. This family was already introduced in paper \cite{Bianchi:2021},
on the basis of previous
results of Gauntlet, Martelli, Sparks and Waldram
\cite{Gauntlett:2004yd}.}  The
distinctive features of the family $\operatorname{Met}(\mathcal{FV})$ are the
following:
\begin{description}
  \item[a)] the
metrics in the family  $\operatorname{Met}(\mathcal{FV})$ are \textit{parameterized by a
single function of one variable $\mathcal{FK}(\mathfrak{v})$};
  \item[b)] for any choice of $\mathcal{FK}(\mathfrak{v})$, the
 corresponding  metric $g_{\mathcal{FK}}$ is \textit{K\"ahlerian};
  \item[c)] for any choice of $\mathcal{FK}(\mathfrak{v})$, the
  \textit{isometry group} of $g_{\mathcal{FK}}$ is $\mathrm{SU(2)\times
  U(1)}$ and the underlying $4$-manifold $\mathcal{M}_{B}$ has cohomogenity one;
  \item[d)]  $\operatorname{Met}(\mathcal{FV})$ includes both \textit{the Kronheimer metric} for the smooth surface $\mathbb{F}_2$ and
{for the weighted
projective plane $ \mathbb{WP}[1,1,2]$.}
\end{description}
Relying on the  AMSY formalism it is straightforward to impose the
condition that $g_{\mathcal{FK}}$ is  an Einstein metric. This
yields a linear differential equation for the function
$\mathcal{FK}(\mathfrak{v})$ whose general integral contains two
parameters (integration constants)
\begin{equation}\label{lammis}
    0\leq\lambda_1 < \lambda_2 < \infty,
\end{equation}
so that one arrives at a two-parameter subclass of K\"ahler Einstein
metrics
\begin{equation}\label{sottopiatto}
    \operatorname{Met}(\mathcal{FV})_{KE} \, \subset \, \operatorname{Met}(\mathcal{FV})
\end{equation}
each singled out by a choice of the parameters in eq. \eqref{lammis}
and defined on a $4$-manifold that we label with the same
parameters $\M_{B}^{[\lambda_1,\lambda_2]}$ {(actually all these manifolds are homeomorphic to $S^2\times S^2$, but the metrics and the related complex structures are singular; more precisely, the metrics display two conical singularities, as we show in section  \ref{singularity}). These metrics already appeared in \cite{Gibbons-Pope} and also  \cite{Gauntlett:2004yd},
see eq.~(4.4) there (a change of coordinates is necessary to compare
the two families of metrics).} \footnote{We thank Dario
Martelli who called our attention to 
  the relation of these $4$-dimensional KE
metrics with the family of $5$-dimensional Sasaki-Einstein
metrics introduced in eq.(4.1) of  the same paper
\cite{Gauntlett:2004yd}.}

{Actually the family of metrics $\operatorname{Met}(\mathcal{FV})$ is a 
subclass  of a $4$-parameter family $\operatorname{Met}(\mathcal{FV})_{ext}$ of extremal
K\"ahler metrics in four real dimensions
that was derived by Calabi in 1982 \cite{extremcalabbo}}.\footnote{We thank Miguel Abreu for informing us of this fact. Extremal metrics are by definition those that extremize
Calabi's functional $\int\mathcal{R}_s[g]^2 \operatorname{vol}(g)$, where $\mathcal{R}_s$ is the
scalar invariant curvature; the notion was introduced by Calabi in \cite{extremcalabbo}.  For a modern ntroduction see e.g.~\cite{Szeke}. }

{The family
$\operatorname{Met}(\mathcal{FV})_{ext}$ was discussed  by Abreu in
\cite{abreu2009toric} using the AMSY
 action-angle coordinate formalism.
In section
\ref{varposympo} we explain the precise form of the family
$\operatorname{Met}(\mathcal{FV})_{ext}$, and using the global constraints on the
periods of the Ricci 2-form characterizing the $\mathbb{F}_2$
surface, we show that the 4-parameter family contains  two disjoint
subfamilies:
\begin{equation}\label{pastronchio}
    \operatorname{Met}(\mathcal{FV})_{ext} \,\supset \, \operatorname{Met}(\mathcal{FV})_{KE} \bigcup
    \operatorname{Met}(\mathcal{FV})_{ext\mathbb{F}_2}  \end{equation}
where  $\operatorname{Met}(\mathcal{FV})_{KE}$ is the  afore-mentioned family 
of KE metrics on the manifolds $\M_{B}^{[\lambda_1,\lambda_2]}$ with
two conical singularities, while 
$\operatorname{Met}(\mathcal{FV})_{ext\mathbb{F}_2}$ is a $2$-parameter
family of extremal K\"ahler metrics on the smooth $\mathbb{F}_2$
Hirzebruch surface. A relevant question  
is whether the Kronheimer K\"ahler metric $g^{\mathbb{F}_2}_{Kro}$
on $\mathbb{F}_2$ is extremal or not. The answer is that it belongs
to the family $\operatorname{Met}(\mathcal{FV})$, as we already know, yet it does not
belong to the subclass $\operatorname{Met}(\mathcal{FV})_{ext}$ as we explicitly show
in the sequel.}
 
{The good  news is that the Calabi Ansatz, originally formulated in
the standard complex formalism, has a particularly simple and
elegant transcription into the AMSY formalism, leading to compact
general formulas for the Ricci-flat metrics on the 2-parameter class
of KE manifolds $\M_{B}^{[\lambda_1,\lambda_2]}$. This allows us to write an explicit exact D3-brane
solution of Type IIB supergravity for each given manifold.}

{It is in conjunction with the construction of this Ricci-flat metric
that another important question arises. One might think that, in
view of the relation between the KE metrics
$g^{[\lambda_1,\lambda_2]}_{4-KE}$ and the Sasaki-Einstein metrics
$g^{[\lambda_1,\lambda_2]}_{5-Sasaki}$ introduced in
\cite{Gauntlett:2004yd}, the Ricci-flat metric obtained from the
Calabi Ansatz (in action-angle coordinates) with
$g^{[\lambda_1,\lambda_2]}_{4-KE}$ as an input, and the Ricci-flat
metric obtained from the metric cone on the Sasaki-Einstein
$g^{[\lambda_1,\lambda_2]}_{5-Sasaki}$ should be the same metric,
modulo change of coordinates. Actually this natural guess is
false and we demonstrate it explicitly in section \ref{sasacco} 
with direct and indirect arguments that utilize invariants
constructed with the Weyl tensor.
So one cannot freely assume that the Sasaki-Einstein metrics
$g^{[\lambda_1,\lambda_2]}_{5-Sasaki}$ of \cite{Gauntlett:2004yd} can
be utilized to determine the gauge-theory dual to D3-brane solution
of supergravity constructed via Calabi Ansatz.}
 
Notwithstanding the relation between
$g^{[\lambda_1,\lambda_2]}_{4-KE}$ and
$g^{[\lambda_1,\lambda_2]}_{5-Sasaki}$ the reversed problem of the
dual pair of theories is still there. We have, by construction, the
exact classical solution of supergravity based on the Ricci-flat
metric. {In order to find the other member of the pair, namely the
corresponding 4-dimensional gauge theory, we should be able 
to derive the spectrum of the theory from a
suitable quiver and a holomorphic superpotential.} This part of the problem is open.
\paragraph{The geometry of the 4-folds.}
From a geometric point of view, it should be stressed that all
4-manifolds in the $\operatorname{Met}(\mathcal{FV})$ family we find in our analysis
that are homeomorphic (not necessarily diffeomorphic) to the product
$ {S}^2\times  {S}^2$. Concerning the complex structure,
for some choices  of the $\mathcal{FK}(\mathfrak{v})$ function we
obtain the second Hirzebruch surface $\mathbb{F}_2$; for other
choices we have the aforementioned a 2-parameter family of KE
manifolds, which of course must be singular. Our analysis in section
\ref{singularity} shows indeed that they have a conical singularity
as we have already stressed.
\subsection*{Further issues}
\paragraph{About $(2,1)$-forms.} \label{pirilla} According to the result proved by Ito and Reid
\cite{itoriddo,crawthesis,CrawIshii} and based on the concept of age
the conjugacy classes in the group $\Gamma$, \footnote{For a recent
review of these matters within a general framework of applications
to brane gauge theories see \cite{Bruzzo:2017fwj,noietmarcovaldo}.}
the homology cycles of $Y^\Gamma$ are all algebraic, so  that  all notrivial
 cohomology groups are    of type $(q,q)$. There
is a correspondence between the cohomology classes of type $(q,q)$
and the discrete group conjugacy classes of  age  $q$:
\begin{equation}\begin{array}{rcl}
\dim H^{1,1}(Y^\Gamma )& =
    & \# \, \mbox{ of junior conjugacy classes in $\Gamma$;} \\[3pt]
\dim H^{2,2}(Y^\Gamma)& =
    & \# \, \mbox{ of senior conjugacy classes in $\Gamma$.}
     \end{array}\label{vecchioni}\end{equation}
Note that $H^{3,3}=0$ as $Y^\Gamma $ is noncompact.
\par
In \cite{Bianchi:2021} it was emphasized that the absence of
harmonic forms of type $(2,1)$, i.e.,  the absence of  infinitesimal deformations of
the complex structure,   is   a serious obstacle  to the construction of
supergravity D3-brane solutions based on $Y^\Gamma $ that have
transverse 3-form fluxes.
\par
It was pointed out in \cite{Bianchi:2021} that by means of the gaussian
integration of certain scalar fields predicted by the McKay $\Gamma$
quiver, and clearly distinguished from the other scalar fields on a group
theoretical basis, one gets a new quiver diagram that is not
directly associated with a discrete group, yet   follows from the
McKay $\Gamma$  quiver  in a unique way. In this way the group
theoretical approach allows one to identify deformations of the
superpotential and hence allows for ``deformations'' of
the crepant resolution. By construction, these deformed varieties will have
nontrivial harmonic (2,1) forms.
\par
In this way, one goes beyond the McKay correspondence. Both
physically and mathematically this is quite interesting and provides
a new viewpoint on several results, some of them well known in the
literature. Most of the latter are about  cyclic groups $\Gamma$
and rely on the powerful tools of toric geometry. Yet the
generalized Kronheimer construction applies also to non-abelian
groups $\Gamma\subset \mathrm{SU(3)}$ and so do  Ito-Reid's results.
Hence available mass-deformations are encoded also in the McKay
quivers of non-abelian groups $\Gamma$ and one might explore the
geometry of the transverse manifolds emerging in these cases.
\par
In section \ref{21forms} we give a negative solution to this problem
for what concerns the varieties we consider in this paper, for which
the group $H^{2,1}$  vanishes. As one may expect, we find that the
manifolds do not carry nontrivial self-dual $(2,1)$-forms. From the
point of view of supergravity, this means that we can only build
classical D3-brane solutions with a 5-form flux but no 3-form
fluxes. 
\paragraph{Sasaki-Einstein manifolds.}
Given the exact   explicit solution for the Ricci-flat metric on the
total spaces of the canonical bundles $\operatorname{tot}(\mathcal
K_{\M_{B}^{[\lambda_1,\lambda_2]}})$, we can investigate their
behavior at large distances from the corresponding base manifold $
\M_{B}^{[\lambda_1,\lambda_2]}$. In principle this procedure should
single out Sasaki-Einstein $5$-manifold ${\widetilde{\M_{SE}}}$
which, at this point, we expect to be different from those
considered \cite{Gauntlett:2004yd}. The ${\widetilde{\M_{SE}}}$
might have some singularities, typically corresponding to the
modding out of some discrete group $\widetilde{\Gamma}$  related to
the original ${\Gamma}$ in the ancestor orbifold
$\mathbb{C}^3/\Gamma$.
\section{D3-solutions of Type IIB Supergravity}
For the reader's convenience in this section
we concisely collect the main formulas related with the
D3-brane solution of Type IIB supergravity.
For more explanations  the reader is referred to section 2 of
\cite{Bianchi:2021}.
\par
We separate the ten coordinates of space-time into the following
subsets:\footnote{Latin indices are always frame indices referring to the vielbein
formalism. Furthermore we distinguish the 4 directions of the brane
volume by using Latin letters from the beginning of the alphabet
while the 3 complex transversal directions are denoted by Latin letters from
the middle and the end of the alphabet. For the coordinate indices
we utilize Greek letters and we do exactly the reverse: early Greek
letters $\alpha,\beta,\gamma,\delta,\dots$ refer to the 3 complex transverse
directions while Greek letters from the second half of the alphabet
$\mu,\nu,\rho,\sigma,\dots$ refer to the D3 brane world volume
directions as it is customary in $D=4$ field theories. }
\begin{equation}
 x^M = \left \{ \begin{array}{rll}
x^\mu, & \mu =0,\dots,3& \mbox{real coordinates of the 3-brane world volume}   \\
y^\tau, & \alpha=1,2,3 & \mbox{complex coordinates of the $Y$
variety}   \
\end{array} \right.
\label{coordisplit}
\end{equation}
and we make the following Ansatz for the metric:
\begin{align}
\label{ansazzo}
 \text{ds}^2_{[10]}&= H(\pmb{y},\bar{\pmb{y}})^{-\frac{1}{2}}\left
(-\eta_{\mu\nu}\,dx^\mu\otimes dx^\nu \right
)+H(\pmb{y},\bar{\pmb{y}})^{\frac{1}{2}} \, \left(
\mathrm{\mathbf{g}}^{\mathrm{RFK}}_{\alpha{\bar\beta}} \, dy^\alpha \otimes d\bar y^{{\bar\beta}}\right) \,
\\
 \text{ds}^2_{Y}&= \mathrm{\mathbf{g}}_{\alpha{\bar\beta}}^{\mathrm{RFK}} \, dy^\alpha \otimes d\bar y^{{\bar\beta}}\\
\eta_{\mu\nu}&= {\rm diag}(+,-,-,-)
\end{align}
where  {$\mathrm{\mathbf{g}}^{\mathrm{RFK}}$} is the K\"ahler metric of
the  manifold $Y$:
\begin{equation}
  \mathrm{\mathbf{g}}_{\alpha \bar{\beta}}^{\mathrm{RFK}} = \partial_\alpha \,
  \partial_{\bar{\beta}} \,
  \mathcal{K}^{\mathrm{RFK}}\left(\pmb{y},\bar{\pmb{y}}\right),
\label{m6defi}
\end{equation}
the real function
$\mathcal{K}^{\mathrm{RFK}}\left(\pmb{y},\bar{\pmb{y}}\right)$ being
a suitable K\"ahler potential. It follows that
$$
{\rm det}(g_{[10]})= H(\pmb{y},\bar{\pmb{y}}) \, {\rm det}(\mathrm{\mathbf{g}^{\mathrm{RFK}}}).
$$
Actually
the  formalism which is best suited for our aims  is the AMSY symplectic one,
rather than using holomorphic coordinates. In terms
of  the {vielbein} the Ansatz \eqref{ansazzo} corresponds to
\begin{equation}
V^{A}= \left \{\begin{array}{rcll}
  V^a  & = & H(\pmb{y},\bar{\pmb{y}})^{-1/4} \, dx^a & a=0,1,2,3\\
  V^\ell& = & H(\pmb{y},\bar{\pmb{y}})^{1/4} \,
  \mathbf{e}^\ell & \ell = 4,5,6,7,8,9
\end{array}\right.
\label{splittoviel}
\end{equation}
where   $\mathbf{e}^\ell$ are the vielbein $1$-forms of the manifold
$Y$. The structure equations  are (the
hats  denote
quantities computed without the warp factor, i.e., with $H=1$)
\begin{equation}\begin{array}{rcl}
0& = &  d \, \mathbf{e}^i  - \widehat{\omega}^{ij} \, \wedge \, \mathbf{e}^k \, \eta_{jk} \\
\widehat{R}^{ij} & = & d \widehat{\omega}^{ij} -
\widehat{\omega}^{ik}\, \wedge \, \widehat{\omega}^{\ell j} \,
\eta_{k\ell} = \widehat{R}^{ij} _{\phantom{ij}\ell m } \,
\mathbf{e}^\ell \,\wedge \, \mathbf{e}^m .
\end{array}\label{structeque}\end{equation}
The relevant property  of the $Y$ metric that we use in solving
Einstein equations is that it is Ricci-flat:
\begin{equation}
 \widehat{R}^{im}_{\phantom{ij}\ell m } = 0 . \label{ricciflatto}
\end{equation}
To derive our solution and discuss its
supersymmetry properties we need the explicit form of the spin connection
for the full $10$-dimensional metric \eqref{ansazzo} and  the
corresponding Ricci tensor. From the torsion equation one can
uniquely determine the solution:
\begin{equation}\begin{array}{rcl}
\omega^{ab} & = & 0  \\
\omega^{a\ell} & = &  \frac{1}{4}  \, H^{-3/2} \, dx^a  \eta^{\ell k} \, \partial_k \, H  \\
\omega^{\ell m} & = & \widehat{\omega}^{\ell m} + \Delta\omega^{\ell
m} \quad ; \quad \Delta\omega^{\ell m} = - \frac{1}{2}  \, H^{-1} \,
\mathbf{e}^{\ell }  \, \eta^{mk} \,\partial_k H
\end{array}\label{spinconnect} \end{equation}
Inserting this result into the definition of the curvature $2$-form
we obtain
\begin{equation}\begin{array}{rcl}
R^{a}_{b} & = & -  \frac{1}{8}\, \left [ H^{-3/2} \Box_{\mathbf{g}}
\, H - H^{-5/2} \,
\partial_k H\partial^k H \right] \, \delta^a_b  \\
R^{a}_{\ell} & = & 0 \\
R_\ell^m   &=& \frac{1}{8}H^{-3/2} \Box_{\mathbf{g}} H\delta_\ell^m
             - \frac{1}{8} H^{-5/2}\partial_s H\partial^s
             H\delta_\ell^m
             +\frac{1}{4} H^{-5/2} \partial_\ell H\partial^m H
\end{array}\label{riccius}\end{equation}
where for any function $f\left(\pmb{y},\bar{\pmb{y}}\right)$  on $Y$ the equation
\begin{equation}
\Box_{\mathbf{g}} \, f\left(\pmb{y},\bar{\pmb{y}}\right) \, =
\,\frac{1}{\sqrt{\mathrm{det}\mathbf{g}}}\, \left( \partial_\alpha
\left(\sqrt{\mathrm{det}\mathbf{g}}\,\,
\mathbf{g}^{\alpha{\bar\beta}} \,
\partial_{{\bar\beta}} \,f \right) \right) \label{laplacious}
\end{equation}
defines  the Laplace--Beltrami operator  with
respect to the  Ricci-flat metric \eqref{m6defi}; we
have omitted the superscript $\mathrm{RFK}$  just for simplicity ---
on the supergravity side of the correspondence we shall   only use
the Ricci-flat metric and there will be  no ambiguity.
\par
The equations of motion for the scalar fields $\varphi$ and
$C_{[0]}$ and for the 3-form field strengths $F^{NS}_{[3]}$ and
$F^{RR}_{[3]}$ can be better analyzed using the complex notation.
Defining, as it is explained in \cite{Bianchi:2021} above:
\begin{equation}\begin{array}{rcl}
{\mathcal{H}}_\pm & = & \pm 2 \,e^{-\varphi/2} F^{NS}_{[3]} + {\rm
i} 2 \,e^{\varphi/2} \,F^{RR}_{[3]} \quad \Rightarrow \quad
\overline{{\mathcal{H}}_+} = -\, \mathcal{H}_- \\
P & =& \frac 1 2 \, d\varphi -{\rm i} \frac 12 \, e^\varphi \,
F_{[1]}^{RR}
\end{array} \label{Psc} \end{equation}
but also setting in our Ansatz
\begin{equation}
  \varphi=0 \quad ; \quad C_{[0]}=0
\label{zerodilat}
\end{equation}
we reduce the equations for the complex 3-forms to
\begin{equation}\begin{array}{rcl}
   {\mathcal{H}}_+ \, \wedge \, \star
  {\mathcal{H}}_+=0  \\
d \star {\mathcal{H}}_+ =
   {\rm i} \, {F}_{[5]}^{RR}\,
  \wedge \, {\mathcal{H}}_+  \end{array}\label{hodge2formeq}
\end{equation}
while  the equation for the 5-form becomes
\begin{equation}
d\star F^{RR}_{[5]} = {\rm i} \, \frac 18 \, {\mathcal{H}}_+ \wedge
{\mathcal{H}}_- \label{f5equazia}
\end{equation}
The Ansatz for the complex 3-forms of type IIB supergravity is
given below and is inspired by what was done in
\cite{Bertolini:2002pr,Bertolini:2001ma} in the case where $Y
= \mathbb{C}\times \mathrm{ALE}_\Gamma$:
\begin{equation}
{\mathcal{H}}_+ = \Omega^{(2,1)}  \label{hpmposiz}
\end{equation}
where $\Omega^{(2,1)}$   lives on $Y$ and satisfies
 \begin{equation}\label{gerundio}
    \star_{\mathbf{g}} \, \mathrm{Q}^{(2,1)} \, = - i  \, \mathrm{Q}^{(2,1)}
\end{equation}
As   shown in \cite{Bianchi:2021} this guarantees that
\begin{equation}\label{ciarlatano}
    {\mathcal{H}}_+ \wedge \star_{10}\, {\mathcal{H}}_+ = 0.
\end{equation}
The Ansatz for $F_{[5]}^{RR}$ is
\begin{equation}\begin{array}{rcl}
F_{[5]}^{RR} & = & \alpha \left( U + \star_{10}\, U \right)   \\
U & =  & d \left( H^{-1} \, \mbox{Vol}_{\mathbb{R}^{(1,3)}} \right)
\end{array} \label{f5ansaz} \end{equation}
where $\alpha$ is a constant to be determined later. By construction
$F_{[5]}^{RR}$ is self-dual and its equation of motion is trivially
satisfied. What is not guaranteed is that also the Bianchi identity
is fulfilled. Imposing it results into a differential equation for
the function $H\left(\pmb{y},\bar{\pmb{y}}\right)$. Indeed
we obtain
\begin{equation}\begin{array}{rcl}\label{scolopio}
    d \, F_{[5]}^{RR}& = & \alpha \,\Box_{\mathbf{g}} \, H(\pmb{y},\bar{\pmb{y}}) \, \times
    \,\mbox{Vol}_{Y}
\end{array}\end{equation}
where
\begin{equation}\begin{array}{rcl}\label{supercaffelatte}
    \mbox{Vol}_{Y} & =  & \sqrt{\mbox{det} \, \mathbf{g}}  \,
    \frac{1}{(3!)^2}\,
    \epsilon_{\alpha\beta\gamma}\,
    dy^\alpha \wedge dy^\beta \wedge dy^\gamma \, \wedge \epsilon_{\bar{\alpha}\bar{\beta}\bar{\gamma}}\,
    d\bar{y}^{\bar{\alpha}} \wedge d\bar{y}^{\bar{\beta}} \wedge
    d\bar{y}^{\bar{\gamma}}
\end{array}\end{equation}is the volume form of the transverse six-dimensional
manifold \textit{i.e.} the total space of the canonical bundle
$K\left[\M_B\right]$. With our Ansatz we obtain
\begin{equation}\begin{array}{rcl}
\frac 1 8 \,{ \mathcal{H}}_+  \, \wedge \, {
  \mathcal{H}}_- &= & \mathbb{J}\left(\pmb{y},\bar{\pmb{y}}\right)
  \, \times \,
  \mbox{Vol}_{Y}  \\
\mathbb{J}\left(\pmb{y},\bar{\pmb{y}}\right) & = & - \, \frac{1}{72
\, \sqrt{\mbox{det}\, \mathbf{g}}} \, \, \,
\Omega_{\alpha\beta\bar{\eta}}\,\, \bar{\Omega}_{\bar{\delta}
\bar{\theta}\gamma} \,\, \epsilon^{\alpha\beta\gamma} \,\,
\epsilon^{\bar{\eta}\bar{\delta}\bar{\theta}}\label{tempra}
\end{array}\end{equation}
and we conclude that
\begin{equation}
  \Box_\mathbf{g}\,  H = - \frac
  {1}{\alpha}\, \mathbb{J}\left(\pmb{y},\bar{\pmb{y}}\right)
\label{diffe}
\end{equation}
This is the main differential equation to which the entire
construction of the D3-brane solution can be reduced. In
\cite{Bianchi:2021} it was  shown that the parameter $\alpha$ is
determined by Einstein's equations and is fixed to $\alpha=1$. With
this value the field equations for the complex three forms
simplify and reduce to the condition that $\Omega^{2,1}$ should
be closed, and then,  being anti-selfdual also co-closed, namely harmonic:
\begin{equation}\label{cannalotto}
    \tilde{\Omega}^{(2,1)} = \star_{\mathbf{g}} {\Omega}^{(2,1)} =
    -\, {\rm i}\,{\Omega}^{(2,1)} \quad ;
    \quad d\star_{\mathbf{g}} {\Omega}^{(2,1)} =0 \quad ;
    \quad d  {\Omega}^{(2,1)} =0
\end{equation}
In other words the solution of type IIB supergravity with 3-form
fluxes exists \textit{if and only if} the transverse space admits
\textit{closed and imaginary anti-self-dual forms} $\Omega^{(2,1)}$,
as we already stated.
\par
Summarizing, in order to construct a D3-brane solution of type IIB
supergravity we need:
\begin{description}
  \item[a)] to find a Ricci-flat K\"ahler metric $\mathbf{g}_{RFK}$ on the transverse 6D space $Y$;
  \item[b)] to verify if in the background of the metric
  $\mathbf{g}_{RFK}$ there exists a nonvanishing linear space of
  anti-self-dual (2,1)-forms $\Omega^{(2,1)}$. In the case of a
  positive answer,  the
  3-form $\mathcal{H}_+$ will be a linear combination of such forms;
 otherwise it will be zero.
  \item[c)] to solve the Laplacian equation for the harmonic function
  $H$ which is homogeneous if there are no 3-form fluxes, otherwise it
  is inhomogeneous as   in eqn.~\eqref{diffe}.
\end{description}
In the next section we describe the AMSY symplectic formalism that
will be propedeutic to the derivation of the KE family of
4D manifolds and later on to the derivation of the Ricci-flat metrics
on their canonical bundles.

\section{The AMSY symplectic formulation}
\label{amysone} Following the discussions and elaborations of
\cite{Bianchi:2021} based on \cite{abreu,Martelli:2005tp,Bykov:2017mgc}, given the K\"ahler potential of a toric
complex $n$-dimensional K\"ahler manifold
$\mathcal{K}(|z_1|,..,|z_n|)$,  where {$z_i=e^{x_i+i\Theta_i}$} are the
complex coordinates, introducing the moment variables
\begin{equation}\label{momentini2}
    \mu^i  =
    \partial_{x_i}\mathcal{K}
\end{equation}
we can obtain the so named symplectic potential by means of the
Legendre transform:
\begin{equation}\label{legendretr}
    {G}\left(\mu_i\right)
    =\sum_{i}^n x_i \,\mu^i \,    - \, \mathcal{K}(|z_1|,..,|z_n|)
\end{equation}
where one assumes that $\mathcal{K}$ only depends on the modules
of the $z$ coordinates to achieve $\mathrm{U}(1)^n$ invariance.
The main issue involved in the use of eqn.~\eqref{legendretr} is the
inversion transformation that expresses the coordinates $x_i$ in
terms of the  moments $\mu^i$. Once this is done one can calculate
the metric in moment variables utilizing the Hessian:
\begin{equation}\label{hessiano}
    {G}_{ij} = \frac{\partial^2}{\partial\mu^i \partial\mu^j} {G}\left(\mu\right)
\end{equation}
and its matrix inverse. Call the $n$ angles by $ \Theta_i $.
Complex coordinates better adapted to the complex structure tensor can be
defined a
\begin{equation}\label{spigolini}
    u_i= e^{z_i} = \exp[x_i \,+\, \mathrm{i} \Theta_i]
\end{equation}
The K\"ahler 2-form has the following universal structure:
\begin{equation}\label{uniKal2}
  \mathbb{K} \,= \,  \sum_{i=1}^{n} \, \mathrm{d}\mu^i\wedge
  \mathrm{d}\Theta_i
\end{equation}
and the metric is expressed as
\begin{equation}\label{sympametra}
    ds^2_{symp} = {\mathbf{G}}_{ij} d\mu^i \, d\mu^j \, + \, {\mathbf{G}}^{-1}_{ij}d\Theta^i \, d\Theta^j
\end{equation}
\section{K\"ahler metrics    with $\mathrm{SU(2)\times U(1)} $ isometry} \label{varposympo}
In this paper we
are interested, to begin with, in   K\"ahler metrics in
two complex dimensions,  $n=2$, where the complex
coordinates  $u,v$ enter the K\"ahler potential
$\mathcal{K}_0(\varpi)$ only through the  real combination
\begin{equation}\label{invarpi}
   \varpi =\left(1+\mid u\mid^2\right)^2 \, \mid
    v\mid^2,
\end{equation}
which guarantees invariance under $\mathrm{SU(2)}\times
\mathrm{U(1)}$ transformations realized  as
\begin{equation}\begin{array}{rcl}
\text{if}\  \mathbf{g} &=& \left(
          \begin{array}{cc}
            a & b \\
            c & d \\
          \end{array}
        \right) \, \in \, \mathrm{SU(2)}\quad \text{then}  \quad
    \mathbf{g}\left(u,v\right) =\left(\frac{a \, u + b}{c\, u +
    d}, \quad v \, \left(c \,u+d\right)^{2}\right) ;
     \\[3pt]
\text{if}\   \mathbf{g} &=& \exp(i\,\theta_1) \, \in \, \mathrm{U(1)}
\quad  \text{then} \quad \mathbf{g}\left(u,v \right) =\left(u, \quad
\exp(i\,\theta_1)\, v  \right).  \label{ciabattabuona}
\end{array}\end{equation}
The above realization of the isometry  captures the idea that, at least locally, the manifold  is an $S^2$ fibration
over $S^2$ ($u$ being a coordinate on the base and $v$ a fiber coordinate),  although their global topology
might be different and have some kind of singularities.
\par
Two cases are of particular interest within such a framework, namely
\begin{description}
  \item[a)]   the singular weighted projective plane $\mathbb{WP}[1,1,2]$;
  \item[b)]  the  second Hirzebruch surface
  $\mathbb{F}_2$.
\end{description}
In the case of the singular variety $\mathbb{WP}[1,1,2]$ we have a
nonK\"ahler--Einstein  metric that emerges from a partial
resolution of the $\mathbb{C}^3/\mathbb{Z}_4$ singularity within the
generalized Kronheimer construction (see
\cite{noietmarcovaldo},\cite{Bianchi:2021}) whose explicit
K\"ahler potential is the following one:
\begin{equation}\label{K0WP112}
    \mathcal{K}_0^{Kr}\left(\varpi\right)= \frac{9}{4} \left(\frac{3 \varpi +\sqrt{\varpi  (\varpi +8)}}{\varpi +\sqrt{\varpi
   (\varpi +8)}}+\log \left(\varpi +\sqrt{\varpi  (\varpi +8)}+4\right)-2-4 \log
   (2)\right)
\end{equation}
On the other hand the K\"ahler potential \eqref{K0WP112} is the
particular case $\alpha = 0$ of a one parameter family of K\"ahler
potentials obtained from the Kronheimer construction:
\begin{equation}\begin{array}{rcl}
  K^{Kr}[\varpi,\alpha] &=& - \displaystyle\frac{9}{16} \left(-4 (\alpha +1) \log \left[\sqrt{\alpha ^2+6
\alpha  \varpi
   +\varpi ^2+8 \varpi }+3 \alpha +\varpi +4\right]\right. \\[5pt]
   &&\displaystyle \left.-\frac{4 \left(\alpha
   \left(\sqrt{\alpha ^2+6 \alpha  \varpi +\varpi  (\varpi +8)}+2 \varpi
   +1\right)+\sqrt{\alpha ^2+6 \alpha  \varpi +\varpi  (\varpi +8)}+\alpha ^2+3
   \varpi \right)}{\sqrt{\alpha ^2+6 \alpha  \varpi +\varpi  (\varpi +8)}+\alpha
   +\varpi }\right. \\[8pt]
   &&\displaystyle\left. +4 \alpha  \log  \sqrt\frac{\sqrt{\alpha ^2+6 \alpha  \varpi
   +\varpi  (\varpi +8)}+\alpha +\varpi }{\sqrt{\varpi }} +8+16 \log 2\right)
\end{array} \label{genKron} \end{equation}
that for $\alpha > 0$ generate \textit{bona fide} K\"ahler metrics
on the second Hirzebruch surface $\mathbb{F}_2$.
\subsection{A family of   4D K\"ahler metrics}
Having mentioned the two explicit examples of
eqns.~\eqref{K0WP112},\eqref{genKron}, now, using the AMSY approach,
we discuss in more general terms  a class of real 4D K\"ahler
manifolds, that we call $\M_B$. These are endowed with a metric
invariant under the $\mathrm{SU(2)\times U(1)}$ isometry group
acting as   in eqn.~\eqref{ciabattabuona}. The class of these
manifolds is singled out  by the above assumption that, in  the
complex formalism, their K\"ahler potential $\mathcal{K}_0(\varpi)$
is  a function only of the invariant $\varpi$ defined in
eqn.~\eqref{invarpi}. Then the explicit form of the K\"ahler
potential $\mathcal{K}_0(\varpi)$ cannot be worked out analytically
in all cases since the inverse Legendre transform involves the roots
of higher order algebraic equations; yet, using the
$\varpi$-dependence assumption, the K\"ahler metric can be
explicitly worked out in the symplectic coordinates and has a simple
and very elegant form -- actually the metric depends on a single
function of one variable $\mathcal{FK}(\vv )$ which encodes all the
geometric properties and substitutes $\mathcal{K}_0(\varpi)$. Posing
all the questions in this symplectic language allows  one  to
calculate all the geometric properties of the spaces in the class
under consideration and leads also to new results and to a more
systematic overview of the already known cases. We choose to treat
the matter in general, by utilizing a \textit{local approach} where
we discuss the differential equations in a given open dense
coordinate patch $u,v$, and we address the question of its global
topological and algebraic structure  only a posteriori, once the
metric as been found in the considered chart, just as one typically
does in General Relativity.
\par
The symplectic structure of the metric on $\M_B$ is
exhibited in the following way:
\begin{equation}\label{trottolina}
    ds_{\M_B}^2 = \mathbf{g}_{\M_B|\mu\nu}
    \, dq^\mu \, dq^\nu \quad ; \quad q^\mu  =   \left\{\mathfrak{u},\vv ,\phi,\tau\right\} \quad ;
    \quad \mathbf{g}_{\M_B} =\left(\begin{array}{c|c}
    \mathbf{G}_{\M_B} & \mathbf{0}_{2\times 2} \\
    \hline
    \mathbf{0}_{2\times 2} &
    \mathbf{G}^{-1}_{\M_B}
    \end{array}\right)
\end{equation}
where the Hessian $\mathbf{G}_{\M_B}$ is defined by:
\begin{align}\mathbf{G}_{\M_B}  =
    \partial_{\mu^i}\,\partial_{\mu^j} \,
    {G}_{\M_B} \quad ; \quad \mu^i =
    \left\{\mathfrak{u},\vv \right\}\label{reducedhessian}
\end{align}
and
\begin{equation}
{G}_{\M_B}= {G}_0(\mathfrak{u},\vv ) \, + \,
\mathcal{D}(\vv )\label{GBsymplectic} \end{equation}
\begin{equation}
 G_0\left(\mathfrak{u},\vv \right) =  \left(\vv -\frac{\mathfrak{u}}{2}\right) \log (2
   \vv -\mathfrak{u})+\frac{1}{2} \mathfrak{u} \log
   (\mathfrak{u})-\frac{1}{2} \vv  \log (\vv ) \label{lupetto}
\end{equation}
The specific structure \eqref{GBsymplectic},\eqref{lupetto} is the
counterpart  within the symplectic formalism, via Legendre transform,
of the assumption that
the K\"ahler potential $\mathcal{K}_0(\varpi)$ depends only on the
$\varpi$ variable.
\par
After noting this important   point, we go back  to  the discussion
of $\M_B$  geometry and   stress that with the given
isometries its Riemannian structure is completely encoded in the
boundary function $ \mathcal{D}(\vv )$. All the other items
in the construction are as follows. For the K\"ahler form we have
\begin{equation}\label{kallerformMB}
    \mathbb{K}^{\M_B} = 2\, \left( \, d\mathfrak{u} \wedge d\phi + d\vv  \wedge d\tau   \right) \,
    = \, \mathbf{K}_{\mu\nu}^{\M_B} \, dq^\mu \wedge
    dq^\nu \quad ; \quad \mathbf{K}^{\M_B} = \left(\begin{array}{c|c}
    \mathbf{0}_{2\times 2} & \mathbf{1}_{2\times 2} \\
    \hline
    - \mathbf{1}_{2\times 2} &
    \mathbf{0}_{2\times 2}
    \end{array}\right)
\end{equation}
and for the complex structure we obtain
\begin{equation}\label{complestruc2}
\mathfrak{J}^{\M_B} = \mathbf{K}^{\M_B}\,
\mathbf{g}^{-1}_{\M_B} = \,\left(\begin{array}{c|c}
\mathbf{0}_{2\times 2} & \mathbf{G}_{\M_B} \\
\hline - \mathbf{G}^{-1}_{\M_B} & \mathbf{0}_{2\times 2}
\end{array}\right)
\end{equation}
Explicitly the $2\times 2 $ Hessian is the following:
\begin{equation}\begin{array}{rcl}\label{GMB}
 \mathbf{G}_{\M_B}& = &    \left(
\begin{array}{cc}
 -\frac{\vv }{\mathfrak{u}^2-2 \mathfrak{u} \vv } &
   \frac{1}{\mathfrak{u}-2 \vv } \\
 \frac{1}{\mathfrak{u}-2 \vv } & \frac{-2 \vv
   (\mathfrak{u}-2 \vv ) \mathcal{D}''(\vv )+\mathfrak{u}+2
   \vv }{2 \vv  (2 \vv -\mathfrak{u})} \\
\end{array}
\right) \\
\mathbf{G}_{\M_B}^{-1}& = &\left(
\begin{array}{cc}
 \frac{\mathfrak{u} \left(-2 \vv  (\mathfrak{u}-2
   \vv ) \mathcal{D}''(\vv )+\mathfrak{u}+2
   \vv \right)}{\vv  \left(2 \vv
   \mathcal{D}''(\vv )+1\right)} & \frac{2 \mathfrak{u}}{2
   \vv  \mathcal{D}''(\vv )+1} \\
 \frac{2 \mathfrak{u}}{2 \vv  \mathcal{D}''(\vv )+1} &
   \frac{2 \vv }{2 \vv  \mathcal{D}''(\vv )+1}
   \\
\end{array}
\right)
\end{array}\end{equation}
The family of metrics \eqref{trottolina} is parameterized by the
choice of a unique one-variable function:
\begin{equation}\label{fungemistero}
    \mathit{f}(\vv )\equiv\mathcal{D}''(\vv )
\end{equation}
and is   worth being considered in its  own  right.
\subsection{The inverse Legendre transform}
\label{reconstruczia} Before proceeding further with the analysis of
this class of metrics, it is convenient to consider the
inverse Legendre transform and see how one    reconstructs  the
K\"ahler potential on $\M_B$.
The inverse Legendre transform provides the K\"ahler potential
through the formula:
\begin{equation}\begin{array}{rcl}\label{K0funzia}
    \mathcal{K}_0 = x_u \, \mathfrak{u} \,+ \,  x_v \, \vv
    \, - \, G_{\M_B}\left(\mathfrak{u},\vv \right)
\end{array}\end{equation}
where $G_{\M_B}\left(\mathfrak{u},\vv \right)$ is
the base manifold symplectic potential defined in
eqn.~\eqref{GBsymplectic}, and
\begin{equation}\label{caramboliere}
    x_u =
    \partial_{\mathfrak{u}}\,G_{\M_B}\left(\mathfrak{u},\vv \right)
    \quad ; \quad x_v =
    \partial_{\vv }\,G_{\M_B}\left(\mathfrak{u},\vv \right)
\end{equation}
which explicitly yields:
\begin{equation}\label{inversioneB}
  x_u=  \frac{1}{2} \left(\log
   \left(\mathfrak{u}\right)-\log \left(2
   \vv -\mathfrak{u}\right)\right)\quad ; \quad x_v=
   \mathcal{D}'\left(\vv \right)+\log \left(2
   \vv -\mathfrak{u}\right)-\frac{1}{2} \log
   \left(\vv \right)+\frac{1}{2}
\end{equation}
Using eqn.~\eqref{inversioneB}) in eqn.~\eqref{K0funzia} we immediately
obtain the explicit form of the base-manifold K\"ahler potential as
a function of the moment $\vv $:
\begin{equation}\label{K0inv0}
    \mathcal{K}_0 = \mathfrak{K}_0\left(\vv \right)=\vv
   \left(\mathcal{D}'\left(\vv \right)+\frac{1}{2}\right)-\mathcal{D}\left(\vv \right)
\end{equation}
The problem is that we need the K\"ahler potential $\mathcal{K}_0$
as a  function of the invariant $\varpi$. Utilizing eqn.
\eqref{inversioneB} it is fairly easy to obtain the expression of
$\varpi$ in terms of the moment $\vv $ for a generic
function $\mathcal{D}(\vv )$ that codifies the geometry of
the base-manifold, obtaining
\begin{equation}\label{omegadiv0}
    \varpi=\left(1+\exp\left[2\,x_u\right]\right)^2
    \,\exp\left[2\,x_u\right] =
    \Omega\left(\vv \right)= 4
    \vv  \exp\left[\,2\,
    \partial_{\vv }\mathcal{D}\left(\vv \right)+1\right]
\end{equation}
If  one is able to invert the function
$\Omega\left(\vv \right)$, the original K\"ahler potential
of the base-manifold can be written as:
\begin{equation}\label{compostadimele}
    \mathcal{K}_0 \left(\varpi\right) =\mathfrak{K}_0\, \circ
    \,
    \Omega^{-1}\left(\varpi\right)
\end{equation}
The inverse function
$\Omega^{-1}\left(\varpi\right)$ can be written explictly
in some simple cases,  but not always, and
this inversion is the main reason why certain K\"ahler metrics can
be much more easily found in the AMSY symplectic formalism which
deals only with real variables than in the complex formalism. Since
nothing good comes without paying a price, the metrics found in the
symplectic approach require that the ranges of the variables
$\mathfrak{u}$ and $\vv $ should be determined, since it is
just in those  ranges that the topology and algebraic structure of
the underlying manifold is hidden; indeed the ranges of
$\mathfrak{u}$ and $\vv $ define a convex closed
\textit{polytope} in the $\mathbb{R}^2$ plane that encodes very
precious information about the structure of the underlying manifold.

\section{The Ricci tensor and the Ricci form}  Calculating the Ricci tensor
for the family of metrics \eqref{trottolina} we obtain the following
structure:
\begin{equation}\label{ricciodimare}
\mathrm{Ric}_{\mu\nu}^{\M_B} =
\left(\begin{array}{c|c}
\mathbf{P}_\mathrm{U} & \mathbf{0}_{2\times 2} \\
\hline \mathbf{0}_{2\times 2} & \mathbf{P}_\mathrm{D}
\end{array}
\right)
\end{equation}
The expressions for $\mathbf{P}_\mathrm{U}$ and
$\mathbf{P}_\mathrm{D}$ are quite lengthy and we omit them. We
rather consider the Ricci 2-form defined by:
\begin{equation}\label{riccioforma}
\mathbb{R}\text{ic}_{\M_B} =
\mathbf{Ric}_{\mu\nu}^{\M_B} \, dq^\mu
    \wedge dq^\nu
\end{equation}
where:
\begin{equation}\label{riccioformcompo}
    \mathbf{Ric}^{\M_B}=\mathrm{Ric}^{\M_B} \,
    \mathfrak{J}^{\M_B} = \left(\begin{array}{c|c}
    \mathbf{0}_{2\times 2} & \mathbf{R} \\
    \hline
    - \mathbf{R}^T & \mathbf{0}_{2\times 2}
    \end{array}
     \right)
\end{equation}
and\begin{equation}\begin{array}{rcl}
\mathbf{R} &=& \left(
\begin{array}{cc}
r_{11} & r_{12} \\
r_{21} & r_{22} \\
\end{array}
\right)
       \\
     r_{11} &=& \frac{2 \vv  \left(\vv  \mathit{f}'(\vv )+2 \vv
   \mathit{f}(\vv )^2+\mathit{f}(\vv )\right)-1}{2 \vv  (2
   \vv  \mathit{f}(\vv )+1)^2}  \\
     r_{12} &=& 0  \\
     r_{21} &=& \displaystyle\frac{2 \mathfrak{u} \vv  \left(\vv ^2 (2 \vv
   \mathit{f}(\vv )+1) \mathit{f}''(\vv )-4 \vv ^3
   \mathit{f}'(\vv )^2+4 \vv  \mathit{f}'(\vv )-4
   \vv ^2 \mathit{f}(\vv )^3-2 \vv
   \mathit{f}(\vv )^2+3 \mathit{f}(\vv )\right)+\mathfrak{u}}{2
   \vv ^2 (2 \vv  \mathit{f}(\vv )+1)^3}  \\[5pt]
     r_{22} &=& \displaystyle \frac{\mathit{f}(\vv ) \left(2 \vv ^2 \left(\vv
   \mathit{f}''(\vv )+\mathit{f}'(\vv )\right)+3\right)+\vv
   \left(\vv  \mathit{f}''(\vv )-4 \vv ^2
   \mathit{f}'(\vv )^2+5 \mathit{f}'(\vv )\right)+2 \vv
   \mathit{f}(\vv )^2}{(2 \vv  \mathit{f}(\vv )+1)^3}  \\
     \mathbf{R}^T&=& \mathbf{P}_\mathrm{D} \,
     \mathbf{G}^{-1}_{\M_B} \label{riccopovero}
   \end{array}\end{equation}
 The last equation is not a definition but rather a consistency
constraint (the Ricci  tensor must be  skew-symmetric).

\paragraph{A two-parameter family of KE metrics for
$\M_B$. }\label{famigliaKE} An interesting and   legitimate question
is whether this family of cohomogeneity one metrics that we named
$\operatorname{Met}(\mathcal{FV})$ in the introduction, contains KE ones. The answer
is positive, and they  make up a two-parameter subfamily. As we
already claimed in the introduction and as it will be shown in the
next section, such a KE family $\operatorname{Met}(\mathcal{FV})_{ext}$ is a
subfamily of a 4-parameter family of extremal metrics
$\operatorname{Met}(\mathcal{FV})_{ext}$. Let us first directly retrieve the KE
family from the appropriate differential constraint. A metric is KE
if the Ricci 2-form is proportional to the K\"ahler 2-form:
\begin{equation}\label{kallopietra}
     \mathbb{R}\text{ic}^{\M_B} = \frac{\mathit{k}}{4} \, \mathbb{K}^{\M_B}
\end{equation}
where $\mathit{k}$ is a  constant. This amounts to requiring  that
the $2\times 2$ matrix $\mathbf{R}$ displayed in
eqn.~\eqref{riccopovero} be proportional via $\frac{\mathit{k}}{4}$
to the identity matrix $\mathbf{1}_{2\times2}$. This condition
implies differential constraints on the function $\mathit{f(\vv )}$
that are uniquely solved by the following function:
\begin{equation}\label{miraculo}
\mathit{f(\vv )} = \frac{-3 \beta +\mathit{k} \vv ^3+3 \vv ^2}{-2
\mathit{k} \vv ^4+6 \vv ^3+6 \beta  \vv }
\end{equation}
the parameter $\beta$ being the additional integration constant,
while $k$ is defined by equation \eqref{kallopietra}.  To retrieve
the original symplectic potential $\mathcal{D}(\vv )$ one has just
to perform a double integration in the variable $\vv $. The explicit
calculation of the integral requires a summation over the three
roots $\lambda_{1,2,3}$ of the following cubic polynomial:
\begin{equation}\label{racete}
    P(x)=x^3\, -\, \frac{3 x^2}{\mathit{k}}\, -\, \frac{3 \beta
    }{\mathit{k}}
\end{equation}
whose main feature is the absence of the linear term. Hence a
beautiful way of parameterizing the family of KE metrics is achieved
by using as parameters two of the three roots of the polynomial
\eqref{racete}. Let us call the independent roots $\lambda_1$ and
$\lambda_2$. The polynomial \eqref{racete} is reproduced by setting:
\begin{equation}\label{rinominopara}
    \mathit{k}\,= \, \frac{3 \left(\lambda _1+\lambda
   _2\right)}{\lambda _1^2+\lambda _2 \lambda _1+\lambda _2^2},\quad \quad \beta
   =  -\frac{\lambda _1^2 \lambda _2^2}{\lambda _1^2+\lambda _2
   \lambda _1+\lambda _2^2},\quad \lambda _3= -\frac{\lambda _1 \lambda
   _2}{\lambda _1+\lambda _2}
\end{equation}
Substituting \eqref{rinominopara} in eqn.~\eqref{miraculo} we obtain
\begin{equation}\begin{array}{rcl}\label{gamellino}
  \mathit{f}(\vv )& = & -\frac{\lambda _1 \vv ^2
   \left(\lambda _2+\vv \right)+\lambda _2 \vv ^2
   \left(\lambda _2+\vv \right)+\lambda _1^2 \left(\lambda
   _2^2+\vv ^2\right)}{2 \vv
   \left(\vv -\lambda _1\right) \left(\vv -\lambda
   _2\right) \left(\lambda _2 \vv +\lambda _1 \left(\lambda
   _2+\vv \right)\right)}
\end{array}\end{equation}
which is completely symmetrical in the exchange of the two
independent roots $\lambda_1,\lambda_2$. Utilizing the expression
\eqref{gamellino} the double integration is easily performed, and we
obtain the explicit result, where we   omitted   irrelevant linear
terms:
\begin{equation}\begin{array}{rcl}\label{sympaKE}
  \mathcal{D}^{KE}(\vv )&=& -\frac{\left(\lambda _1^2+\lambda _2 \lambda _1+\lambda _2^2\right)
   \left(\vv -\lambda _1\right) \log \left(\vv -\lambda
   _1\right)}{\lambda _1^2+\lambda _2 \lambda _1-2 \lambda _2^2} \\[3pt]
   &&  \hskip20mm -\frac{\left(\lambda
   _1^2+\lambda _2 \lambda _1+\lambda _2^2\right) \left(\vv -\lambda
   _2\right) \log \left(\vv -\lambda _2\right)}{-2 \lambda _1^2+\lambda _2
   \lambda _1+\lambda _2^2} \\[3pt]
   &&+\frac{\left(\lambda _1^2+\lambda _2 \lambda _1+\lambda
   _2^2\right) \left(\lambda _2 \vv +\lambda _1 \left(\lambda
   _2+\vv \right)\right) \log \left(\lambda _2 \vv +\lambda _1
   \left(\lambda _2+\vv \right)\right)}{\left(\lambda _1+\lambda _2\right)
   \left(2 \lambda _1^2+5 \lambda _2 \lambda _1+2 \lambda _2^2\right)}-\frac{1}{2}
   \vv  \log (\vv )
\end{array}\end{equation}
Comparing with  the original papers on the AMSY formalism
\cite{abreu,Martelli:2005tp},  we note that the full symplectic
potential for the 4-manifold $\M_B$ has precisely the structure of
what is  there called \textit{natural symplectic potential}
\begin{equation}\label{naturalia}
    G_{natural}= \sum_{\ell=1}^r \, c_\ell  \,
    \mathit{p}_\ell\left(\mathfrak{u},\vv \right)\times \log
    \left[\mathit{p}_\ell\left(\mathfrak{u},\vv \right)\right]
    \quad ;\quad
    \mathit{p}_\ell\left(\mathfrak{u},\vv \right) =
    \text{linear functions of the moments}
\end{equation}
{(with $r=4$ in our case).} The only difference is that in
\cite{abreu} the coefficients $c_\ell$ are all equal while here they
differ one from the other in a precise way that depends on the
parameters $\lambda_1$, $\lambda_2$ defining the metric and the
argument of the logarithms. As we are going to discuss later on, the
same thing  happens also for the nonKE metric on the second
Hirzebruch surface $\mathbb{F}_2$ derived from the Kronheimer
construction.
\par
Next we turn to a general discussion of the properties of the
metrics in $\operatorname{Met}(\mathcal{FV})$ and to the organization of the latter
into special subfamilies. This will also clarify the precise
location of the KE metrics in the general landscape.
\section{Properties of the   family of metrics}
Let us then perform a complete study of the considered class of
4-dimensional metrics.  We do not start from  a given manifold but
rather from the family of metrics $
\operatorname{Met}(\mathcal{FV})$ parameterized by the choice of the function
$\mathcal{FK}(\vv )$ of one variable $\vv $, given explicitly in
coordinate form. The first tasks we are confronted with are the
definition of the maximal extensions of our coordinates, and the
search of possible singularities in the metric and/or in the
Riemannian curvature, which happens to be the cleanest probing tool.
Secondly we can calculate integrals of the Ricci and K\"ahler
2-forms. All this information is easily computed since everything
reduces to the evaluation of a few integral-differential functionals
of the function $\mathcal{FK}(\vv )$. Thirdly, we can construct
geodesics relatively to the given metric and try to explore their
behavior. This is probably the finest and most accurate tool to
visualize the geometry of a manifold. We can also integrate the
complex structure and  find explicitly the complex coordinates.
\par
We  begin by observing that all the metrics deriving from the
symplectic potential defined by eq.s
\eqref{GBsymplectic},\eqref{lupetto}\footnote{In this section which
deals only with the base manifold $\M_B$ and where there is no risk
of confusion we drop the suffix $0$ in the moment variables, in
order to make formulas simpler.} admit a general form which we
display below:
\begin{equation}\label{metrauniversala}
    ds^2_{\M_B}= \frac{d\vv ^2}{\mathcal{FK}(\vv )}\, + \,\mathcal{FK}(\vv ) \left[d\phi  (1-\cos \theta )+d\tau
   \right]^2\,+\,\vv
   \underbrace{\left(d\phi^2 \sin ^2\theta +d\theta
   ^2\right)}_{S^2 \, \text{metric}}
\end{equation}
where we have defined
\begin{equation}\label{FKfunzione}
   \mathcal{FK}(\vv )=\frac{2 \vv }{2 \vv  \mathcal{D}''(\vv )+1}
\end{equation}
This expression for the metric is obtained performing a convenient
change of variable:
\begin{equation}\label{cambiovariabile}
    \mathfrak{u}\, \rightarrow \, \left(1 -\cos\theta\right)\,\vv \quad ;
    \quad \theta \, \in \, \left[0,\pi\right]
\end{equation}
which automatically takes into account that $\mathfrak{u}\leq 2\,\vv
$. Furthermore this  change of variables clearly reveals that all
the three dimensional sections of $\M_B$ obtained by fixing $\vv
=\text{const}$ are $S^1$ fibrations on $S^2$ which is consistent
with the isometry $\mathrm{SU(2)} \times \mathrm{U(1)}$. Indeed all
the spaces $\M_B$ have cohomogeneity equal to one and the moment
variable $\vv $ is the only one whose dependence is not fixed by
isometries.
\par
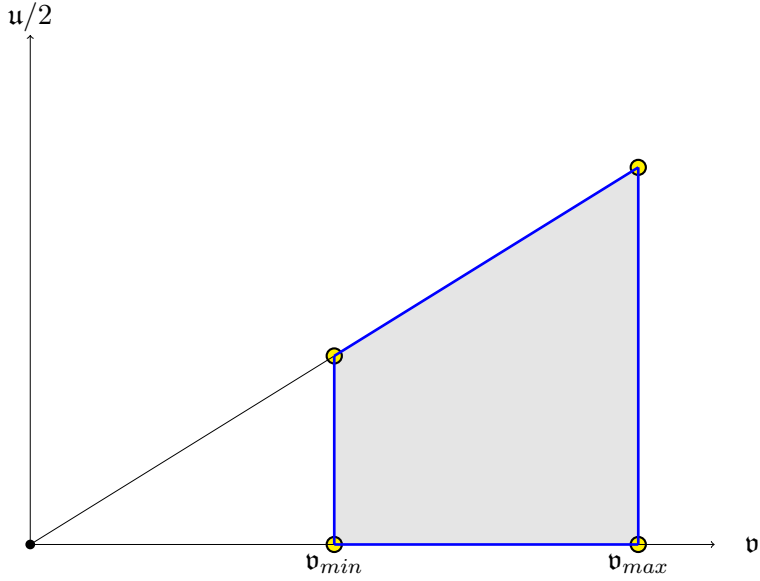
\begin{figure}
\begin{center}
\begin{tikzpicture}[scale=0.50]
\path [fill=gray,opacity=0.2] (-4,-7.5) to (4,-7.5) to  (4,2.5) to
(-4,-2.5) to (-4,-7.5); \node at (4,-8) {$\vv _{max}$}; \node at
(-4,-8) {$ \vv _{min}$}; \draw [thick] [fill=yellow] (-4,-7.5)
circle (0.2 cm) ; \draw [thick] [fill=yellow] (4,-7.5) circle (0.2
cm) ; \draw [thick] [fill=yellow] (4,2.5) circle (0.2 cm); \draw
[thick] [fill=yellow] (-4,-2.5) circle (0.2 cm); \draw [thick]
[fill=black] (-12,-7.5) circle (0.1 cm); \draw [black,line
width=0.03] (-12,-7.5)--(4,2.5); \draw [black,line width=0.03][->]
(-12,-7.5)--(6,-7.5); \draw [black,line width=0.03][->]
(-12,-7.5)--(-12,6);
 \draw [blue,line width=1](-4,-7.5) --(4,-7.5);
\draw[blue,line width=1](-4,-7.5)--(-4,-2.5); \draw[blue,line
width=1](-4,-2.5)--(4,2.5); \draw[blue,line
width=1](4,-7.5)--(4,2.5); \node at (7,-7.5) {$\vv $}; \node at
(-12,6.5) {$\mathfrak{u}/2$};
\end{tikzpicture}
\caption{\label{politoppo} The universal polytope in the $\vv
,\frac{\mathfrak{u}}{2}$ plane for all the metrics of the $\M_B$
manifolds considered in this paper and defined in equation
\eqref{metrauniversala}}
\end{center}
\end{figure}

The next important point is that the metric \eqref{metrauniversala}
is positive definite only in the interval of the positive $\vv
$-axis where $\mathcal{FK}(\vv )\geq 0$. Let us name the lower and
upper endpoints of such  interval $\vv _{min}$ and $\vv _{max}$,
respectively. If the interval $\left[\vv _{min},\vv _{max}\right]$
is finite, then the space $\M_{B}$ is compact and the domain of the
coordinates $\mathfrak{u},\vv $ is provided by the trapezoidal
polytope displayed in Figure \ref{politoppo}.
\par
Our two main examples, which both correspond to the same universal
polytope of Figure  \ref{politoppo}, are provided by the case of the
one-parameter family of \textit{Kronheimer metrics} on the
$\mathbb{F}_2$-surface, studied in
\cite{Bianchi:2021,noietmarcovaldo}, whose K\"ahler potential was
recalled in eqn.~\eqref{genKron}) and by the family
$\operatorname{Met}(\mathcal{FV})_{ext}$ of \textit{extremal K\"ahler metrics} due to
Calabi and mentioned in the introduction. In addition, within the
first class, we have the degenerate case where the parameter
$\alpha$ goes to zero and the trapezoid degenerates into a triangle.
That case corresponds to the singular space $\M_B =
\mathbb{WP}[1,1,2]$ (a weighted projective plane).
\par
Extremal metrics are defined in the present cohomogeneity one case
by the differential equation (see \cite{abreu2009toric}):
\begin{equation}\label{extremality}
    \frac{\partial^2}{\partial \vv^2} \, \mathcal{R}_s(\vv)\, = \,0
\end{equation}
where $\mathcal{R}_s(\vv)$ is the scalar curvature.
\begin{table}[htb!]
\begin{equation}
\begin{array}{|l|l|l|l|}
\hline
\null&\null&\null&\null\\
 \mathcal{FK}^{\mathbb{F}_2}_{Kro}(\vv ) \, =\,
\frac{\left(1024 \vv ^2-81 \alpha ^2\right) (32 \vv -9 (3 \alpha
+4))}{16 \left(81 \alpha ^2+1024 \vv ^2-576 (3 \alpha +4) \vv
\right)} & \vv _{min} \, =\, \frac{9 \alpha }{32} & \vv _{max} \,
=\, &\null\\
\null&\null&\frac{9}{32}(3 \alpha +4)
&\alpha >0 \\
\null&\null&\null&\null\\
\hline
\null&\null&\null&\null\\
\mathcal{FK}^{\mathbb{WP}[1,1,2]}_{Kro}(\vv ) \, = \, \frac{\vv  (8
\vv -9)}{4 \vv -9} &
\vv _{min} \, =\,0 &\vv _{max} \,= \, \frac{9}{8} & \alpha =0\\
\null&\null&\null&\null\\
\hline
\null&\null&\null&\null\\
\mathcal{FK}_{ext}(\vv )\, = \, \frac{-\mathcal{A}+8 \mathcal{C}
\mathfrak{v}^3-16 \mathcal{D}
   \mathfrak{v}^4+4 \mathfrak{v}^2-2 \mathfrak{v}
   \mathcal{B}}{4 \mathfrak{v}}&\vv _{min}\, =\, \lambda^r_1 &\vv _{max} \,= \,  \lambda^r_2 & 0 < \lambda^r_1 < \lambda^r_2\\
\null&\null&\null&\null\\
\hline
\null&\null&\null&\null\\
\underbrace{\mathcal{FK}^{KE}(\vv )}_{\mathcal{B}=\mathcal{D}=0} \,
= \, -\frac{\left(\vv -\lambda _1\right) \left(\vv -\lambda
   _2\right) \left(\lambda _2 \vv +\lambda _1 \left(\lambda
   _2+\vv \right)\right)}{\left(\lambda _1^2+\lambda _2 \lambda
   _1+\lambda _2^2\right) \vv } &  \vv _{min} \, =\, \lambda_1 &  \vv _{max} \, = \,
       \lambda_2 &  0 < \lambda_1 < \lambda_2 \\
\null&\null&\null&\null\\
\hline
\null&\null&\null&\null\\
\mathcal{FK}^{KE}_{0}(\vv ) \, = \,
\frac{\vv(\lambda_2-\vv)}{\lambda_2}
 &  \vv _{min} \, =\,0 &  \vv _{max} \, = \,
       \lambda_2 &    \lambda_2  >0\\
\null&\null&\null&\null\\ \hline
\null&\null&\null&\null\\
\mathcal{FK}^{cone}(\vv ) \, = \,  \vv   & \vv _{min} \, =\, 0 &
\vv_{max} \,= \,
       \infty &  \null\\
\null&\null&\null&\null\\
\hline
\null&\null&\null&\null\\
 \mathcal{FK}^{\mathbb{F}_2}_{ext}(\vv)\, = \,
\frac{(\mathit{a}-\mathfrak{v}) (\mathit{b}-\mathfrak{v})
\left(\mathit{a}^2 (3 \mathit{b}-\mathfrak{v})+\mathit{a}
\left(\mathit{b}^2+4 \mathit{b} \mathfrak{v}+3
\mathfrak{v}^2\right)+\mathit{b} \mathfrak{v}
(\mathit{b}+\mathfrak{v})\right)}{\mathfrak{v} \left(\mathit{a}^3+3
\mathit{a}^2 \mathit{b}-3 \mathit{a}
\mathit{b}^2-\mathit{b}^3\right)}& \vv _{min} \, =\,\mathit{a} &
\vv_{max} \, = \,
       \mathit{b} & \mathit{b}>\mathit{a}>0 \\
\null&\null&\null&\null\\
\hline
     \end{array}
\end{equation}
\caption{\label{casoni} Notable choices for the function
$\mathcal{FK}$.}
\end{table}
Inserting in eq.\eqref{extremality} the expression of
$\mathcal{R}_s$ calculated later in eq.\eqref{conturbante} we obtain
the following linear differential equation of order four:
\begin{equation}\label{quartdifeq}
    \frac{\mathfrak{v}^2 \left(-\mathcal{F}\mathcal{K}^{(3)}(\mathfrak{v})\right)+2
   \mathfrak{v} \mathcal{F}\mathcal{K}''(\mathfrak{v})-2
   \mathcal{F}\mathcal{K}'(\mathfrak{v})+2}{\mathfrak{v}^3}-\frac{1}{2}
   \mathcal{F}\mathcal{K}^{(4)}(\mathfrak{v})\, = \, 0
\end{equation}
whose general integral contains four integration constants (we name
them $\mathcal{A},\mathcal{B},\mathcal{C},\mathcal{D}$) and can be
written as follows:
\begin{equation}\label{AbreuCalabfam}
    \mathcal{FK}_{ext}(\vv) \, = \, \frac{-\mathcal{A}+8 \,\mathcal{C} \mathfrak{v}^3-16 \,\mathcal{D} \mathfrak{v}^4+4
   \mathfrak{v}^2-2 \mathfrak{v}\, \mathcal{B}}{4 \mathfrak{v}}
\end{equation}
The explicit expression \eqref{AbreuCalabfam} is very much inspiring
and useful. The function $\mathcal{FK}_{ext}(\vv)$ is rational and
it is the quotient of a quartic polynomial with a fixed coefficient
of the quadratic term divided by the linear polynomial $4\,\vv$. A
convenient way of parameterizing the entire family of metrics is
therefore in terms of the four roots
$\lambda_1,\lambda_2,\lambda_3,\lambda_4$, as we did in section
\ref{famigliaKE} for the cubic polynomial of eq.\eqref{racete} (see
eq. \eqref{rinominopara}). Combining eq.\eqref{miraculo} with eq.s
\eqref{fungemistero} and \eqref{FKfunzione} we obtain the expression
of the function $\mathcal{FK}^{KE}$ corresponding to the  KE
metrics:
\begin{equation}\label{pagnatta}
    \mathcal{FK}^{KE}(\vv) \, = \,\frac{3 \beta -k \mathfrak{v}^3+3 \mathfrak{v}^2}{3 \mathfrak{v}}
\end{equation}
Comparing eq.~\eqref{pagnatta} with eq.~\eqref{AbreuCalabfam} we see
that the KE metrics belong to the family of extremal metrics  and are
singled out by the constraint
\begin{equation}\label{immergokello}
    \mathcal{A}\, = \, -4\,\beta \quad ; \quad \mathcal{B} \, = \, 0
    \quad ; \quad \mathcal{C}\, = \, \frac{k}{6} \quad ; \quad
    \mathcal{D} \, = \, 0
\end{equation}
The most relevant aspect of the above eq.~\eqref{immergokello} is
the suppression of the quartic and linear terms
($\mathcal{D}\,=\,\mathcal{B}\, = \, 0$), which fixes the number of
free roots to two, as we know, the third being fixed in terms of
$\lambda_1,\lambda_2$.
 
We have summarized the relevant choices of the function
$\mathcal{FK}(\vv)$ in Table \ref{casoni}. Inspecting this  table
we see that the functions $\mathcal{FK}_{Kro}^{\mathbb{F}_2}(\vv )$
and $\mathcal{FK}^{KE}(\vv )$ show strict similarities but also a
difference which is expected to account for different topologies. In
both cases the function $\mathcal{FK}(\vv )$ is the ratio of a cubic
polynomial having three real roots, two positive and one negative,
and of a denominator that has no zeros in the $\left[\vv _{min},\vv
_{max}\right]$ interval. In the KE case there is a simple pole at
$\vv =0$ while for $\mathbb{F}_2$ (which is not KE) the denominator
has two zeros and therefore $\mathcal{FK}(\vv )$ has two simple
poles  at
\begin{equation}\label{baldrino}
  \vv _{poles} =  \frac{9}{32} \left[\left(3 \alpha +4\right)\pm 2 \sqrt{2} \sqrt{\alpha ^2+3 \alpha
   +2}\right]
\end{equation}
These poles are  out of the interval
$\left[\vv_{min},\vv_{max}\right]$ for any positive $\alpha >0$,
namely these poles  do not correspond to points of  the manifold
$\M_B$, just as it is the case for the single pole $\vv =0$ in the
KE case. One also notes that the function
$\mathcal{FK}^{\mathbb{F}_2}_{ext}(\vv)$ cannot be reduced to the
form \eqref{AbreuCalabfam} by any choice of the parameters
$\mathcal{A},\mathcal{B},\mathcal{C},\mathcal{D}$, so that the
smooth K\"ahler metric induced on the second Hirzebruch surface by
the Kronheimer construction (\cite{Bianchi:2021,noietmarcovaldo})
is not an extremal metric. A consistency check comes from the
evaluation on $\mathcal{FK}^{\mathbb{F}_2}_{ext}(\vv)$ of the scalar
curvature provided by eq.\eqref{conturbante}. In this case the
scalar curvature is in no way linear in $\vv$, being a rational
function of degree $6$ in the numerator and of degree $7$ in the
denominator. The same is true of the limiting case $\alpha \to 0$.
The last   case ($\mathcal{FK}(\mathfrak v)=\mathfrak v$)
corresponds to a metric cone on the $3$-sphere, i.e., $\C^2/\Z_2$
with a flat metric. The case $\mathcal{FK}_0^{KE}$ will be discussed
in Section \ref{singularity}.
\par
Finally in table \ref{casoni} we observe the choice of the function
\begin{equation}\label{candela}
\mathcal{FK}^{\mathbb{F}_2}_{ext}(\vv)\, = \,
\frac{(\mathit{a}-\mathfrak{v}) (\mathit{b}-\mathfrak{v})
\left(\mathit{a}^2 (3 \mathit{b}-\mathfrak{v})+\mathit{a}
\left(\mathit{b}^2+4 \mathit{b} \mathfrak{v}+3
\mathfrak{v}^2\right)+\mathit{b} \mathfrak{v}
(\mathit{b}+\mathfrak{v})\right)}{\mathfrak{v} \left(\mathit{a}^3+3
\mathit{a}^2 \mathit{b}-3 \mathit{a}
\mathit{b}^2-\mathit{b}^3\right)}
\end{equation}
That above in eq.\eqref{candela} is a particular case of the general
case $\mathcal{FK}_{ext}(\vv )$, corresponding to the following
choice of the parameters:
\begin{eqnarray}\label{cromotalpa}
 \mathcal{A}& =&  -\frac{4 \mathit{a}^2 \mathit{b}^2 (3
   \mathit{a}+\mathit{b})}{\mathit{a}^3+3 \mathit{a}^2 \mathit{b}-3 \mathit{a}
   \mathit{b}^2-\mathit{b}^3}\quad ; \quad \mathcal{B}\, = \, \frac{8 \mathit{a}^3
   \mathit{b}}{\mathit{a}^3+3 \mathit{a}^2 \mathit{b}-3 \mathit{a}
   \mathit{b}^2-\mathit{b}^3}\nonumber\\
 \mathcal{C}& =& -\frac{2 \mathit{a}^2}{\mathit{a}^3+3
   \mathit{a}^2 \mathit{b}-3 \mathit{a} \mathit{b}^2-\mathit{b}^3}\quad ; \quad
   \mathcal{D}\, = \,
   \frac{3 \mathit{a}+\mathit{b}}{4 \left(-\mathit{a}^3-3 \mathit{a}^2 \mathit{b}+3
   \mathit{a} \mathit{b}^2+\mathit{b}^3\right)}
\end{eqnarray}
where the parameters $\mathit{a},\mathit{b}$ are real, positive and
naturally ordered $\mathit{b}>\mathit{a}>0$. Wherefrom does the
special form \eqref{cromotalpa} originate? We claim that the metric
defined by the function \eqref{candela} is a smooth metric on the
smooth $\mathbb{F}_2$ surface. The algebraic constraints that reduce
the four parameters
$\mathcal{A},\mathcal{B},\mathcal{C},\mathcal{D}$ to the form
\eqref{cromotalpa} are derived from the conditions, already
preliminarily  discussed in \cite{noietmarcovaldo} and specifically
worked out in section 8.2.2 of \cite{Bianchi:2021}, on the
periods of the Ricci two-form localized on the standard toric
homology cycles $C_1$ and $C_2$ of$\mathbb{F}_2$. Utilizing eq.s
\eqref{ricciuto} and \eqref{fantastilione3} derived in the next
section \ref{quattrogambe} we have (see eq.(8.14) of
\cite{Bianchi:2021}):
\begin{equation}\label{cracchiato}
    \mathbb{R}ic\mid_{C_1} \, = \,\mathfrak{A}(\vv) \, \sin[\theta] \,
  d\theta\wedge d\phi \quad ; \quad \mathbb{R}ic\mid_{C_2} \, = \, \mathfrak{C}(\vv) \,
  d\theta\wedge d\phi
\end{equation}
the relevant functions being given in \eqref{fantastilione3}. The
conditions on the periods are as follows:
\begin{equation}\label{craniodibronzo}
   \frac{1}{2\pi}\, \int_{C_1}\mathbb{R} \, = \, 0 \quad ; \quad
   \frac{1}{2\pi}\, \mathbb{R} \, = \, 2
\end{equation}
that, as shown in section 8.2.2 of are automatically verified in
\cite{Bianchi:2021}) are automatically verified by the Kronheimer
metric.
\par
The two conditions \eqref{craniodibronzo} become two statements on
the functions $\mathfrak{A}(\vv),\mathcal{D}(\vv)$ defined in eq.
\eqref{fantastilione3} and completely determined in terms of the
function $\mathcal{FK}(\vv)$ and its derivatives:
\begin{equation}\label{parcondicio}
    \mathfrak{A}[\vv_{min}]\, = \, 0 \quad ; \quad
    \int_{\vv_{min}}^{\vv_{max}}\mathfrak{D}[\vv]d\vv \, = \, 2
\end{equation}
The result provided in eq. \eqref{candela} corresponding  to the
parameter choice \eqref{cromotalpa} is deduced in the following way.
First we re-parameterize the function \eqref{AbreuCalabfam} in terms
of the four roots of the quartic polynomial appearing in the
numerator that we name $\mu_1,\mu_2,\mu_3,\mu_4$ obtaining:
\begin{align}\label{torriano}
&\mathcal{FK}_{ext}(\vv) \, = \, \\
&\frac{\mu _1 \left(\mathfrak{v} \left(\mu _3+\mathfrak{v}\right)
\left(\mu
   _4+\mathfrak{v}\right)+\mu _2 \left(\mu _3 \left(\mathfrak{v}-\mu
   _4\right)+\mathfrak{v} \left(\mu _4+\mathfrak{v}\right)\right)\right)}{2
   \mathfrak{v} \left(\mathfrak{v}-\mu _1\right) \left(\mathfrak{v}-\mu _2\right)
   \left(\mathfrak{v}-\mu _3\right) \left(\mathfrak{v}-\mu
   _4\right)} +\\
&\frac{\mu _2 \left(\mu _3+\mathfrak{v}\right) \left(\mu
   _4+\mathfrak{v}\right)+\mathfrak{v} \left(\mathfrak{v} \left(\mu
   _4-\mathfrak{v}\right)+\mu _3 \left(\mu _4+\mathfrak{v}\right)\right)}{2
   \left(\mathfrak{v}-\mu _1\right) \left(\mathfrak{v}-\mu _2\right)
   \left(\mathfrak{v}-\mu _3\right) \left(\mathfrak{v}-\mu _4\right)}
\end{align}
Secondly we rename $\mu_2 = a$,$\mu_3 =b$ deciding that
$0<a<b<\infty$ and we calculate the two conditions
\eqref{parcondicio} using the function $\mathcal{FK}_{ext}(\vv)$ in
eq.\eqref{torriano} as an input. We get a system of quadratic
algebraic equations for the remaining roots $\mu_1,\mu_4$ that has
the following solutions
\begin{align}
&\mu_2 \, = \, \mathit{a} \quad ; \quad \mu_3 \, = \, \mathit{b} \\
&\mu _1 \, = \, \frac{\mathit{a}^2-\left(\mathit{b}^2\pm
\sqrt{\mathit{a}^4-44
   \mathit{a}^3 \mathit{b}-10 \mathit{a}^2 \mathit{b}^2+4 \mathit{a}
   \mathit{b}^3+\mathit{b}^4}\right)-4 \mathit{a} \mathit{b}}{6 \mathit{a}+2
   \mathit{b}}\\
&\mu _4 \, = \, \frac{\mathit{a}^2-\left(\mathit{b}^2\mp
   \sqrt{\mathit{a}^4-44 \mathit{a}^3 \mathit{b}-10 \mathit{a}^2 \mathit{b}^2+4
   \mathit{a} \mathit{b}^3+\mathit{b}^4}\right)-4 \mathit{a} \mathit{b}}{6
   \mathit{a}+2 \mathit{b}} \label{coffele}
\end{align}
Substitution of eq.\eqref{coffele} into eq. \eqref{torriano}
produces the function $\mathcal{FK}^{\mathbb{F}_2}_{ext}(\vv)$
presented in \eqref{torriano} and recalled in table \ref{casoni}.
Furthermore
  long as the roots $\mu _1,\mu_4$ as given above roots are complex conjugate of
each other or, being real, do not fall in the interval $[a,b]$, the
K\"ahler metric generated by the function
$\mathcal{FK}^{F_2}_{ext}(\vv)$ is smooth and well defined on the
second Hirzebruch surface $\mathbb{F}_2$. The domain where this
happens in the plane $\mathit{a},\mathit{b}$ can be easily studied
looking at the discriminant under the square root in
\eqref{coffele}.
\subsection{Vielbein formalism and the curvature 2-form of
$\M_B$}\label{quattrogambe} The metric \eqref{metrauniversala} is in
diagonal form so it is easy to write a set of vierbein 1-forms.
Indeed if we set
\begin{equation}\label{vierbeine}
    \mathbf{e}^i = \left\{\frac{d\vv }{\sqrt{\mathcal{FK}(\vv )}},\,\sqrt{\mathcal{FK}(
   \vv )}\, \left[d\phi  (1-\cos \theta )+d\tau
   \right],\,\sqrt{\vv }\,
   d\theta ,\,\sqrt{\vv }\, d\phi \,\sin \theta\right\}
\end{equation}
the line element \eqref{metrauniversala} reads
\begin{equation}\label{vilbone}
    ds^2_B = \sum_{i=1}^4 \, \mathbf{e}^i \otimes \mathbf{e}^i
\end{equation}
Furthermore we can calculate the matrix vielbein and its  inverse
quite  easily, obtaining:
\begin{equation}\begin{array}{rcl}
  \mathbf{e}^i &\ = & E^i_\mu \, dy^\mu \quad ; \quad y^\mu \, = \{\vv ,\theta ,\phi ,\tau \}   \\
  E^i_\mu &=& \left(
\begin{array}{cccc}
 \frac{1}{\sqrt{\mathcal{FK}(\vv )}} & 0 & 0 & 0 \\
 0 & 0 & \sqrt{\mathcal{FK}(\vv )} (1-\cos \theta ) &
   \sqrt{\mathcal{FK}(\vv )} \\
 0 & \sqrt{\vv } & 0 & 0 \\
 0 & 0 & \sqrt{\vv } \sin \theta  & 0 \\
\end{array}
\right)   \\
  E^\nu_j &=& \left(
\begin{array}{cccc}
 \sqrt{\mathcal{FK}(\vv )} & 0 & 0 & 0 \\
 0 & 0 & \frac{1}{\sqrt{\vv }} & 0 \\
 0 & 0 & 0 & \frac{\csc (\theta )}{\sqrt{\vv }} \\
 0 & \frac{1}{\sqrt{\mathcal{FK}(\vv )}} & 0 & \frac{(\cos \theta -1)
   \csc \theta }{\sqrt{\vv }} \\
\end{array}
\right)
\end{array}\end{equation}
By means of the {\sc mathematica} package {\sc
Vielbgrav23}\footnote{{\sc Vielbgrav23} is a MATHEMATICA  package for the calculation of the spin connection the
curvature 2-form and the intrinsic components of the Riemann tensor
in vielbein formalism. Constantly updated, it was originally written by
one us (P.F.), almost thirty years ago. It can be furnished upon
request and it will be at disposal on the the De Gruyter site for
the readers of the forthcoming book \cite{nuovogruppo}.} we can
easily calculate the Levi-Civita spin connection and the curvature
2-form from the definitions
\begin{equation}\label{radicchione}
    0 = \mathfrak{T}^i= d\mathbf{e}^i \, + \, \omega^{ij} \, \wedge \,
    \mathbf{e}^j \quad ; \quad \mathfrak{R}^{ij}=d\omega^{ij}
    \, + \, \omega^{ik} \, \wedge \, \omega^{kj} =
    \mathcal{R}^{ij}_{\phantom{ik}k\ell} \,\mathbf{e}^k \, \wedge \,
    \mathbf{e}^\ell
\end{equation}
obtaining
\begin{equation}\begin{array}{rcl}\label{Rdueforma}
  \mathfrak{R}^{12} &=& -\frac{\mathcal{FK}''(\vv )}{2} \mathbf{e}^{1}\wedge
  \mathbf{e}^{2}
   -\frac{
   \left(\vv
   \mathcal{FK}'(\vv )-\mathcal{FK}(\vv )\right)}{2 \vv ^2}\,\mathbf{e}^{3}\wedge
   \mathbf{e}^{4}  \\[5pt]
  \mathfrak{R}^{13} &=& -\frac{\left(\vv
   \mathcal{FK}'(\vv )-\mathcal{FK}(\vv )\right)}{4
   \vv ^2}\, \mathbf{e}^{1}\wedge \mathbf{e}^{3} \, -\frac{\left(\vv
   \mathcal{FK}'(\vv )-\mathcal{FK}(\vv )\right)}{4 \vv ^2}\, \mathbf{e}^{2}\wedge \mathbf{e}^{4}  \\[5pt]
  \mathfrak{R}^{14} &=& \frac{\left(\vv
   \mathcal{FK}'(\vv )-\mathcal{FK}(\vv )\right)}{4
   \vv ^2}\, \mathbf{e}^{2}\wedge \mathbf{e}^{3}  \, -\, \frac{\left(\vv
   \mathcal{FK}'(\vv )-\mathcal{FK}(\vv )\right)}{4 \vv ^2} \, \mathbf{e}^{1}\wedge \mathbf{e}^{4}  \\[5pt]
  \mathfrak{R}^{23} &=& \frac{ \left(\vv
   \mathcal{FK}'(\vv )-\mathcal{FK}(\vv )\right)}{4
   \vv ^2}\, \mathbf{e}^{1}\wedge \mathbf{e}^{4}\, -\,\frac{\left(\vv
   \mathcal{FK}'(\vv )-\mathcal{FK}(\vv )\right)}{4 \vv ^2}\,\mathbf{e}^{2}\wedge \mathbf{e}^{3}   \\[5pt]
  \mathfrak{R}^{24}&=& -\frac{\left(\vv
   \mathcal{FK}'(\vv )-\mathcal{FK}(\vv )\right)}{4
   \vv ^2}\,\mathbf{e}^{1}\wedge \mathbf{e}^{3} \, -\, \frac{\left(\vv
   \mathcal{FK}'(\vv )-\mathcal{FK}(\vv )\right)}{4 \vv ^2} \, \mathbf{e}^{2}\wedge \mathbf{e}^{4}  \\[5pt]
  \mathfrak{R}^{34} &=& \frac{
   (\vv -\mathcal{FK}(\vv ))}{\vv ^2}\, \mathbf{e}^{3}\wedge \mathbf{e}^{4}\, -\frac{ \left(\vv
   \mathcal{FK}'(\vv )-\mathcal{FK}(\vv )\right)}{2
   \vv ^2} \, \mathbf{e}^{1}
   \wedge \mathbf{e}^{2}
\end{array}\end{equation}
Equation \eqref{Rdueforma} shows that the   Riemann tensor
$\mathcal{R}^{ij}_{\phantom{ik}k\ell}$ is constructed  in terms of
only three functions:
\begin{equation}\label{trefunzie}
    \mathcal{CF}_1(\vv ) = \mathcal{FK}''(\vv ) \quad ; \quad
    \mathcal{CF}_2(\vv )= \frac{\left(\vv
   \mathcal{FK}'(\vv )-\mathcal{FK}(\vv )\right)}{
   \vv ^2} \quad ; \quad \mathcal{CF}_3(\vv )=\frac{(\vv -\mathcal{FK}(\vv ))}{\vv ^2}
\end{equation}
If  these functions  are regular in the interval $\left[\vv
_{min},\vv _{max}\right]$ the Riemann tensor is well defined and
finite in the entire polytope of Figure \ref{politoppo} and $\M_B$
should be a smooth compact manifold. From the expression
\eqref{Rdueforma} the {\sc Mathematica Code}  immediately derives
the Riemann and Ricci tensors and the curvature scalar. This latter
reads as follows:
\begin{equation}\label{conturbante}
    \mathcal{R}_s \, = \, -\frac{\mathfrak{v} \mathcal{F}\mathcal{K}''(\mathfrak{v})+2
   \mathcal{F}\mathcal{K}'(\mathfrak{v})-2}{2 \mathfrak{v}}
\end{equation}
and its form was used above to define the extremal metrics.
Similarly in the anholonomic vielbein basis, the Ricci tensor takes
the following form:
\begin{eqnarray}\label{riccetto}
   &&\mathcal{R}_{ij} \, = \, \nonumber\\
   &&\left(
\begin{array}{cccc}
 \frac{\mathcal{F}\mathcal{K}(\mathfrak{v})-\mathfrak{v} \left(\mathfrak{v}
   \mathcal{F}\mathcal{K}''(\mathfrak{v})+\mathcal{F}\mathcal{K}'(\mathfrak{v})\right)}
   {4 \mathfrak{v}^2} & 0 & 0 & 0 \\
 0 & \frac{\mathcal{F}\mathcal{K}(\mathfrak{v})-\mathfrak{v} \left(\mathfrak{v}
   \mathcal{F}\mathcal{K}''(\mathfrak{v})+\mathcal{F}\mathcal{K}'(\mathfrak{v})\right)}
   {4 \mathfrak{v}^2} & 0 & 0 \\
 0 & 0 & -\frac{\mathfrak{v}
   \left(\mathcal{F}\mathcal{K}'(\mathfrak{v})-2\right)
   +\mathcal{F}\mathcal{K}(\mathfrak{v})}{4 \mathfrak{v}^2} & 0 \\
 0 & 0 & 0 & -\frac{\mathfrak{v}
   \left(\mathcal{F}\mathcal{K}'(\mathfrak{v})-2\right)
   +\mathcal{F}\mathcal{K}(\mathfrak{v})}{4 \mathfrak{v}^2} \\
\end{array}
\right)
\end{eqnarray}
All the metrics in the considered family are of cohomogeneity one
and have the same isometry, furthermore they are all K\"ahler and
share the same K\"ahler 2-form that can be written as it follows:
\begin{equation}\label{kalleroduef}
    \mathbb{K} = d\mathfrak{u}\wedge d\phi \, + \, d\vv  \wedge
    d\tau = \mathbf{e}^1 \wedge \mathbf{e}^2 \, + \, \mathbf{e}^3 \wedge
    \mathbf{e}^4 = \frac{1}{2} \, \mathfrak{J}_{ij}\, \mathbf{e}^i \, \wedge \,
    \mathbf{e}^j
\end{equation}
where:
\begin{equation}\label{complessostrut}
    \mathfrak{J}^i_{\phantom{i}j} = \left(
\begin{array}{cccc}
 0 & 1 & 0 & 0 \\
 -1 & 0 & 0 & 0 \\
 0 & 0 & 0 & 1 \\
 0 & 0 & -1 & 0 \\
\end{array}
\right)=  \delta^{ik} \, \mathfrak{J}_{kj}
\end{equation}
is the complex structure in flat indices. Utilizing
$\mathfrak{J}^i_{\phantom{i}j}$ the Ricci 2-form is defined by:
\begin{equation}\label{canzellero}
    \mathbb{R}\mathrm{ic} = \mathbb{R}_{ij} \, \mathbf{e}^i \wedge \mathbf{e}^j
    \quad ; \quad \mathbb{R}_{ij}= \mathcal{R}_{i\ell} \, \mathfrak{J}^\ell_{\phantom{\ell}j}
\end{equation}
and explicitly one obtains:
\begin{equation}\label{ricciuto}
  \mathbb{R}\mathrm{ic} \, = \,   \mathfrak{A}(\vv) \, \sin[\theta] \,
  d\theta\wedge d\phi \, + \, \mathfrak{B}(\vv) \,\left(1\, -\,
  \cos[\theta] \right) d\vv\wedge d\phi + \mathfrak{C}(\vv) \, d\vv\wedge
  d\tau
\end{equation}
In eq. \eqref{ricciuto} the functions of $\vv$ are the following
ones:
\begin{align}
\mathfrak{A}(\vv) &= -\frac{\mathfrak{v}
\left(\mathcal{F}\mathcal{K}'(\mathfrak{v})-2\right)+\mathcal{F}\mathcal{K}(\mathfrak{v})}{2 \mathfrak{v}}\\
\mathfrak{B}(\vv) &=
-\frac{\mathcal{F}\mathcal{K}(\mathfrak{v})-\mathfrak{v}
\left(\mathfrak{v}
\mathcal{F}\mathcal{K}''(\mathfrak{v})+\mathcal{F}\mathcal{K}'(\mathfrak{v})\right)}
{2 \mathfrak{v}^2} \\
\mathfrak{C}(\vv) &=  \frac{\mathfrak{v}^2
\left(-\mathcal{F}\mathcal{K}''(\mathfrak{v})\right)-\mathfrak{v}
\mathcal{F}\mathcal{K}'(\mathfrak{v})+\mathcal{F}\mathcal{K}(\mathfrak{v})}{2
\mathfrak{v}^2} \label{fantastilione3}
\end{align}
\par
\begin{figure}
\centering
\includegraphics[width=8cm]{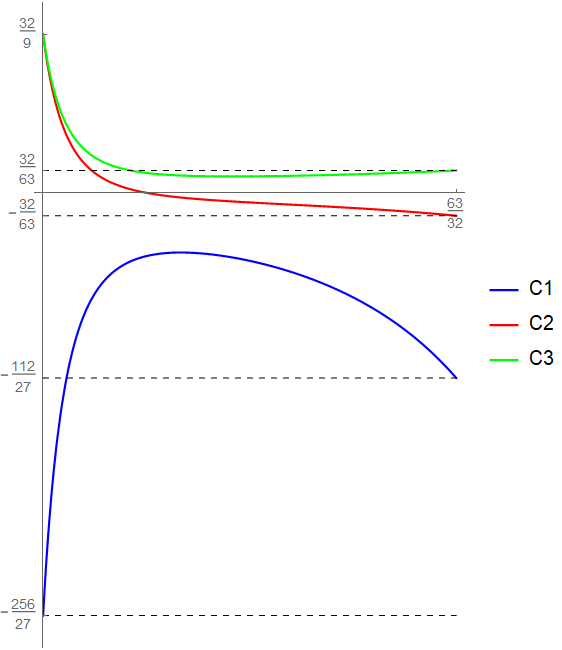}\hskip10mm\includegraphics[width=8cm]{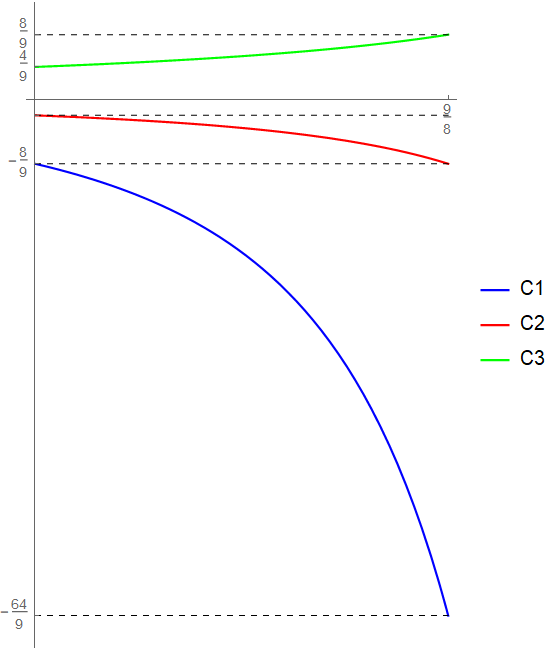}
\caption{\label{plotto123}  A (left):  Plot of the three functions
$\mathcal{CF}^{\mathbb{F}_2}_{1,2,3}(\vv )$ entering the intrinsic
Riemann curvature tensor for the   ``Kronheimer'' metric on
$\mathbb{F}_2$ with the choice of the parameter $\alpha = 1$. B
(right): Plot of the three functions
$\mathcal{CF}^{\mathbb{WW}_{112}}_{1,2,3}(\vv )$ entering the
intrinsic Riemann curvature tensor for the  Kronheimer metric on
$\mathbb{WP}[1,1,2]$ with the choice of the parameter $\alpha = 0$.
Comparing this picture with the one on the left we see the
discontinuity. In all smooth cases the functions
$\mathcal{CF}^{\mathbb{F}_2}_{2,3}$ attain the same value in the
lower endpoint of the interval while for the singular case of the
weighted projective space, the initial values of
$\mathcal{CF}^{\mathbb{WP}[1,1,2]}_{2,3}(\vv )$ are different.}
\end{figure}

\paragraph{The   $\mathbb{F}_2$ Kronheimer case.}
In the   $\mathbb{F}_2$ case with the ``Kronheimer''  metric we
have:
\begin{equation}\begin{array}{rcl}
  \mathcal{CF}^{\mathbb{F}_2}_1(\vv ) &=& \frac{331776 (\alpha +1) (\alpha +2) \left(729 \alpha ^2 (3 \alpha +4)+16384
   \vv ^3-3888 \alpha ^2 \vv \right)}{\left(81 \alpha ^2+1024
   \vv ^2-576 (3 \alpha +4) \vv \right)^3}  \\[5pt]
\mathcal{CF}^{\mathbb{F}_2}_2(\vv )&=& -\frac{9}{32
   \vv ^2 \left(81 \alpha ^2+1024 \vv ^2-576 (3 \alpha +4)
   \vv \right)^2}\, \times \, \
\left(6561 \alpha ^4 (3 \alpha +4)+1048576 (3 \alpha +4)
   \vv ^4\right. \\[5pt]
   &&\left.-1179648 \alpha ^2 \vv ^3
    +497664 \alpha ^2 (3 \alpha
   +4) \vv ^2-93312 \alpha ^2 (3 \alpha +4)^2 \vv \right) \\[5pt]
\mathcal{CF}^{\mathbb{F}_2}_3(\vv ) &=&
\displaystyle \frac{\vv -\displaystyle\frac{\left(1024 \vv ^2-81 \alpha
^2\right) (32
   \vv -9 (3 \alpha +4))}{16 \left(81 \alpha ^2+1024
   \vv ^2-576 (3 \alpha +4) \vv \right)}}{\vv ^2}
\end{array}\end{equation}
The three functions
$\mathcal{CF}^{\mathbb{F}_2}_{1,2,3}(\vv )$ are
smooth in the interval $\left(\frac{9 \alpha }{32},\frac{9}{32} (3
\alpha +4) \right)$ and they are defined at the endpoints: see for
instance Figure \ref{plotto123}A.
\par
\begin{figure}
\centering
\includegraphics[width=9cm]{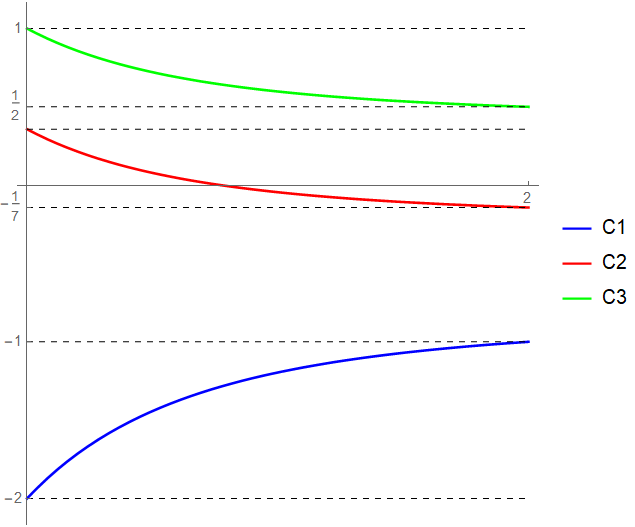}
\caption{\label{KEfunzie} Plot of the three functions
$\mathcal{CF}^{KE}_{1,2,3}(\vv )$ entering the intrinsic
Riemann curvature tensor for the  KE metric  with the
choice of the parameter $\lambda_1 = 1$, $\lambda_2 = 2$. }
\end{figure}
\begin{figure}
\centering
\includegraphics[width=7.3cm]{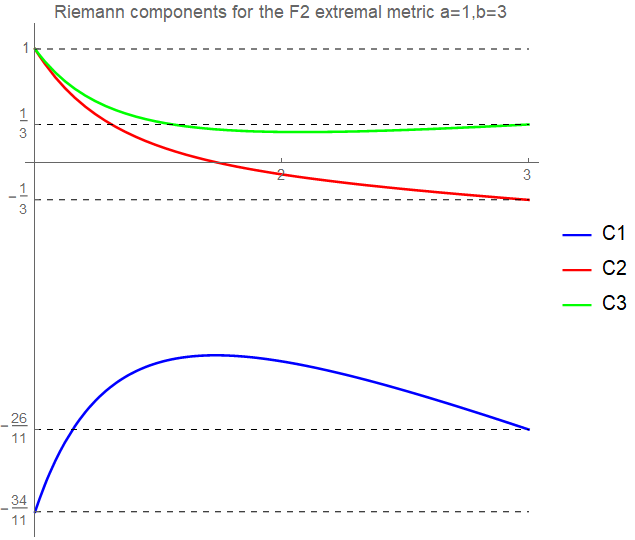}
\includegraphics[width=7cm]{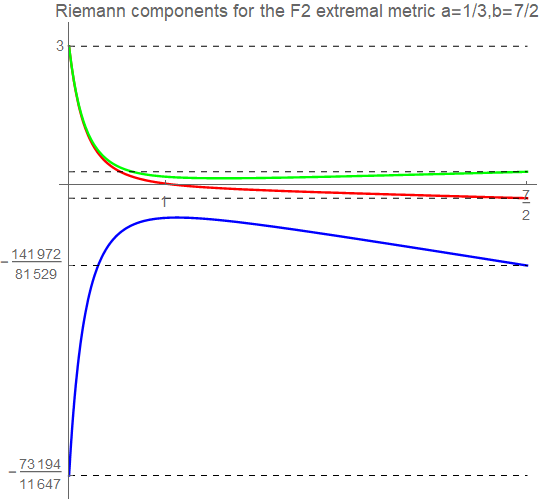}
\caption{\label{RiefunzioF2ext} Plot of the three functions
$\mathcal{CF}^{F2ext}_{1,2,3}(\vv )$ entering the intrinsic Riemann
curvature tensor for the  extremal K\"ahler metrics on $F_2$ with
with two different  choices of the parameter $\mathit{a} = 1$,
$\mathit{b} = 2$ and $\mathit{a} = 1/3$, $\mathit{b} = 7/2$. }
\end{figure}

Indeed the values of the three functions at the
endpoints   are
\begin{equation}\begin{array}{rcl}\label{parbino}
 \mathcal{CF}^{\mathbb{F}_2}_{1,2,3}\left(\vv _{min}\right)& = &  \left\{-\frac{128 (\alpha +1)}{9 \alpha  (\alpha +2)},\frac{32}{9 \alpha
   },\frac{32}{9 \alpha }\right\} \\
   \mathcal{CF}^{\mathbb{F}_2}_{1,2,3}\left(\vv _{max}\right)& = &\left\{-\frac{32 (3 \alpha +4)}{9 \left(\alpha ^2+3 \alpha
   +2\right)},-\frac{32}{27 \alpha +36},\frac{32}{9 (3 \alpha +4)}\right\}
\end{array}\end{equation}
 The singularity which might be developed by the space
corresponding to the value $\alpha=0$ is evident from eqns.~\eqref{parbino}.
The intrinsic components of the Riemann curvature
seem to have a singularity in the lower endpoint of the interval,
for $\alpha = 0$.
\paragraph{The case of the singular manifold $\mathbb{WP}[1,1,2]$.}
In the previous section we utilized the wording \textit{seem to have
a singularity} for the components of the Riemann curvature in the
case of the space $\mathbb{WP}[1,1,2]$ since  such a singularity in
the curvature actually does not exist. The space
$\mathbb{WP}[1,1,2]$ has indeed a singularity at $\vv  \, = \, 0$
but it is very mild since the intrinsic components of the Riemann
curvature are well-behaved in $\vv  = 0$ and have a finite limit. It
depends on the way one does the limit $\alpha \rightarrow 0$. If we
first compute the value of the curvature $2$-form at the  endpoints
for generic $\alpha$ and then we do the limit $\alpha \to 0$ we see
the singularity that is evident from equations \eqref{parbino}. On
the other hand, if we first reduce the function $\mathcal{FK}(\vv )$
to its $\alpha=0$ form we obtain:
\begin{equation}\label{wp112FK}
    \mathcal{FK}^{\mathbb{WP}[1,1,2]}(\vv )\, =
    \,\frac{\vv  (8 \vv -9)}{4 \vv -9}
\end{equation}
and the corresponding functions appearing in the curvature are:
\begin{equation}\label{funzie123WP}
\mathcal{CF}^{\mathbb{WP}[1,1,2]}_{1,2,3}\left(\vv \right)\,
= \, \left\{\frac{648}{(4 \vv -9)^3},-\frac{18}{(9-4
   \vv )^2},\frac{4}{9-4 \vv }\right\}
\end{equation}
which are perfectly regular in the interval $\left[0,9/8\right]$ and
have finite value at the endpoints (see Figure \ref{plotto123wp}B).
\paragraph{The case of the KE  manifolds.}
In the case of the KE metrics the function
$\mathcal{FK}(\vv )$ is
\begin{equation}\label{felino}
    \mathcal{FK}^{KE}(\vv )= -\frac{\left(\vv -\lambda _1\right) \left(\vv -\lambda
   _2\right) \left(\lambda _1 \lambda _2+\left(\lambda _1+\lambda _2\right)
   \vv \right)}{\left(\lambda _1^2+\lambda _2 \lambda _1+\lambda
   _2^2\right) \vv }
\end{equation}
and the corresponding functions entering the intrinsic components of
the Riemann curvature are
\begin{equation}\label{perdinci}
\mathcal{CF}^{KE}_{1,2,3}\left(\vv \right)=
\left\{-\frac{2 \left(\lambda _1^2 \lambda _2^2+\left(\lambda
   _1+\lambda _2\right) \vv ^3\right)}{\left(\lambda
   _1^2+\lambda _2 \lambda _1+\lambda _2^2\right)
   \vv ^3},\frac{2 \lambda _1^2 \lambda
   _2^2-\left(\lambda _1+\lambda _2\right) \vv ^3}{2
   \left(\lambda _1^2+\lambda _2 \lambda _1+\lambda _2^2\right)
   \vv ^3},\frac{\lambda _1^2 \lambda _2^2+\left(\lambda
   _1+\lambda _2\right) \vv ^3}{\left(\lambda
   _1^2+\lambda _2 \lambda _1+\lambda _2^2\right)
   \vv ^3}\right\}
\end{equation}
and the interval of variability of the moment coordinate
$\vv $ is the following
$\vv \in\left[\lambda_1,\lambda_2\right]$. Correspondingly
the  boundary values are
\begin{equation}\begin{array}{rcl}\label{pergiovetonante}
 \mathcal{CF}^{KE}_{1,2,3}\left(\vv _{min}\right)&=&\left\{-\frac{2}{\lambda _1},\frac{1}{\lambda _1}-\frac{3
   \left(\lambda _1+\lambda _2\right)}{2 \left(\lambda
   _1^2+\lambda _2 \lambda _1+\lambda
   _2^2\right)},\frac{1}{\lambda _1}\right\}  \\
  \mathcal{CF}^{KE}_{1,2,3}\left(\vv _{max}\right) &=&
  \left\{-\frac{2}{\lambda _2},\frac{1}{\lambda _2}-\frac{3
   \left(\lambda _1+\lambda _2\right)}{2 \left(\lambda
   _1^2+\lambda _2 \lambda _1+\lambda
   _2^2\right)},\frac{1}{\lambda _2}\right\} \\
\end{array}\end{equation}
We can use the case $\lambda_1=1,\lambda=2$ as a standard example.
In this case the behavior of the three functions is displayed in
Figure \ref{KEfunzie}
\paragraph{The case of the extremal K\"ahler metric on the second Hizebruch surface  $\mathbb{F}_2$.}
Finally we consider the case of the extremal metric on $F_2$
discussed in the previous pages and defined by the function
\eqref{torriano}. In this case the three functions \eqref{trefunzie}
parameterizing the curvature 2-form and hence the intrinsic
components of the Riemann tensor are the following ones:
\begin{eqnarray}
\label{piripicchio}
  \mathcal{CF}^{F2ext}_1 &=& \frac{2 \mathit{a}^2 \mathit{b}^2 (3 \mathit{a}+\mathit{b})-8 \mathit{a}^2
   \mathfrak{v}^3+6 \mathfrak{v}^4 (3 \mathit{a}+\mathit{b})}{\mathfrak{v}^3
   (\mathit{a}-\mathit{b}) \left(\mathit{a}^2+4 \mathit{a}
   \mathit{b}+\mathit{b}^2\right)} \\
  \mathcal{CF}^{F2ext}_2 &=& \frac{\mathit{a}^3 \mathit{b} (2 \mathfrak{v}-3 \mathit{b})-\mathit{a}^2
   \left(\mathit{b}^3+2 \mathfrak{v}^3\right)+3 \mathit{a} \mathfrak{v}^4+\mathit{b}
   \mathfrak{v}^4}{\mathfrak{v}^3 (\mathit{a}-\mathit{b}) \left(\mathit{a}^2+4
   \mathit{a} \mathit{b}+\mathit{b}^2\right)}\\
 \mathcal{CF}^{F2ext}_3 &=& \frac{4 \mathit{a}^3 \mathit{b} \mathfrak{v}-\mathit{a}^2 \mathit{b}^2 (3
   \mathit{a}+\mathit{b})+4 \mathit{a}^2 \mathfrak{v}^3+\mathfrak{v}^4 (-(3
   \mathit{a}+\mathit{b}))}{\mathfrak{v}^3 (\mathit{a}-\mathit{b})
   \left(\mathit{a}^2+4 \mathit{a} \mathit{b}+\mathit{b}^2\right)}
\end{eqnarray}
A plot of the three functions for the extremal $\mathbb{F}_2$
metric, to be compared with the analogous plot relative to the
Kronheimer metric (Fig. \ref{plotto123}A) on the same manifold is
shown in Fig.\ref{RiefunzioF2ext}.
\subsection{The complex structure and its integration}
\begin{figure}
\centering
\includegraphics[width=5cm]{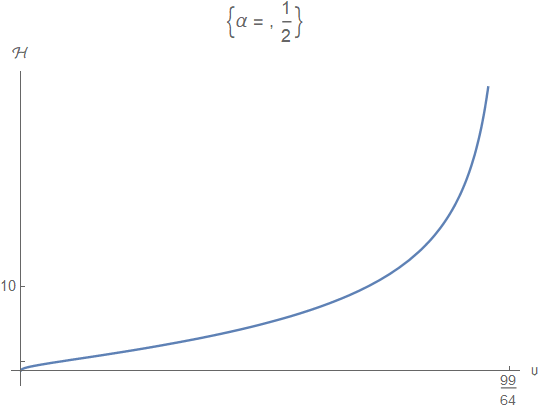}
\includegraphics[width=5cm]{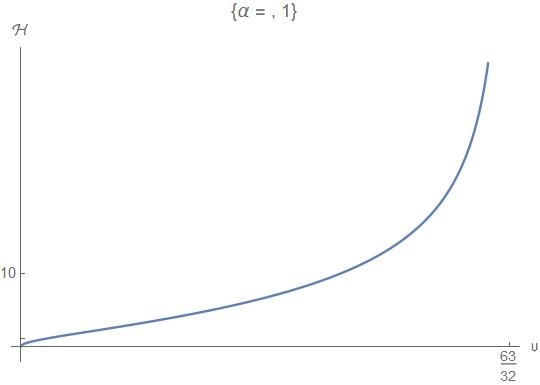}
\includegraphics[width=5cm]{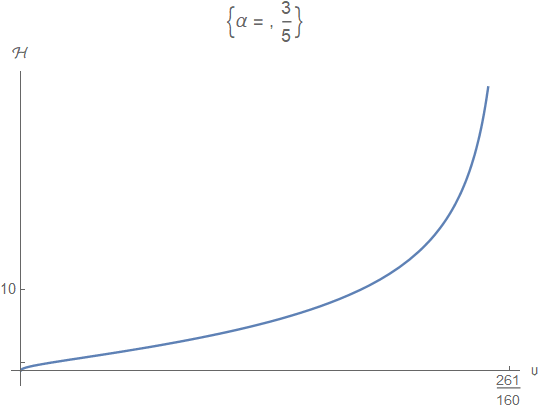}
\caption{\label{Hvfunzion} Plot of  three examples of the $
{H}^{\mathbb{F}_2}_{Kro}(\vv )$ function for three different choices
of the parameter $\alpha$.}
\end{figure}

\begin{figure}
\centering
\includegraphics[width=8cm]{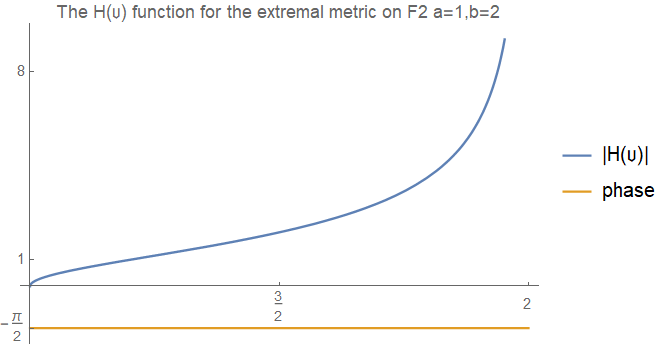}
\includegraphics[width=8cm]{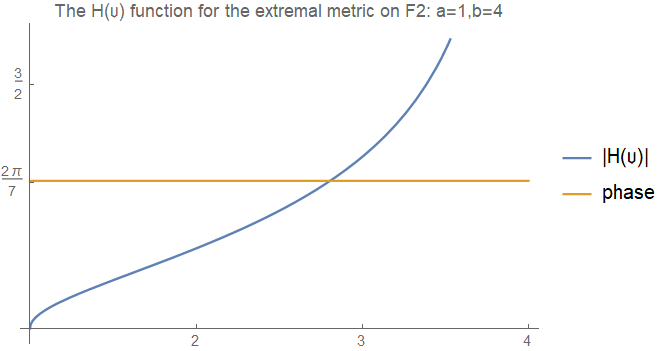}
\caption{\label{calzedonna} Plot of two examples of the $
{H}^{\mathbb{F}_2}_{ext}(\vv )$ function for two different choices
of the parameters $a,b$.}
\end{figure}

We can
easily convert the complex structure into curved indices using the
vierbein and its inverse:
\begin{equation}\label{curvJW}
    \mathfrak{J}^\mu_{\phantom{\mu}\nu}=E^\mu_i \, \mathfrak{J}^i_{\phantom{i}j}
    \,E^j_\nu =\left(
\begin{array}{cccc}
 0 & 0 & -\,\mathcal{FK}(\vv ) \left(\cos \theta +1\right) &
   \mathcal{FK}(\vv ) \\
 0 & 0 & \sin \theta  & 0 \\
 0 & -\csc \theta  & 0 & 0 \\
 -\frac{1}{\mathcal{FK}(\vv )} & \tan \frac{\theta
   }{2} & 0 & 0 \\
\end{array}
\right)
\end{equation}
Since $\mathfrak{J}^2 = - \,\mathbf{1}_{4 \times 4}$ the
eigenvalues of $\mathfrak{J}$ are $\pm i$ and the eigenvectors are
the rows of the following matrix:
\begin{equation}\label{autovettori}
    \mathfrak{a}^i_\mu = \left(
\begin{array}{cccc}
 \frac{i}{\mathcal{FK}(\vv )} & -i \tan \frac{\theta
   }{2} & 0 & 1 \\
 0 & i \csc \theta & 1 & 0 \\
 -\frac{i}{\mathcal{FK}(\vv )} & i \tan \frac{\theta
   }{2} & 0 & 1 \\
 0 & -i \csc \theta & 1 & 0 \\
\end{array}
\right)
\end{equation}
We obtain the eigen-differentials by defining:
\begin{equation}\label{eigendif}
\mathrm{da}^i \,= \mathit{i} \, \mathfrak{a}^i_\mu \, dx^\mu
\quad
    ;\quad dx^\mu =\{d\vv ,\, d\theta,\,d\phi ,\, d\tau \}
\end{equation}
The essential thing is that the eigen-differentials are all closed
and that the first two are the complex conjugate of the second two:
\begin{equation}\label{radicalione}
    \mathrm{d}\mathrm{da}^i \,= \,0 \quad (i=1,\dots,4) \quad ;
    \quad \mathrm{da}^1 =\overline{\mathrm{da}^3} \quad ;
    \quad \mathrm{da}^2 =\overline{\mathrm{da}^4}
\end{equation}
This allows us to define the two complex variables $u$ and $v$, by
setting:
\begin{equation}\begin{array}{rcl}\label{dudv}
 \mathrm{da}^3 &=& \mathrm{d}\log[v] \\
 \mathrm{da}^4 &=& \mathrm{d}\log[u]
 \end{array}\end{equation}
In this way one obtains the universal result:
\begin{equation}\label{uvcoordi}
    u =  e^{i \phi }\, \tan \,\frac{\theta}{2} \quad ; \quad v = \frac{1}{2} e^{i \tau } \left(\cos \theta +1\right)\, H(\vv )
\end{equation}
where:
\begin{equation}\label{vfunzia}
  H(\vv )=\exp\left[\int \frac{1}{\mathcal{FK}(\vv )} \,
  d\vv \right]
\end{equation}
Hence the whole difference between the various spaces is encoded in
the properties of the function $H(\vv )$ which is obviously
defined up to a multiplicative  constant due to the additive
integration constant in the exponential.
\paragraph{The
function $H(\vv )$ for the Kronheimer metric on the smooth
$\mathbb{F}_2$ surface.} In the case of the Kronheimer metric on
$\mathbb{F}_2$ we obtain
\begin{equation}\label{HF2}
    H^{\mathbb{F}_2}_{Kro}(\vv ) =\mathit{i}
    \sqrt{\frac{1024 \vv ^2-81 \alpha ^2}{-32 \vv +9 (3
   \alpha +4)}}\,
\end{equation}
The factor $\mathit{i}$ can always be reabsorbed into a shift of
$\pi/2$ of the phase $\tau$ and the function
$\mathcal{H}^{\mathbb{F}_2}(\vv )$ is positive definite in
the finite interval $\left[ \frac{9 \alpha }{32} \, , \,
\frac{9}{32} (3 \alpha +4)\right]$ and goes from 0 to $+\infty$ for
all positive values of $\alpha >0$. See Figure  \ref{Hvfunzion} for
some examples.
What is   important is the monotonic behavior of the
function $ {H}^{\mathbb{F}_2}(\vv )$, which  guarantees that the two coordinates
$u,v$ describe a copy of $\mathbb{C}^2$ and hence define a dense open
chart in the compact manifold $\mathbb{F}_2$.
\paragraph{The
function $H(\vv )$ for the extremal Calabi metric on the smooth
$\mathbb{F}_2$ surface.} Just for comparison we can consider the
$H(\vv )$ function also for the extremal metrics on $F_2$ defined by
the function in \eqref{candela}. We obtain:
\begin{equation}\label{qurioso}
    H^{\mathbb{F}_2}_{ext}(\vv )\,  =\,\frac{\exp \left(-\frac{(\mathit{a}-\mathit{b}) (3 \mathit{a}+\mathit{b}) \tan
   ^{-1}\left(\frac{-\mathit{a}^2+4 \mathit{a} \mathit{b}+6 \mathit{a}
   \mathfrak{v}+\mathit{b}^2+2 \mathit{b} \mathfrak{v}}{\sqrt{-\mathit{a}^4+44
   \mathit{a}^3 \mathit{b}+10 \mathit{a}^2 \mathit{b}^2-4 \mathit{a}
   \mathit{b}^3-\mathit{b}^4}}\right)}{\sqrt{-\mathit{a}^4+44 \mathit{a}^3
   \mathit{b}+10 \mathit{a}^2 \mathit{b}^2-4 \mathit{a}
   \mathit{b}^3-\mathit{b}^4}}\right)}{\sqrt{\frac{\mathit{b}-\mathfrak{v}}{\mathit{a}
   -\mathfrak{v}}}}
\end{equation}
The behavior is absolutely similar as one can see in
Fig.\ref{calzedonna}.

\begin{figure}
\centering
\includegraphics[width=10cm]{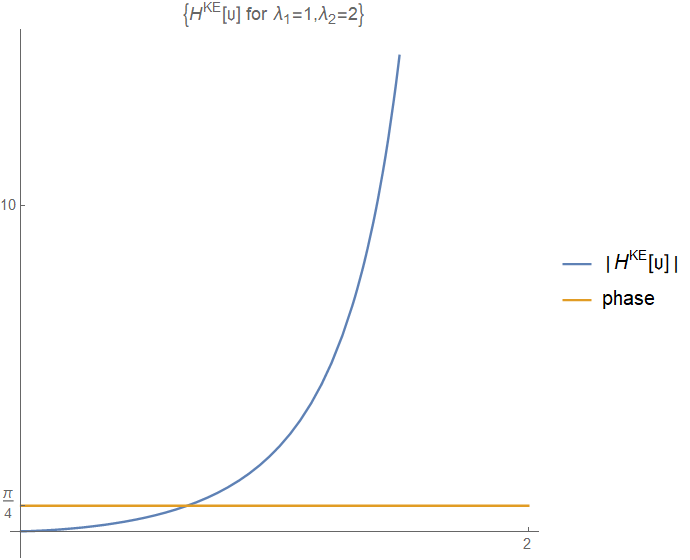}
\caption{\label{tanti7} Plot of  the $H(\vv )$ function in
the KE case with the choice
$\lambda_1=1,\lambda_2=2$.}
\end{figure}

\paragraph{The
function $H(\vv )$ for the KE metrics.} In the
case of the KE metrics in an equally easy way we
obtain  the following result:
\begin{multline}\label{HvKE}
   H^{KE}(\vv ) =
      \exp\left[-\left(\lambda _1^2+\lambda _2 \lambda _1+\lambda _2^2\right)
   \left(\frac{\log \left(\vv -\lambda _1\right)}{\lambda
   _1^2+\lambda _2 \lambda _1-2 \lambda _2^2}\right. \right. \\ \left. \left. +\frac{\log
   \left(\vv -\lambda _2\right)}{-2 \lambda _1^2+\lambda
   _2 \lambda _1+\lambda _2^2}-\frac{\log \left(\lambda _2
   \vv +\lambda _1 \left(\lambda
   _2+\vv \right)\right)}{2 \lambda _1^2+5 \lambda _2
   \lambda _1+2 \lambda _2^2}\right)\right]
\end{multline}
The structure of the function is similar to that of the   $\mathbb{F}_2$ case, since there is a zero of the function in
the lower limit $\vv \rightarrow\lambda_1$ and a pole in the
upper limit $\vv \rightarrow\lambda_2$, yet this time the
exponents of the pole and of the zero are rational numbers depending on the choice of the
roots $\lambda_{1,2}$;
 similarly it happens for the third factor associated with the third root
which is located out of the basic polytope (see  Figure \ref{politoppo}).
Our canonical example $\lambda_1=1,\lambda_2=2$ helps to illustrate
the general case; with this choice we obtain
\begin{equation}\label{carolinus}
     H^{KE}(\vv )\mid_{\lambda _1=1,\lambda _2=2} =
     \frac{(\vv -1)^{7/5} (3
   \vv +2)^{7/20}}{(\vv -2)^{7/4}}=
   e^{\mathit{i}\frac{\pi}{4}} \times \underbrace{\frac{(\vv -1)^{7/5} (3
   \vv +2)^{7/20}}{(2-\vv )^{7/4}}}_{\mathcal{H}^{KE}(\vv )}
\end{equation}
where, once again, the constant phase factor can be reabsorbed by a
constant shift of the angular variable $\tau$ and what remains  of
$\mathcal{H}^{KE}(\vv )$ is a positive definite function of $\vv $
in the interval $[1,2]$ that has the same feature  of  its analogue
in the   $\mathbb{F}_2$ cases, namely it maps, smoothly and
monotonically, the finite interval $(1,2)$ into the infinite
interval $(0,+\infty)$. The behavior of this function is displayed
in Figure  \ref{tanti7}.
\subsection{The structure of $\M_3$ and the conical singularity}
\label{singularity}   Let us
anticipate the main argument which we will develop   further
on. The two real manifolds defined by the restriction to the
dense chart $\mathfrak{u},\vv ,\phi,\tau$, of the surface
$\mathbb{F}_2$ and of  the manifold $\M_B^{KE}$ are fully
analogous. Cutting the compact four manifold into $\vv  \, =
\,\text{const}$ slices we always obtain the same result, namely a
three manifold $\M_3$ with the structure of a circle
fibration on $S^2$:
\begin{equation}\label{frullatore}
\M_B  \, \supset \,\M_3 \,
\stackrel{\pi}{\longrightarrow} \, S^2
   \quad ; \quad \forall p \in S^2 \quad \pi^{-1}(p) \sim
   S^1
\end{equation}
The metric on $\M_3$ is the standard one for fibrations:
\begin{equation}\label{m3metric}
    ds^2_{\M_3}=\vv
    \, \left(d\phi^2 \sin ^2\theta +d\theta
   ^2\right)\, + \,\mathcal{FK}(\vv ) \left[d\phi  (1-\cos \theta )
    +d\tau\right]^2
\end{equation}

The easiest way to understand $\M_3$ is to
study its intrinsic curvature by using the dreibein formalism.
Referring to equation \eqref{m3metric} we introduce the following
dreiben 1-forms:
\begin{equation}\label{trebanni}
    \pmb{\epsilon}^1 = \sqrt{\vv }\,d\theta \quad ;
    \quad \pmb{\epsilon}^2= \sqrt{\vv }\,\sin \theta d\phi
    \quad ; \quad \pmb{\epsilon}^3 = \sqrt{\mathcal{FK}(\vv )} \left[d\phi
   (1-\cos \theta )+d\tau \right]
\end{equation}
The fixed parameter $\vv $ plays the role of the squared
radius of the sphere $S^2$ while
$\sqrt{\mathcal{FK}(\vv )}$ weights the contribution of the
circle fiber defined over each point $p\in S^2$. At the
endpoints of the intervals $\mathcal{FK}(\vv _{min})\, =
\,\mathcal{FK}(\vv _{max})\, = 0$  the fiber shrinks to
zero.

\begin{figure}
\centering
\includegraphics[width=10cm]{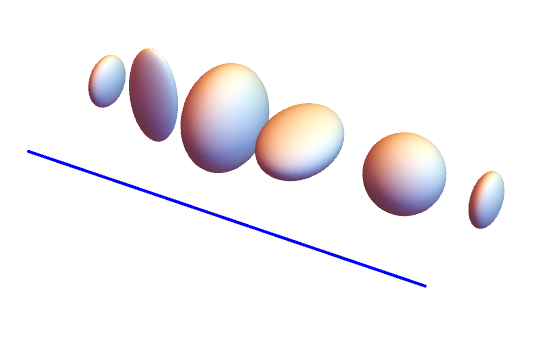}
\caption{\label{concetto}   A conceptual picture of the
$\M_B$ spaces that include also the second Hirzebruch
surface. The finite blue segment represent the
$\vv $-variable varying from its minimum to its maximum
value. Over each point of the line we have a three dimensional space
$\M_3$ which is homeomorphic to a 3-sphere but is
variously deformed at each different value $\vv $. At the
initial and final points of the blue segment the three dimensional
space degenerates into an $S^2$ sphere. Graphically we
represent the deformed 3-sphere as an ellipsoid and the
2-sphere as a flat filled circle.}
\end{figure}

Using the standard formulas of differential geometry and once again
the {\sc mathematica} packgage {\sc Vielbgrav23} we calculate the
spin connection and the curvature 2-form. We obtain:
\begin{equation}\label{gundashapur}
    \mathfrak{R}=\left(
\begin{array}{cc||c}
 0 & \frac{ (4
   \vv -3
   \mathcal{FK}(\vv ))}{4
   \vv ^2}\,\, \pmb{\epsilon}^1\, \wedge \,\pmb{\epsilon}^2 &
   \frac{
   \mathcal{FK}(\vv )}{4
   \vv ^2}\, \,\pmb{\epsilon}^1\, \wedge \,\pmb{\epsilon}^3 \\
-\, \frac{ (4
   \vv -3
   \mathcal{FK}(\vv ))}{4
   \vv ^2}\,\, \pmb{\epsilon}^1\, \wedge \,\pmb{\epsilon}^2 & 0 &
   \frac{
   \mathcal{FK}(\vv )}{4
   \vv ^2} \,\, \pmb{\epsilon}^2\, \wedge \,\pmb{\epsilon}^3 \\
   \hline
 -\frac{
   \mathcal{FK}(\vv )}{4
   \vv ^2}\,\,\pmb{\epsilon}^1\, \wedge \,\pmb{\epsilon}^3 &
    -\frac{
   \mathcal{FK}(\vv )}{4
   \vv ^2} \, \, \pmb{\epsilon}^2\, \wedge \,\pmb{\epsilon}^3 & 0 \\
\end{array}
\right)
\end{equation}
The Riemann curvature 2-form in flat indices has
constant components and if the coefficient $\frac{ (4
   \vv -3
   \mathcal{FK}(\vv ))}{4
   \vv ^2}$ were equal to the coefficient $\frac{
   \mathcal{FK}(\vv )}{4
   \vv ^2}$ the 2-form in eqn.~\eqref{gundashapur} would be the
   standard Riemann 2-curvature of the homogeneous space
   $\mathrm{SO(4)/SO(3)}$, namely the the 3-sphere
   $S^3$. What we learn from this easy calculation is that
    every section $\vv  = \text{constant}$ of
   $\M_B$ is homeomorphic to a 3-sphere endowed with a
   metric that is not the maximal symmetric one with isometry $\mathrm{SU(2)
   \times SU(2)}$ but a slightly deformed one with isometry $\mathrm{SU(2)
   \times U(1)}$: in other words we deal with a 3-sphere deformed
into the 3-dimensional analogue of an ellipsoid. At the endpoints of
the $\vv $-interval the ellipsoid degenerates into a sphere
since the third dreibein $\pmb{\epsilon}$ vanishes. A conceptual
picture of the full space $\M_B$ is provided in picture
Figure  \ref{concetto}.
\paragraph{Global properties of $\M_3$.}
Expanding on the global properties of $\M_3$, we describe it as a magnetic monopole bundle over $S^2$,  and prove that
the corresponding monopole strength   is $n=2$.
We start from the definition of the action of the ${\rm SU}(2) $ isometry \eqref{ciabattabuona} and
describe the 2-sphere $S^2$ spanned by $\theta$ and $\phi$ as $\mathbb{CP}^1$ with projective coordinates $U^0,\,U^1$:
\begin{equation}
    U^0 =   r \sin\left(\frac{\theta}{2}\right)\,e^{i\,\frac{\gamma+\phi}{2}}\,, \quad   U^1 =
      r \cos\left(\frac{\theta}{2}\right)\,e^{i\,\frac{\gamma-\phi}{2}}\label{U0U1}
\end{equation}
where $0\le \theta \le \pi$, $0\le \phi<2\pi$, $0\le \gamma<4\pi$.
In the North patch $\mathcal{U}_N$, $U^1\neq 0$ and the sphere is spanned by the stereographic
coordinate $u_N\ =  U^0/U^1$, while in the south patch $\mathcal{U}_S$, $U^0\neq 0$ and the stereographic
coordinate $u_S =  U^1/U^0$. The transformation properties \eqref{ciabattabuona} define a line bundle
whose local trivializations about the two poles are:
\begin{align}
    \phi^{-1}_N(\mathcal{U}_N)&=\left(u_N,\,v_N\right)= \left(\frac{U^0}{U^1},\,\xi \, (U^1)^2\right)\,,\\
      \phi^{-1}_S(\mathcal{U}_S)&=\left(u_S,\,v_S\right)=\left(\frac{U^1}{U^0},\,\xi \, (U^0)^2\right)\,,
\end{align}
where $\xi$ is a complex number in the fiber not depending on the patch.
As $(U^0,\,U^1)$ transform linearly under the
an ${\rm SU}(2)$-transformation:
\begin{equation}
 \left(\begin{matrix}U^0\cr U^1\end{matrix}\right)\rightarrow \left(\begin{matrix}d & c\cr b & a\end{matrix}\right)
  \left(\begin{matrix}U^0\cr U^1\end{matrix}\right)\,,
\end{equation}
the fiber coordinate $v$ transforms so that $(1+|u_N|^2)^2 |v_N|^2$ and $(1+|u_S|^2)^2 |v_S|^2$,
in $\mathcal{U}_N$ and $\mathcal{U}_S$, respectively, are invariant.
The transition function on the fiber reads, at the equator $\theta=\pi/2$:
\begin{equation}
    t_{NS}=\left(\frac{U^1}{U^0}\right)^2=e^{-2i\phi}=e^{-in\,\phi}\,,
\end{equation}
implying that the ${\rm U}(1)$-bundle associated with the phase of $v$ (i.e. the  submanifold
of the K\"ahler-Einstein space at constant $|v|$), is a monopole bundle with monopole strength $n=2$.
This has to be contrasted with the Hopf-fibering description of $S^3$, for which the local trivializations
have fiber components $U^1/|U^1|$ and $U^0/|U^0|$ in the two patches, respectively,
and $t_{NS}$ at the equator is $U^1/U^0=e^{-i\phi}$. In this case,  the monopole strength is $n=1$.
 {We can verify that the manifold at constant $|v|$ is a Lens space $S^3/\mathbb{Z}_2$ also by direct inspection of the metric.
 This is done in Appendix \ref{AA}.}
\paragraph{Conical singularities   and regularity of $\mathbb{F}_2$.}
{Let us analyze the exact form of the singularities (when they are present). The restriction of the metric to
a fiber spanned by $\vv $ and $\tau\in (0,2\pi)$ is }
\begin{equation}
ds^2=\frac{d\vv ^2}{\mathcal{FK}(\vv )}+\mathcal{FK}(\vv )\,d\tau^2\,.\label{metfib}
\end{equation}
Let  $\lambda$ denote one of the two  roots $\lambda_1,\,\lambda_2$
of $\mathcal{FK}(\vv )$. Close to $\lambda$, to first order in $\vv
$, in the KE case, the metric \eqref{metfib} is flat and features a
deficit angle signalling a conifold singularity. This singularity is
absent in the $\mathbb{F}_2$ cases, as expected. To show this   let
us Taylor expand $\mathcal{FK}(\vv )$ about $\lambda$:
\begin{equation}
    \mathcal{FK}(\vv )= \mathcal{FK}'(\lambda)(\vv -\lambda)+O((\vv -\lambda)^2).
\end{equation}
We can verify that:
\begin{equation}
\begin{array}{lclcl}
\mbox{KE $\phantom{Kronheimer}$}&:&\mathcal{FK}_{KE}'(\lambda_1)=
 \frac{(\lambda_2-\lambda_1)(\lambda_1+2 \lambda_2)}{\lambda_1^2+
 \lambda_2^2+\lambda_1\lambda_2} &;&
 \mathcal{FK}'_{KE}(\lambda_2)=\frac{(\lambda_1-\lambda_2)(2\lambda_1+ \lambda_2)}{\lambda_1^2+ \lambda_2^2+\lambda_1\lambda_2}\,,   \\[2pt]
\mbox{$\mathbb{F}_2$
Kronheimer}&:&\mathcal{FK}_{\mathbb{F}_2|Kro}'\left(\frac{9\alpha}{32}\right)=2&
; &
\mathcal{FK}_{\mathbb{F}_2|Kro}'\left(\frac{9(3\alpha+4)}{32}\right)=-2 \\
\mbox{$\mathbb{F}_2$
Extremal}&:&\mathcal{FK}_{\mathbb{F}_2|ext}'(\mathit{a}) = 2 & ;& \mathcal{FK}_{\mathbb{F}_2|ext}'(\mathit{b})=-2 \\
\end{array}
\end{equation}
Next we replace the first-order expansion  of this function in the fiber metric:
\begin{equation}
ds^2=\frac{d\vv ^2}{\mathcal{FK}'(\lambda)(\vv -\lambda)}+\mathcal{FK}'(\lambda)(\vv -\lambda)\,d\tau^2\,,\label{metfib2}
\end{equation}
and write it as a flat metric in polar coordinates:
\begin{equation}
ds^2=dr^2+\beta^2\,r^2\,d\tau^2\,.\label{metfib3}
\end{equation}
One can easily verify that:
\begin{equation}
    r=2\,\sqrt{\frac{\vv -\lambda}{\mathcal{FK}'(\lambda)}}\,\,,\,\,\,\beta=\frac{|\mathcal{FK}'(\lambda)|}{2}\,.
\end{equation}
Defining $\tilde{\varphi}= \beta \,\tau$, we can write the fiber metric as follows:
$$ds^2=dr^2+r^2\,d\tilde{\varphi}^2\,.$$
Now the polar angle varies in the range:
$\tilde{\varphi}\in [0,\,2\pi\,\beta]$.
If $\beta<1$ we have a deficit angle:
$$\Delta\phi=2\pi (1-\beta)\,.$$
\par
Let us see what this implies in the various possible cases of Table
\ref{casoni}.
\begin{enumerate} \item
In the case of the $\mathbb{F}_2$ manifold one has
$|\mathcal{FK}'(\lambda)|=2$ and $\beta=1$, both for the Kronheimer
metric and for the extremal one of the Calabi family, so there is no
conical singularity, as expected.
\item In the case of $\mathbb{WP}[1,1,2]$ we have
$$\mathcal{FK}(\lambda) = \frac{32\lambda^2-144\lambda+81}{(4\lambda-9)^2}.$$
For the limiting value $\lambda=0$ we obtain $\beta=\frac12$, i.e., a $\C^2/\Z_2$ singularity,
while for $\lambda=\frac98$ we have $\beta=1$, i.e., no singularity, as we expected as
$\mathbb{WP}[1,1,2]$ is an orbifold $\P^2/\Z_2$ with one singular point.
\item In the KE manifold case considering $\lambda=\lambda_1$, we have:
\begin{equation}
    \mathcal{FK}'(\lambda_1)=\frac{(\lambda_2-\lambda_1)(\lambda_1+2 \lambda_2)}{\lambda_1^2+
    \lambda_2^2+\lambda_1\lambda_2}=-1+\frac{3\lambda_2^2}{\lambda_1^2+ \lambda_2^2
    +\lambda_1\lambda_2}<-1+\frac{3\lambda_2^2}{ \lambda_2^2}=2\,\Rightarrow\,\,\beta<1\,,
\end{equation}
and $|\mathcal{FK}'(\lambda_2)|<|\mathcal{FK}'(\lambda_1)|$, so that $\beta<1$ also at $\lambda_2$.
The manifold has two conical singularities, both in the same fiber of the projection to one of the $S^2$'s.
One of the singularities will be an orbifold singularity of type $\C^2/\Z_n$ if the corresponding value
of $\beta$ is
\begin{equation}\label{betaint} \beta = 1 - \frac1n. \end{equation}
It is interesting to note that when this happens, the form of the function $\mathcal{FK}$
does not allow the other singularity to be of this type as as well, as the corresponding integer $m$ should satisty
$$ m = \frac{4n}{2+5n \pm \sqrt{9n^2+12n-12}}$$
which is not satisfied by any pair $(m,n)$ where both $m$, $n$ are integers greater than 1;
so the singular fiber can never be a football or a spindle.
\item
We   discuss the case $\lambda_1=0$. In this case
\begin{equation}
    \mathcal{FK}^0(\mathfrak{v}) =\frac{\mathfrak{v}(\lambda_2-\mathfrak{v})}{\lambda_2}\,.
\end{equation}
If we focus on the fiber metric:
\begin{equation}
ds^2=\frac{d\vv ^2}{\mathcal{FK}_0(\vv )}+\mathcal{FK}_0(\vv )\,d\tau^2\,.\label{metfib0}
\end{equation}
We can easily verify that in the coordinates $\tilde{\theta}\in [0,\pi]$ and $\tilde{\varphi}\in [0,\pi)$ defined by
$$
\mathfrak{v}(\tilde{\theta})=R^2\,\sin^2\left(\frac{\tilde{\theta} }{2}\right)
\le R^2=\lambda_2\,\,,\,\,\,\tilde{\varphi}=\frac{\tau}{2}\,,$$
where $R = \sqrt{\lambda_2}$, the fiber metric \eqref{metfib0} becomes
 \begin{equation}
     ds^2=R^2\,\left(d\tilde{\theta}^2+\sin^2(\tilde{\theta})\,d\tilde{\varphi}^2\right)\,.
 \end{equation}
 Since $\tilde{\varphi}=\tau/2\in\,[0,\pi)$, the fiber is the splindle $S^2/\mathbb{Z}_2$.
 {Topologically, the entire 4-manifold is still $S^2\times S^2$.}
\item
For $\mathcal{FK}(\mathfrak v)=\mathfrak v$ we   get $\beta=\frac12$,
in accordance with the fact that the variety in this case is $\C^2/\Z_2$.
 \end{enumerate}
\par
 {In the cases 3 and 4 the singular locus is of the form $S^2\times p_\pm$, where $p_\pm$ are the ``poles''
 of the fibers of the projection $S^2\times S^2\to S$. In complex geometric terms, it is a pair of divisors, both
 isomorphic to $\P^1$.}
\par
 \section{Complex structures}
{In this section we study the complex structures corresponding to
the KE case, i.e., the cases  corresponding to the function
$\mathcal{FK}^{KE}$. These are singular KE manifolds of complex
dimension 2,   homeomorphic to $S^2\times S^2$.} To make the
analysis completely quantitative, let us choose a value of the
parameter $\alpha$ and two values of $\lambda_1, \lambda_2$ so that
the basic polytope becomes exactly identical in the two cases.
\par We choose the value $\alpha = \frac{4}{9}$ so that the
endpoints of the interval in the pure $\mathbb{F}_2$ Kronheimer case
are:
\begin{equation}\label{coriaceo}
    \vv _{min} = \frac{1}{8} \quad ; \quad \vv _{max} =
    \frac{3}{2}
\end{equation}
and using the previously discussed procedure we obtain the complex
$v$ coordinate for the $\mathbb{F}_2$ pure case:
\begin{equation}\label{vcF2-1}
    v_{\null_{\mathbb{F}_2}} = \exp\left[\mathit{i} \,\left(\tau \,
    + \frac{\pi}{4}\right)\right]\times \frac{1+\cos \theta}{2}
    \times \sqrt{\frac{64 \vv ^2 -1 }{3\, - \, 2\, \vv }}
\end{equation}
In the same way we obtain the complex $v$ variable for the K\"ahler
Einstein case:
\begin{equation}\label{vcF2-2}
    v_{\null_{KE}} = \exp\left[\mathit{i} \,\left(\tau \,
    + \frac{157}{154}\,\pi\right)\right]\times \frac{1+\cos \theta}{2}
    \times \frac{\left(\vv -\frac{1}{8}\right)^{157/275}
   \left(\frac{3 \vv }{2}+\frac{1}{8}
   \left(\vv +\frac{3}{2}\right)\right)^{157/350}}{\left(
   \frac{3}{2}-\vv \right)^{157/154}}
\end{equation}
Obviously there is no holomorphic way of writing $v_{\null_{KE}}$ in
terms of  $v_{\null_{\mathbb{F}_2}}$ or viceversa. This can be
immediately seen in the following way. Taking the ratio of the two
coordinates $v$ we obtain:
\begin{equation}\label{lombricorosso}
   \left(\mathit{i}\right)^{\frac{237}{154}} \times \frac{ v_{\null_{\mathbb{F}_2}}}{ v_{\null_{KE}}} =
   4 \times \frac{ 2^{1877/3850} (3-2 \vv )^{40/77} \sqrt{64 \vv ^2-1}}{(8
   \vv -1)^{157/275} (26 \vv +3)^{157/350}}
\end{equation}
 Moreover with some manipulations we can write
\begin{equation}\label{ilota}
    \vv =  \frac{1}{64} \left(-\varpi_{\null_{\mathbb{F}_2}}\,\pm \,\sqrt{\varpi_{\null_{\mathbb{F}_2}} ^2
    \,+192 \,\varpi_{\null_{\mathbb{F}_2}}\, +\,64}
   \right)
\end{equation}
Inserting eqn.~\eqref{ilota} into eqn.~\eqref{lombricorosso} we
see that the complex coordinate
$v_{\null_{KE}}$ is not a holomorphic function of the complex
coordinates $u_{\null_{\mathbb{F}_2}},v_{\null_{\mathbb{F}_2}}$.
\par
\begin{figure}
\centering
\includegraphics[width=10cm]{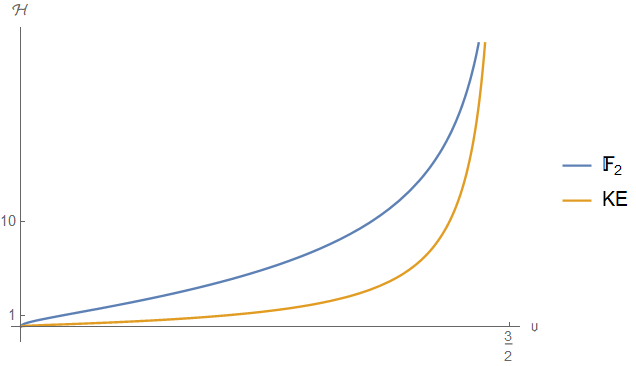}
\caption{\label{comparativo} Comparison  of  the plots of
$\mathcal{H}^{\mathbb{F}_2}(\vv )$ for the pure
$\mathbb{F}_2$ case with $\mathcal{H}^{KE}(\vv )$ for the
KE case, when they are calibrated to insist on the
same interval $\left[\frac 18 , \frac 32\right]$}
\end{figure}

This argument   can be used for all values of $\alpha$. We can
always choose the independent roots $\lambda_{1,2}$ so that the
interval of the moment variable $\vv _{min}$,$\vv _{max}$ coincides
in the K\"ahler Einstein case and in the $\mathbb{F}_2$ Kronheimer
or extremal cases. In the dense open chart that we are using
$\mathbb{F}_2$ and the manifold that admits a KE metric  have
different   complex structures. The plot of the two functions
$\mathcal{H}(\vv )$ is displayed in Figure \ref{comparativo}.
\par
The remaining problem is therefore the following. When we utilize
the Kronheimer metric for $\mathbb{F}_2$ written in  real variables
in the open chart provided by the coordinates $\mathfrak{u},\vv
,\phi,\tau$ we know that the closure of such a dense chart is the
second Hirzebruch surface. The same can be said if we utilize the
extremal metric. In the same open real chart we have also a KE
metric: the question is, \textit{What is the closure of such an open
real chart compatible with the KE metric?}. The important point to
keep in mind while trying to answer such a question is that,
topologically, the Hirzebruch surface is just $S^2 \times S^2 $.
What makes this real manifold Hirzebruch is the complex structure
induced by the holomorphic   embedding in $\mathbb{P}^1\times
\mathbb{P}^2$ as an algebraic variety. Yet the complex structure of
$\mathbb{F}_2$ is different and incompatible with the complex
structure compatible with the KE metric defined  in the same open
chart. This must be the guiding principle.
\paragraph{The  homeomorphism with $S^2 \times S^2$.}
\label{omeomorfismo}

{We want to analyze in detail the homeomorphism with the space $S^2\times S^2$
in the KE case.
Prior to that let us stress, once again, that all the manifolds we are considering are KE by construction, as:}
\begin{description}
  \item[a)] There is a candidate K\"ahler form written as
  \begin{equation}\label{cellario1}
    \mathbb{K}= K_{ij} \mathbf{e}^i \wedge \mathbf{e}^j
  \end{equation}
where the one forms $\mathbf{e}^i$ are a tetrad representation of
the considered metric:
\begin{equation}\label{cellario2}
    ds^2_{\M_B}= g_{\mu\nu} dx^\mu \, dx^\nu =
    \delta_{ij} \, \mathbf{e}^i  \mathbf{e}^j
\end{equation}
 \item[b)] The candidate form is closed:
 \begin{equation}\label{cellario3}
    \mathrm{d}\mathbb{K}=0
 \end{equation}
  \item[c)] The component  tensor of the K\"ahler form in flat
  indices $K_{ij}$ satisfies the condition:
  \begin{equation}\label{cellario4}
    K_{ij}\,K_{jk} = - \, \delta_{ik}
  \end{equation}
This guarantees that for each metric we can construct the
corresponding complex structure tensor:
\begin{equation}\label{logocomplex}
    \mathfrak{J}_\nu^{\phantom{\nu}\mu} = {E}^i_\mu \, K_{ij} \,
    \delta^{jk} \, {E}_k^\nu \quad ; \quad \mathfrak{J}^2 = -
    \, \mathrm{Id}
\end{equation}
and by construction the metric is hermitian with respect to that
  complex structure.
\end{description}
The K\"ahler form is the same for all the family of metrics
\eqref{metrauniversala} and each metric chooses the complex
structure with respect to which it is hermitian. In principle these
complex structures are all different, yet some of them might be
compatible, as it is the case for the one parameter family of
metrics on the Hirzebruch surface, that share the same complex
structure and can be described in terms of the same complex
coordinates. However in the previous section we have already shown
that the complex structure selected by one of the KE
metrics is certainly incompatible with that of the Hirzebruch
surface and this removes any possible conceptual clash.
On the other
hand there is no obstacle to the fact that the underlying real
manifold of the Hirzebruch case and of the (singular) KE case
might be homeomorphic and this is what we want to show.
\par
In the next lines we argue how to construct explicitly such
a homeomorphism. First of all, by looking at the metric in
eqn.~\eqref{metrauniversala} we see that the first 2-sphere is
already singled out in the standard coordinates $\theta$ and $\phi$.
As for the second sphere the azimuthal angle is already identified
in the coordinate $\tau$. It remains to be seen that the coordinate
$\vv $ in the finite closed range $\left[\vv _{min}
\, ,\,\vv _{max}\right]$ is in one-to-one continuous
correspondence with a new right ascension angle $\chi$.
\paragraph{Behavior of the function
$\sqrt{\mathcal{FK}(\vv )}$.} To this effect the main point
is that the function $\sqrt{\mathcal{FK}(\vv )}$ should be
upper limited by the value $1$ in the interval
$\left[\vv _{min} \, ,\,\vv _{max}\right]$, it
should grow monotonically from $0$ to a maximum value $a_0 \leq 1$,
attained at $\vv \, =\,\vv _0$ and then it should
decrease monotonically from $a_0$ to $0$ in the second part of the
interval $\left[\vv _0\, , \, \vv _{max}\right]$.
Under such conditions the inverse function $\arcsin$ can be applied
unambiguously to $\sqrt{\mathcal{FK}(\vv )}$ and we can
obtain a one-to-one continuous map between the coordinate
$\vv $ and a new right ascension angle $\chi$. The
homeomorphism is encoded in the following relation where the
function $\mathfrak{h}(\vv )$ is continuous and monotonous
only under the above carefully specified conditions.
\begin{equation}\label{evosido}
\chi = \mathfrak{h}(\vv )= \frac{\pi
\arcsin\left(\sqrt{\mathcal{FK}(\vv )}\right)}{2
\arcsin\left(a_0\right)}\, +\, \Theta
\left(\vv -\vv _0\right) \, \pi\,
\left(1-\frac{\arcsin\left(\sqrt{\mathcal{FK}(\vv )}\right)}{\arcsin\left(a_0\right)}\right)
\end{equation}
In the above formula the symbol $\Theta(x)$ denotes the well known
Heaviside step function that vanishes for $x<0$ and evaluates to $1$
for $x>0$.
\par
\begin{figure}
\centering
\includegraphics[width=7cm]{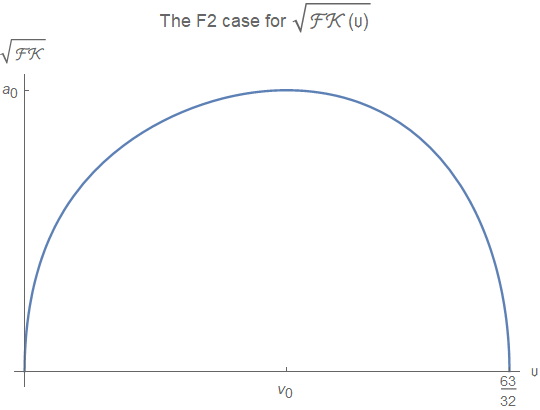}
\includegraphics[width=7cm]{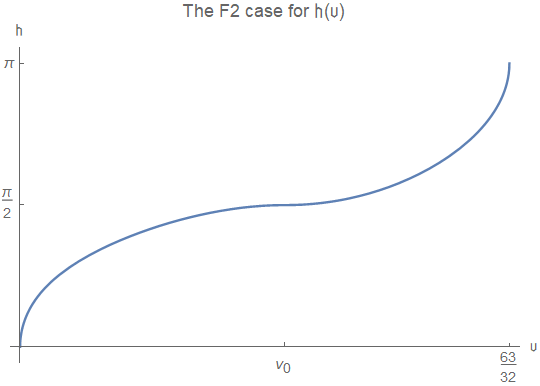}
\caption{\label{F2homeo}  On the left the plot of the function
$\sqrt{\mathcal{FK}^{\mathbb{F}_2}(\vv )}$ corresponding to
$\alpha=1$ and explicitly displayed in eqn.~\eqref{labirinto}. On the
right the corresponding function $\mathfrak{h}(\vv )$
providing the homeomorphism to the right ascension angle.}
\end{figure}
\begin{figure}
\centering
\includegraphics[width=7cm]{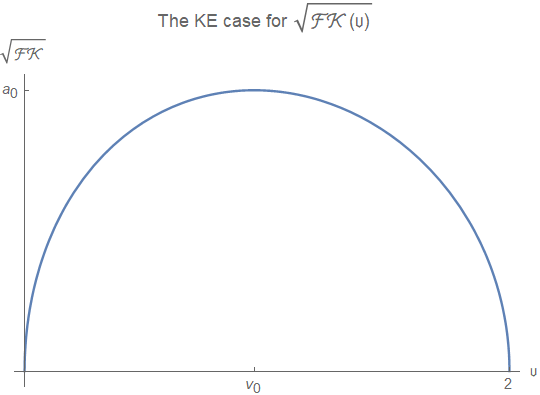}
\includegraphics[width=7cm]{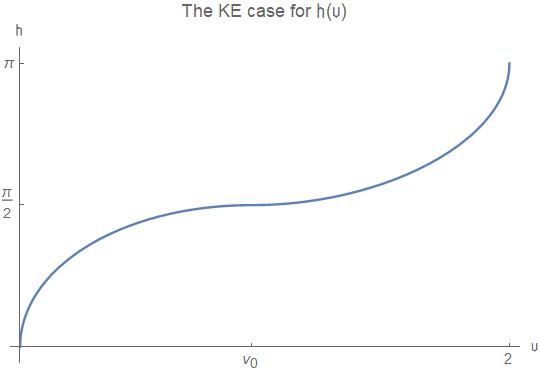}
\caption{\label{KEhomeo}  On the left the plot of the function
$\sqrt{\mathcal{FK}^{KE}(\vv )}$ corresponding to
$\lambda_1=1,\lambda_2=2$ and explicitly displayed in
eqn.~\eqref{otniribal}. On the right the corresponding function
$\mathfrak{h}(\vv )$ providing the homeomorphism to the
right ascension angle.}
\end{figure}

The relevant fact is that both for the case of the Hirzebruch
surface metric and for KE ones the above specified
conditions are verified and the homeomorphism \eqref{evosido} can be
written. We examine in detail one instance of the first and one
instance of the second case, having verified that in each class the
chosen examples represent the behavior of all members of the same
class. For the Hirzebruch case we set the parameter $\alpha=1$ and
we obtain:
\begin{equation}\label{labirinto}
  \mathcal{FK}^{\mathbb{F}_2}\left( \vv \right) =
   \frac{(32 \vv -63) \left(1024 \vv ^2-81\right)}{16 \left(1024
   \vv ^2-4032 \vv +81\right)}
\end{equation}
For the KE case we choose, as we already did in previous sections,
$\lambda_1 =1$, $\lambda_2 = 2$ and we obtain:
\begin{equation}\label{otniribal}
    \mathcal{FK}^{KE}\left( \vv \right) =
    -\frac{(\vv -2) (\vv -1) (3 \vv +2)}{7 \vv }
\end{equation}
The behavior of the function $\sqrt{\mathcal{FK}(\vv )}$ in
the two cases and of the associated homeomorphism on the right
ascension angle is shown in Figure s \ref{F2homeo},\ref{KEhomeo}
On the basis of the above lore in order to explore the behavior of a
function, a vector field or whatever different geometric object in
the neighborhood of the North pole of the second sphere one has at
one's disposal a well-defined transition function from the
coordinates $\vv _N,\tau_N$ to the coordinates
$\vv _S,\tau_S$:
\begin{equation}\begin{array}{rcl}\label{transitionvtau}
    \vv _S& = &
    \mathfrak{h}^{-1}\left(\mathfrak{h}(\vv _N)-\pi\right) \\
    \tau_S &=& -\tau_N+\pi
\end{array}\end{equation}
The inversion of the function $\mathfrak{h}$ is highly nontrivial
since it is a combination of transcendental and algebraic functions,
yet it can always be done numerically, if needed. As for the metric
itself there is no need since the curvature is nonsingular in the
point $\vv _{max}$ and the geodesics are also well behaved
in all limits.
\par
The conclusion is that the underlying manifold of the K\"ahler
Einstein metrics is $S^2 \times S^2$, as in
the   case of $\mathbb F_2$.
\section{Liouville vector field on $\M_B$ and the contact structure
on  $\M_3$} \label{liuvillo} Next we go back to consider
general properties of the metric \eqref{metrauniversala}
that,utilizing eqn.~\eqref{cambiovariabile} we also rewrite in terms
of the coordinates $\mathfrak{u},\vv ,\phi,\tau$:
\begin{equation}\label{metrauniUV}
ds^2_{\M_B} \, =
\,\frac{\mathrm{d}\vv ^2}{\mathcal{FK}(\vv )} +\,
\frac{\mathfrak{u}  (2
\vv -\mathfrak{u})}{\vv }\mathrm{d}\phi^2\, +
\,\frac{(\vv
   \mathrm{d}\mathfrak{u}-\mathfrak{u} \mathrm{d}\vv )^2}{\mathfrak{u} \vv
   (2 \vv -\mathfrak{u})}+\frac{\mathcal{FK}(\vv )
   }{\vv ^2} \,(\mathfrak{u}\, \mathrm{d}\phi
    +\vv \, \mathrm{d}\tau
   )^2
\end{equation}
The corresponding K\"ahler 2-form is provided by equation
\eqref{kalleroduef} that for reader's convenience we copy here
below:
\begin{equation}\label{readerconv}
    \mathbb{K} = d\mathfrak{u}\wedge d\phi \, + \, d\vv  \wedge
    d\tau
\end{equation}
The pair $\left(\M_B,\mathbb{K}\right)$ constitutes a
symplectic manifold, independently from the Riemannian structure
provided by the metric \eqref{metrauniUV}. For symplectic manifolds
there exists the notion of Liouville vector fields (see for instance
\cite{contatoregeiges,mariopietroinpress}) defined as follows. The
vector field $\mathbf{L}\in
\Gamma\left[T\M_B,\M_B\right]$ is   a
Liouville vector field if
\begin{equation}\label{liuvizzo}
    \mathcal{L}_{\mathbf{L}}\,\mathbb{K}= \mathbb{K}
\end{equation}
where $\mathcal{L}_{\mathbf{V}}$ denotes the Lie derivative along
the specified vector field $\mathbf{V}$. Utilizing  Cartan's formula
for the Lie derivative we get:
\begin{equation}\label{zupparancida}
    \mathcal{L}_{\mathbf{L}}\mathbb{K}=\mathit{i}_\mathbf{L}
    \,\mathrm{d}\mathbb{K} \, + \,
    \mathrm{d}\left(\mathit{i}_\mathbf{L}\,\mathbb{K}\right) \,
    =\,\mathrm{d}\left(\mathit{i}_\mathbf{L}\,\mathbb{K}\right)\,= \, \mathbb{K}
\end{equation}
A very simple Liouville field for the symplectic manifold
$\left(\M_B,\mathbb{K}\right)$ is the following one:
\begin{equation}\label{calsasco}
    \mathbf{L}= \mathfrak{u} \, \frac{\partial}{\partial
    \mathfrak{u}}\, + \, \vv  \, \frac{\partial}{\partial \vv }
\end{equation}
as one can immediately verify.
\par
Another result   in symplectic geometry, (see
\cite{contatoregeiges,Etnyre2000,miranda2021,cardona2021integrable,unochiuseinteg,trasversozero,mariopietroinpress})
states that a $(2n+1)$-submanifold $\mathcal{Z} \subset
\M_B$ of a $(2n+2)$-dimensional symplectic  manifold
$(\M_B,\mathbb{K})$ that is transverse to a Liouville field
$\mathbf{L}$ is a contact manifold with contact structure
\begin{equation}\label{particumagno}
    \Omega =\mathit{i}_\mathbf{L}\,\mathbb{K}
\end{equation}
In view of this theorem, the interpretation of the
$\M^{\vv }_3$ manifolds, extensively discussed in
previous sections, that have the topology of $S^3$ and are
all transverse to the Liouville field since they correspond to fixed
values of the coordinate $\vv $, becomes clear. They
constitute the leaves of a foliation of the symplectic manifold
$\left(\M_B,\mathbb{K}\right)$ in diffeomorphic contact
manifolds whose contact form is:
\begin{equation}\label{culattadimaiale}
    \Omega = \mathfrak{u} \, \mathrm{d}\phi \, + \, \vv  \, \mathrm{d}\tau
\end{equation}
On each leave $\vv =\text{const}$ we have:
\begin{equation}\label{contattone}
    \mathrm{d} \Omega=
    \mathrm{d}\mathfrak{u}\wedge\mathrm{d}\phi \quad ; \quad \mathrm{d}
    \Omega\,\wedge \,\Omega = \vv  \,\mathrm{d}\mathfrak{u}\wedge\mathrm{d}\phi
    \wedge \mathrm{d}\tau = \text{const} \times \mathrm{Vol}_3^{\vv }
\end{equation}
\paragraph{The Reeb field and  Beltrami equation.}
It is now interesting to calculate the normalized Reeb field
associated with the contact form $\Omega$. This is possible since
the symplectic  manifold $\left(\M_B,\mathbb{K}\right)$ is
endowed with the Riemannian structure provided by the metric
\eqref{m3metric}. Expanding the one-form $\Omega$ along the
coordinate differentials:
\begin{equation}\label{compitoacasa}
    \Omega= \Omega_\mu \, dy^\mu \quad ; \quad y^\mu =
    \{\theta,\phi,\tau\}
\end{equation}
we find:
\begin{equation}\label{compitoinclasse}
    \Omega_\mu = \{0,\vv  (1-\cos (\theta )),\vv \}
\end{equation}
Utilizing the inverse of the metric tensor defined by the line
element \eqref{m3metric} namely:
\begin{equation}\label{Igg3}
    g^{\mu\nu} = \left(
\begin{array}{ccc}
 \frac{1}{\vv } & 0 & 0 \\
 0 & \frac{\csc ^2(\theta )}{\vv } & \frac{(\cos (\theta )-1) \csc ^2(\theta
   )}{\vv } \\
 0 & \frac{(\cos (\theta )-1) \csc ^2(\theta )}{\vv } &
   \frac{1}{\mathcal{FK}(\vv )}+\frac{\tan ^2\left(\frac{\theta
   }{2}\right)}{\vv } \\
\end{array}
\right)
\end{equation}
we can raise the index of $\Omega_\mu$ and we obtain the components
of a normalized Reeb vector field:
\begin{equation}\label{ribbone}
 U^\mu \,= \,g^{\mu\nu}\Omega_\nu =
 \left\{0,0,\frac{\vv }{\mathcal{FK}(\vv )}\right\}
 \quad \Rightarrow \quad \mathbf{U} \, =
 \,\frac{\vv }{\mathcal{FK}(\vv )}\, \partial_\tau
\end{equation}
such that:
\begin{equation}\label{cresolo}
    \Omega\left(\mathbf{U}\right) = 1 \quad ; \quad
    i_{\mathbf{U}}\,\mathrm{d}\Omega = 0
\end{equation}
It is  a notable fact that the above
  Reeb vector field automatically satisfies Beltrami equation.
Indeed it is known that every contact structure in 3 dimensions admits a
contact form and an associated Reeb field that satisfies Beltrami
equation (see for instance \cite{Etnyre2000,mariopietroinpress}) yet
it is remarkable that the choice of the Liouville vector field
\eqref{liuvizzo} immediately selects a Beltrami Reeb field. The
verification of our statement is almost immediate if we utilize the
formulation of Beltrami equation introduced in \cite{cardone2019}
(see also \cite{mariopietroinpress}), namely:
\begin{equation}\label{miraquestocardo}
    \mathrm{d}\Omega^{\mathbf{U}}=\lambda \, i_\mathbf{U} \, \mathrm{Vol}_3
\end{equation}
where $\mathrm{Vol}_3$ denotes the volume 3-form of the considered
3-manifold, $\Omega^{\mathbf{U}}$ is the contact form that
admits $\mathbf{U}$ as normalized Reeb field and $\lambda \in
\mathbb{R}$ is the Beltrami eigenvalue. In our case the volume form
is:
\begin{equation}\label{grandevaligia}
    \mathrm{Vol}_3 =
    \pmb{\epsilon}^1\wedge\pmb{\epsilon}^2\wedge\pmb{\epsilon}^3 \,
    = \, \vv \, \sqrt{\mathcal{FK}(\vv )} \,
    \sin\theta \,\mathrm{d}\theta\wedge \mathrm{d}\phi \wedge \mathrm{d}\tau
\end{equation}
and equation \eqref{miraquestocardo} is satisfied with eigenvalue:
\begin{equation}\label{suovalore}
    \lambda =- \frac{1}{\vv }.
\end{equation}
\section{Geodesics for the family of manifolds $\M_B$}
In this section we study the general problem of calculating the
geodesics for the class of metrics \eqref{metrauniversala}.
Sometimes the differential system determining the geodesics
is completely integrable and this
allows one to reduce it to first
order and to quadratures, obtaining in this way the complete set of
all geodesics --- leaving apart   the practical problem of inverting
transcendental functions, which possibly can be accomplished with numerical
methods.
An example is  the Kerr metric where there is a
hidden first integral (the Carter constant) which can be revealed by
the use of the Hamilton-Jacobi formulation, and allows for complete
integration.

We show in this section that for any choice of the function
$\mathcal{FK}(\vv )$ the geodesic dynamical system
associated with the metrics \eqref{metrauniversala} is fully
integrable and admits a hidden Carter constant, another first integral in addition to the Hamiltonian,  which allows to
write the full system of geodesic lines for all the metrics in the
class, in particular for the $\mathbb{F}_2$ surface and for the
KE manifolds brought to attention in this
paper.

\subsection{The geodesic equation}
We take for Lagrangian functional the square of the arc length
\begin{equation}\label{lagrangianus}
\mathcal{L}=\frac{1}{2}\,\mathcal{FK}(\vv )
\left(\dot{\phi}  \left(1-\cos\theta
\right)+\dot{\tau}\right)^2+\frac{\dot{\vv }^2}{\mathcal{FK}(\vv )}+\vv
\left(\dot{\phi} ^2 \sin ^2\theta+\dot{\theta} ^2\right).
\end{equation}
As usual, the Euler-Lagrange equations take the standard form if
 the Lagrangian satisfies the
constraint
\begin{equation}\label{barlengo}
    \mathcal{L}\mid_{\text{on geodesics}} = \frac{k}{4}
\end{equation}
for some $k>0$; in this way parameter along the curves is    the arc length $s$.  In a   mechanical analogy, $k$ is the energy.

\paragraph{Cyclic variables and   conserved
momenta.} The   angles $\phi$ and $\tau$ are   cyclic variables
(due to toric symmetry)  which leads
to two first integral of the motion, which we call $\ell,\ m $, and
    can  be represented in  a  synthetic way:
\begin{equation}\label{carneadone}
\begin{pmatrix}   \ell \\    m  \end{pmatrix} =
 \begin{pmatrix}   p_\phi \\ p_\tau \end{pmatrix}
 = \mathfrak{M}  \begin{pmatrix}       \dot{\phi} \\  \dot{\tau} \end{pmatrix} ,
           \quad  \mathfrak{M}= \frac{1}{2}
            \begin{pmatrix}   8 \mathcal{FK}(\vv ) \sin ^4\frac{\theta }{2}+2
   \vv  \sin ^2\theta  & -2 \mathcal{FK}(\vv ) (\cos \theta
   -1) \\
 -2 \mathcal{FK}(\vv ) (\cos \theta -1) & 2 \mathcal{FK}(\vv )
   \end{pmatrix}
\end{equation}

\paragraph{The Hamiltonian.}
We perform the Legendre transform in order to obtain the
hamiltonian $H$:
\begin{equation}\label{hamiltocliffo}
 H =  \dot{\phi} \, p_{\phi}+\, \dot{\tau} p_{\tau }\,+ \dot{\theta}\,p_{\theta }\,  + \,
\dot{\vv } p_{\vv } \, -\, \mathcal{L}
\end{equation}
getting
\begin{equation}\label{quelpazzodidublino}
    H(q,p)= \frac{1}{2} \left(\frac{p_{\tau
   }^2}{\mathcal{FK}(\vv )}+\mathcal{FK}(\vv )
   p_{\vv }^2+\frac{\csc ^2\theta  \left[(\cos \theta -1) p_{\tau
   }+p_{\phi }\right]{}^2+p_{\theta }^2}{\vv }\right)
\end{equation}
where
\begin{equation}\label{pqretti}
    p =\left\{
    p_{\phi},p_{\tau},p_{\theta},p_{\vv }\right\} \quad;
    \quad q \,=\, \left\{
   \phi,\,\tau,\,\theta,\,\vv \right\}
\end{equation}
are the   momenta and coordinates.
\par
As it is  always the case in the geodesic problem, the hamiltonian has the
structure
\begin{equation}\label{parecchio}
    H = g^{ij}(q) \,p_i \, p_j
\end{equation}
having denoted by $g^{ij}(q)$ the inverse metric tensor.

\paragraph{The reduced Lagrangian and the reduced Hamiltonian.}
Having singled out two
first integrals of the motion  $\ell,\mathfrak{m}$, it is convenient to introduce a reduced
lagrangian for the two residual degrees of freedom
$\vv ,\theta$ that, geometrically, correspond to the two
angles of \textit{right ascension} of the 2-spheres composing the
underlying differentiable manifold (see section \ref{omeomorfismo}).
 The reduction of the lagrangian is obtained by replacing the
velocities of the cyclic coordinates $q^{\mathfrak{c}}$ with the
corresponding momenta $p_{\mathfrak{c}}$ that are constant of the
motion, namely $\ell,m$:
\begin{equation}\label{redLag}
    \mathcal{L}_{red} = \frac{\mathcal{FK}(\vv ) \left({\dot \theta} ^2 \,\vv ^2\, +\, \csc ^2\theta
    \,
   (\mathit{m} \cos \theta -\mathit{m}+\ell )^2\right)\, +\, \mathit{m}^2
   \vv \, +\, \vv  \,\dot{\vv }^2}{2 \vv  \mathcal{FK}(\vv )}
\end{equation}
Performing the Legendre transform   we
obtain the reduced hamiltonian
\begin{equation}\begin{array}{rcl}\label{redHam}
    H_{red}  &=&    p_{\vv } \, \dot{\vv }\, + \,p_{\theta } \,
    \dot{\theta} -  \mathcal{L}_{red} \\
& =  & \displaystyle \frac{1}{2} \left(-\frac{\mathit{m}^2}{\mathcal{FK}(\vv )}+\mathcal{FK}(\vv )
   p_{\vv }^2-\frac{\csc ^2\theta  (\mathit{m} \cos \theta -\mathit{m}+\ell
   )^2-p_{\theta }^2}{\vv }\right)
\end{array}\end{equation}
where
\begin{equation}\label{fantaschio}
    p_{\vv } = \frac{\dot{\vv }}{\mathcal{FK}(\vv )}, \quad
    \quad p_{\theta} = \vv  \, \dot{\theta}
\end{equation}
\paragraph{The Carter constant and the reduction to quadratures.}
Considering now the reduced system with four hamiltonian variables
we have a nice surprise: there is an additional function of the $q$
and $p$ that is in involution with the hamiltonian and therefore
constitutes an additional conserved quantity, yielding in this way
the complete integrability of the   system.
 Since it is the analogue of the Carter
constant for the Kerr metric we call it the \textit{Carter
hamiltonian} and we denote it with the letter $\mathcal{C}$:
\begin{equation}\label{carterconst}
    \mathcal{C} =\csc ^2\theta  \, \left(\mathit{m} \cos \theta -\mathit{m}+\ell \right)^2\, - \,p_{\theta }^2
\end{equation}
An immediate calculation shows that the Carter function has
vanishing Poisson bracket with the reduced hamiltonian:
\begin{equation}\label{involuzione}
    \left\{\mathcal{C} \, , \, H_{red} \right\}=\sum_{i=1}^2
    \left(
    \frac{\partial \mathcal{C}}{\partial q^i} \,\frac{\partial H_{red}}{\partial
    p_i}\, - \, \frac{\partial \mathcal{C}}{\partial p_i} \,\frac{\partial H_{red}}{\partial
    q^i}\right ) = 0
\end{equation}
Hence on any solution of the equations motion (that is along
geodesics) both the principal hamiltonian $H_{red}$ and
$\mathcal{C}$ must assume constant values that we call respectively
$\mathcal{E}$ (the energy) and $K$ (the Carter constant):
\begin{equation}\label{theintegrals}
    H_{red} =  \mathcal{E} \quad ; \quad \mathcal{C} \,
    = \, K
\end{equation}
Using equations \eqref{carterconst},\eqref{redHam} we can solve
algebraically eqn. \eqref{theintegrals} for the two momenta
$p_{\vv }$ and $p_{\theta}$ and we get the following two
first order differential equations:
\begin{equation}\begin{array}{rcl}
  \frac{\mathrm{d}\theta}{\mathrm{d}s} &=& \displaystyle \frac{\sqrt{\csc ^2\theta\,
  \left(\cos 2 \theta \, \left(K+\mathit{m}^2\right)-K+3
   \mathit{m}^2+2 \ell ^2+4 \mathit{m} \cos \theta  (\ell -\mathit{m})-4 \mathit{m}
   \ell \right)}}{\sqrt{2} \vv }  \\[5pt]
 \frac{\mathrm{d}\vv }{\mathrm{d}s}&=& \displaystyle \frac{\sqrt{\mathcal{FK}(\vv ) (K+2 \vv  \mathcal{E})+\mathit{m}^2
   \vv }}{\sqrt{\vv }}
\end{array} \label{lucamoro}
 \end{equation}
Eliminating the derivatives with respect to $s$ we finally obtain
the differential equation of the ``orbit''
\begin{equation}\label{pastrogus}
  \frac{d\theta}{d\vv } =  \frac{\sqrt{\csc ^2(\theta ) \left(\cos (2 \theta ) \left(K+\mathit{m}^2\right)-K+3
   \mathit{m}^2+2 \ell ^2+4 \mathit{m} \cos (\theta ) (\ell -\mathit{m})-4 \mathit{m}
   \ell \right)}}{\sqrt{2} \sqrt{\vv } \sqrt{\mathcal{FK}(\vv ) (K+2
   \vv  \mathcal{E})+\mathit{m}^2 \vv }}
\end{equation}
which can be reduced to quadratures:
\begin{equation}\begin{array}{rcl}
     \Lambda\left(\theta\right) &=& \displaystyle \int \frac{d\theta }{\sqrt{\csc ^2(\theta )
    \left(\cos (2 \theta ) \left(K+\mathit{m}^2\right)-K+3
   \mathit{m}^2+2 \ell ^2+4 \mathit{m} \cos (\theta ) (\ell -\mathit{m})-4 \mathit{m}
   \ell \right)}} \\[8pt]
   \Sigma\left(\vv \right)  &=&
 \displaystyle   \int\frac{d\vv }{\sqrt{2} \sqrt{\vv } \sqrt{\mathcal{FK}(\vv ) (K+2
   \vv  \mathcal{E})+\mathit{m}^2 \vv }}
\end{array}\label{arabesque}\end{equation}
The solution of the geodesic problem is provided by giving the
dependence of the variable $\vv $ on  the \textit{right ascension angle} $\theta$  of the first
sphere:
\begin{equation}
    \vv  = \Sigma^{-1}\circ
    \Lambda\left(\theta\right), \quad
    \theta = \Lambda^{-1}\circ
    \Sigma\left(\vv \right)\label{secundus}
\end{equation}
Both functions $\Lambda$ and $\Sigma$ are transcendental and the
inversion problem can be solved only numerically, except for some  special
cases as we are going to illustrate in the next section.
\par
We conclude this  section by noting  that the existence
of the Carter conserved hamiltonian is probably an implicit
consequence of the larger nonabelian isometry of the original
metric. The two first integrals $\ell$, $\mathfrak{m}$ follow from the
toric symmetry $\mathrm{U(1)\times U(1)}$. The Carter constant is
indirectly linked to the extension to $\mathrm{SU(2)}$ of one of the
two $\mathrm{U(1)}$'s. If we had  $\mathrm{SU(2)\times SU(2)}$
isometry, then the metric would be the direct product of two
Fubini-Study metrics. With $\mathrm{SU(2)\times U(1)}$ isometry we
have the hybrid case where one sphere is the fiber and the other is
the base manifold.
\par
\subsection{Irrotational geodesics}
The function $\Lambda(\theta)$ can be calculated explicitly in the
general case ($\ell\leq 0,\mathit(m) \leq 0$) and we obtain:
\begin{equation}\begin{array}{rcl}
  \Lambda(\theta) &=& \frac{N(\theta)}{D(\theta)} \\
  N(\theta) &=& \sqrt{\cos (2 \theta ) \left(K+\mathit{m}^2\right)-K+3 \mathit{m}^2+2 \ell ^2+4
   \mathit{m} \cos (\theta ) (\ell -\mathit{m})-4 \mathit{m} \ell }  \\
   &&\times \left(-\arctan\left[\displaystyle \frac{\sec ^2\left(\frac{\theta }{2}\right) \left(\cos (\theta )
   \left(K+\mathit{m}^2\right)+\mathit{m} (\ell
   -\mathit{m})\right)}{\sqrt{K+\mathit{m}^2} \sqrt{\sec ^4\left(\frac{\theta }{2}\right)
   (\mathit{m} \cos (\theta )-\mathit{m}+\ell )^2-4 K \tan ^2\left(\frac{\theta
   }{2}\right)}}\right]\right) \\[20pt]
   D(\theta) &=& (\cos (\theta )+1) \sqrt{K+\mathit{m}^2} \sqrt{\sec ^4\left(\frac{\theta }{2}\right)
   (\mathit{m} \cos (\theta )-\mathit{m}+\ell )^2-4 K \tan ^2\left(\frac{\theta
   }{2}\right)}
\end{array}\end{equation}
while the integral defining the function $\Sigma(\vv )$  in the
general case  does not evaluate to a combination of known special functions --- neither for f the $\mathbb{F}_2$ metric nor for   KE metrics.
\par
Although it can be done, it is rather cumbersome to write explicit
computer codes for the numerical computation of the function
$\Sigma$ in the general case and for the needed inverse of the
function $\Lambda$. Hence, at this stage it is difficult to present
explicit geodesics with trivial angular momenta. The alternative
is the numerical integration of the pair of first order equations
\eqref{lucamoro} but also here we meet some
difficulties since the differential system is \textit{stiff}\footnote{The term ``stiff'' comes from numerical analysis
and denotes a differential equation or differential system whose numerical solution
is unstable unless the step size is taken to be very small.}
and without special care and an in-depth study of the phase space, the
standard integration routines run into divergences and fail to
provide solutions both for the KE and the Hirzebruch
case. This is not surprising given the analogy with the Kerr metric.
Indeed the study of Kerr geodesics is a wide field, there is a large
variety of types of geodesics and each requires nontrivial
computational efforts to be worked out.
\par
Yet in our case things enormously simplify if we consider
\textit{irrotational geodesics} defined as those where $\ell=
\mathit{m} = 0$ and only the Carter constant $\mathcal C$ and the energy
$\mathcal{E}$ label the curve.   {Geometrically this corresponds to the fact that the
azimuthal angles $\phi,\tau$ span a two-dimensional torus $\mathrm{T}^2$.}
Pursuing the analogy with General Relativity, irrotational geodesics
are the analogues of  the \textit{radial geodesics} utilized   in
cosmology and in the study of the causal structure of spacetimes
where one preserves only time $t$ and radial distance $r$. The
analogues of $t$ and $r$ are in our case the variables $\vv $
and  $\theta$, namely, the two ascension angles of
$S^2\times S^2$.
\par
By suppressing angular momenta
things simplify drastically. The orbit equation \eqref{pastrogus}
reduces to
\begin{equation}\label{raptodattilo}
    \frac{d\theta}{d\vv }= \frac{\sqrt{-K}}{\sqrt{\vv } \sqrt{\mathcal{FK}(\vv ) (K+2 \vv
   \mathcal{E})}},
\end{equation}
which implies
\begin{equation}\label{perdicca}
    \theta =\mathcal{FM}(\vv ,K,\mathcal{E})\, =
    \, \int \, \frac{\sqrt{-K}}{\sqrt{\vv } \sqrt{\mathcal{FK}(\vv ) (K+2 \vv
   \mathcal{E})}} \, \mathrm{d}\vv .
\end{equation}
The good news is that in the KE case (with the choice
 $\lambda_1=1$, $\lambda_2=2$), the integral of eqn.~\eqref{perdicca}
can be explicitly evaluated obtaining
\begin{equation}\begin{array}{rcl}\label{KEFM}
    \mathcal{FM}^{KE}(\vv ,K,\mathcal{E})& = &\frac{N(\vv ,K,\mathcal{E})}{D(\vv ,K,\mathcal{E})} \\
N(\vv ,K,\mathcal{E}) &=& \sqrt{7} \sqrt{-K}
(\vv -2) (\vv -1) \sqrt{\frac{(3 \vv +2)
   (K+4 \mathcal{E})}{K+2 \vv  \mathcal{E}}}    \\
   &&\times \,F\left(\arcsin\left(\frac{\sqrt{-\frac{(3 K-4 \mathcal{E}) (\vv -2)}{K+2 \mathcal{E}
   \vv }}}{2 \sqrt{2}}\right)|\frac{8 (K+2 \mathcal{E})}{3 K-4
   \mathcal{E}}\right) \\[5pt]
D(\vv ,K,\mathcal{E})&=&\sqrt{\vv }
\sqrt{-\frac{(\vv -2) (3 K-4 \mathcal{E})}{K+2 \vv
   \mathcal{E}}} \sqrt{\frac{(\vv -1) (K+4 \mathcal{E})}{K+2 \vv
   \mathcal{E}}}    \\[3pt]
   &&\times \sqrt{-\frac{\left(3 \vv ^3-7 \vv ^2+4\right) (K+2
   \vv  \mathcal{E})}{\vv }}
\end{array}\end{equation}
where by $F(z|h)$ we have denoted the $F$ elliptic function.
\par
\begin{figure}
\centering
\includegraphics[height=8cm]{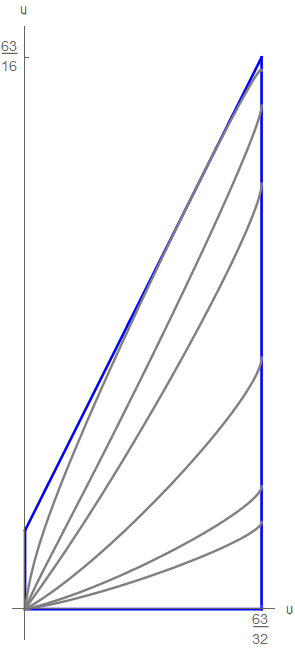}
\includegraphics[width=7cm]{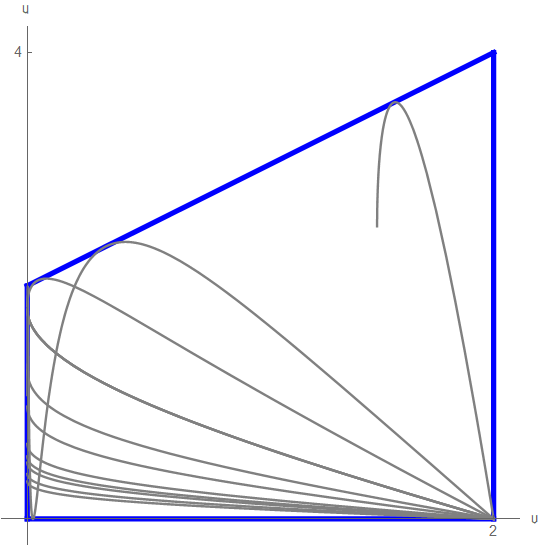}
\caption{\label{giroclando} On the left a plot of some irrotational
geodesics for the case of the Hirzebruch surface. On the right plot
of the same type of geodesics for the KE metric. }
\end{figure}

In the case of the Hirzebruch surface metric, the integral in
eqn.~\eqref{perdicca} does not evaluate to known special functions,
yet it can be  easily computed numerically, allowing one to
  to draw the geodesic curves in the
$\mathfrak{u},\vv $-plane: the parametric form is \begin{equation}\label{paracurva}
 \left\{
   \left(1-\cos\left[\mathcal{FM}(\vv ,K,\mathcal{E})\right]\right)\,\vv \,
   , \, \vv \right\}
\end{equation}
as follows from eqn.~\eqref{cambiovariabile}. Choosing various
different values of the energy and of the Carter constant we obtain
the curves shown in Figure \ref{giroclando}.
In both cases the irrotational geodesics are smooth and  approach  value $\vv _{max}$, that is, the North pole of the
second sphere. The only difference is that in the Hirzebruch case
they reach $\vv _{max}$ at various values of the right
ascension of the first 2-sphere while in the KE case
they tend to reach the North Pole of the second sphere arriving
simultaneously at the North Pole of the first sphere.
\par
\section{The Calabi Ansatz and the AMSY symplectic formalism}\label{trick}
Having studied in some detail the KE base
manifolds $\M_B^{KE}$,  we turn now  to main issue of this
paper, namely, the construction of a Ricci-flat metric on their
canonical bundle. We turn to the
method introduced by Calabi , which however only works   for KE base manifolds.
\par
As we anticipated in the introduction the Calabi ansatz method will
produce a Ricci-flat metric on the canonical bundle
$\operatorname{tot}\left[K\left[\M_{KE}\right]\right]$ that one
might be tempted to consider diffeomorphic to the Ricci-flat metric
on the metric cone over the Sasaki-Einstein manifolds of
\cite{Gauntlett:2004yd}; as we anticipated this is not true,
notwithstanding the very close relation of the K\"ahler Einstein
manifolds discussed above with the 5-dimensional Sasaki-Einstein
manifolds of \cite{Gauntlett:2004yd}. We have so far postponed the
comparison of the KE metrics discussed in the above sections with
the base manifolds of the 5-dimensional SE fibrations of
\cite{Gauntlett:2004yd} since such a comparison will be done more
appropriately just in one stroke together with the comparison of the
6-dimensional Ricci-flat metrics.
\paragraph{Calabi's Ansatz} Calabi's paper
\cite{Calabi-Metriques} introduces the following Ansatz for the local K\"ahler
potential $\mathcal{K}(z,\bar{z},w,\bar{w})$ of a K\"ahler  metric $g_E$
on the total space of a holomorphic vector bundle $E \to \M$, where $\M$ is a compact K\"ahler manifold
satisfying the conditions already stated in the introduction:
 \begin{equation} \label{Ansatz}
    \mathcal{K}(z,\bar{z},w,\bar{w})=\mathcal{K}_0(z,\bar{z})+ U(\lambda)
\end{equation}
where   $\mathcal{K}_0(z,\bar{z})$ is a K\"ahler potential for
$g_{\M}$, ($z^i$, $i=1,\dots \,
\mathrm{dim_\mathbb{C}}\M$) being the complex coordinates
of the base manifold) and $U$ is a function of a real variable $\lambda$,
which we shall identify with the function
\begin{equation}\label{squarnorm}
    \lambda = \mathcal{H}_{\mu\bar\nu}(z,\bar z) \,w^\mu \,  w^{\bar\nu}\, =
    \parallel w \parallel^2
\end{equation}
(the square norm   of a section of the bundle with
respect to a fiber metric $\mathcal{H}_{\mu\bar\nu}(z,\bar z)$). If
$\theta$ is the Chern connection on $E$, canonically determined by
the Hermitian structure $\mathcal{H}$ and the holomorphic structure
of $E$, its local connections forms can be written as
\begin{equation}\label{connetto}
\theta^{\phantom{\nu}\lambda}_\nu=\sum_i dz^{i}
\,L^{\phantom{i|\nu}\lambda}_{i|\nu}
\end{equation}
where
\begin{equation}\label{concof}
 L^{\phantom{i|\nu}\lambda}_{i|\nu}=\sum_{\bar\mu}\mathcal{H}^{\lambda\bar{{\mu}}}\frac{\partial}{\partial z^{i}}
 \mathcal{H}_{\nu\bar{\mu}}
 \qquad ; \qquad [\mathcal{H}^{\lambda\bar{\mu}}]=([\mathcal{H}_{\lambda\bar{\mu}}]^{-1})^{T}
\end{equation}
The curvature 2-form $\Theta$ of the connection $\theta$ is given
by:
\begin{equation}\label{ciccio}
    \Theta^{\phantom{\nu}\lambda}_{\nu}=\sum_{i,\bar\jmath} \,
    dz^{i}\wedge d\bar{z}^{\bar \jmath}
    \,S^{\phantom{i{\bar\jmath}|\nu}\lambda}_{i{\bar\jmath}|\nu}
    \qquad; \qquad
    S^{\phantom{i{\bar\jmath}|\nu}\lambda}_{i{\bar\jmath}|\nu} \, =
    \, \frac{\partial}{\partial
    {\bar z}^{\bar \jmath}}\, L^{\phantom{i|\nu}\lambda}_{i|\nu}
\end{equation}
The K\"ahler metric $g_{E}$ corresponding to the K\"ahler potential
$\mathcal{K}$ can be written  as follows:
\begin{multline}
    \partial\bar{\partial}\mathcal{K}\, =\,
    \sum_{i,{\bar \jmath}}\Bigl[g_{i{\bar \jmath}}+\lambda\, U'(\lambda) \,
    \sum_{\lambda,\nu,\bar\mu} \mathcal{H}_{\sigma\bar{\mu}}
     S^{\phantom{i{\bar\jmath}|\rho}\sigma}_{i{\bar\jmath}|\rho} w^{\rho}
     \bar{ w}^{\bar\mu}\Bigr]dz^{i}d\bar{z}^{\jmath}
     +\sum_{\sigma,\bar\mu}\Bigl[U'(\lambda)\,+\,
     \lambda \, U''(\lambda)\Bigr] \,
      \mathcal{H}_{\sigma\bar{\mu}} \, \triangledown w^{\sigma}\,\triangledown\bar{ w}^{\bar\mu}.
\end{multline}
If $E$ is a line bundle then the above equation reduces to
\begin{equation}\label{pbarppsi}
\partial\bar{\partial}\mathcal{K}\, =\,\sum_{i,\bar{\jmath}} \,[g_{i\bar{\jmath}}+\lambda U'(\lambda)\,
S_{i\bar{\jmath}}]dz^{i}d\bar{z}^{\bar \jmath}+[U'(\lambda )\, +\,
\lambda U''(\lambda)]\mathcal{ H}(z,\bar{z})\triangledown
w\triangledown\bar{ w}
\end{equation}
where $\lambda=\mathcal{H}(z,\bar{z})\,w\bar{w}$ is  the nonnegative
real quantity defined in equation \eqref{squarnorm} and
$\triangledown w$ denotes the covariant derivative of the
fiber-coordinate with respect to Chern connection $\theta$:
\begin{equation}\label{curlandia}
  \triangledown w  =dw \,+ \, \theta \, w
\end{equation}
\subsection{Ricci-flat metrics on canonical bundles} Now we assume
that $E$ is the canonical bundle  $K_{\M}$ of a K\"ahler surface
$\M$ ($\dim_\C \M = 2$). The total space of $K_{\M}$ has vanishing
first Chern class, i.e., it is a noncompact Calabi-Yau manifold, and
we may try to construct explicitly a Ricci-flat metric on it.
Actually, following Calabi, we can reduce the  condition that
$g_{E}$ is Ricci-flat to a differential equation for the function
$U(\lambda)$ introduced in equation \eqref{Ansatz}. Note that under
the present assumptions $S$ is a scalar-valued 2-form on $\M$.
\par
Since our main target is the construction of a Ricci-flat
metric on the space $\operatorname{tot}(\mathcal K_{\M^{KE}_B})$, where
$\M^{KE}_B$ denotes any of the KE manifolds
discussed at length in previous sections, we begin precisely with an
analysis of that case which will allow to derive a general form of
    $U(\lambda)$ as a function of the moment $\mathfrak{w}$ associated with the
$\mathrm{U(1)}$ group acting by phase transformations of the fiber
coordinate $w$,  and of certain coefficients $A,B,F$ that are
determined in terms of the K\"ahler potential $\mathcal{K}_0$ of the
base manifold $\M$. Consistency of the Calabi Ansatz requires that
these coefficients should be constant, which happens in the case of
base manifolds equipped with K\"ahler Einstein metrics. KE metrics
do not exist on Hirzebruch surfaces and the Calabi Ansatz is not
applicable in this case. As we discuss in the sequel, there exists a
Ricci-flat metric on the canonical bundle of a singular blow-down of
$\mathbb{F}_2$, namely the weighted projective plane
$\mathbb{WP}[1,1,2]$, which is known in the AMSY symplectic toric
formalism of \cite{abreu} and \cite{Martelli:2005tp}. If we were
able to do the inverse Legendre transform we might reconstruct the
so far missing K\"ahler potential and get inspiration on possible
generalizations of the Calabi Ansatz. Hence we are going to pay a
lot of attention to both formulations, the K\"ahler one and the
symplectic one.
\subsection{Calabi Ansatz   for 4D K\"ahler  metrics
with $\rm{SU}(2)\times\rm{U}(1)$ isometry} \label{F2Calab}
The Calabi Ansatz can be applied with success or not according to
the structure of the K\"ahler potential $\mathcal{K}_0$  for the
base manifold $\M$ and  the algebraic form of the invariant
combination $\Omega$ of the complex coordinates $u,v$ which is the
only real variable from which the K\"ahler potential
$\mathcal{K}_0=\mathcal{K}_0(\Omega) $ is assumed to depend. On the
other hand   $\Omega$ encodes  the group of isometries which is
imposed on the K\"ahler metric of $\M$.

In the case of the \  metrics discussed in section \ref{varposympo},
having $\rm{SU}(2)\times\rm{U}(1)$ isometry,
the invariant is
chosen to be
\begin{equation}\label{OmegaHirze}
   \Omega =  \varpi
\end{equation}
where $\varpi$ was defined in eqn.~\eqref{invarpi}. This choice
guarantees the isometry of the K\"ahler metric $g_\M$
against the group $\mathrm{SU(2)} \times \mathrm{U(1)}$ with the
action described in eqn.~\eqref{ciabattabuona}. Hence we focus on such
manifolds and we consider a K\"ahler potential for $\M$
that for the time being is a generic function
$\mathcal{K}_0(\varpi)$ of the invariant variable.
In this case  the determinant of the K\"ahler metric
$g_\M$ has  an explicit expression in terms of
$\mathcal{K}_0(\varpi)$
\begin{equation}\label{determio}
    \mathrm{det}\left(g_\M\right) \, =
    \, 2 \varpi  {\mathcal{K}_0}'(\varpi ) \left(\varpi  {\mathcal{K}_0}''(\varpi )+{\mathcal{K}_0}'(\varpi
   )\right)
\end{equation}
while the determinant of the Ricci tensor takes the   form
\begin{equation}\begin{array}{rcl}
\mathrm{det}\left(\mathrm{Ric}_\M\right)&=&
\frac{N_{Ric}}{D_{Ric}} \\
N_{Ric}& = & 2 \left(\varpi ^2  {\mathcal{K}_0}''(\varpi )^2+
{\mathcal{K}_0}'(\varpi )^2+\varpi
    {\mathcal{K}_0}'(\varpi ) \left(\varpi   {\mathcal{K}_0}^{(3)}(\varpi )+4  {\mathcal{K}_0}''(\varpi
   )\right)\right) \times  \\
   && \times \left(-\varpi ^3  {\mathcal{K}_0}''(\varpi )^4+\varpi ^2  {\mathcal{K}_0}'(\varpi
   ) \left(\varpi   {\mathcal{K}_0}^{(3)}(\varpi )- {\mathcal{K}_0}''(\varpi )\right)
    {\mathcal{K}_0}''(\varpi )^2 \right.  \\
   &&\left.+\varpi   {\mathcal{K}_0}'(\varpi )^2 \left(-\varpi ^2
    {\mathcal{K}_0}^{(3)}(\varpi )^2- {\mathcal{K}_0}''(\varpi )^2+\varpi  \left(\varpi
    {\mathcal{K}_0}^{(4)}(\varpi )+2  {\mathcal{K}_0}^{(3)}(\varpi )\right)  {\mathcal{K}_0}''(\varpi
   )\right)\right. \\
   &&\left.+ {\mathcal{K}_0}'(\varpi )^3 \left(3  {\mathcal{K}_0}''(\varpi )+\varpi  \left(\varpi
    {\mathcal{K}_0}^{(4)}(\varpi )+5  {\mathcal{K}_0}^{(3)}(\varpi
    )\right)\right)\right) \\
    D_{Ric}& = &  {\mathcal{K}_0}'(\varpi )^3 \left(\varpi   {\mathcal{K}_0}''(\varpi )+ {\mathcal{K}_0}'(\varpi )\right)^3
\end{array}\end{equation}
and the scalar curvature
\begin{equation}\label{Rscalar}
   R_s=\mathrm{Tr}\left(\mathrm{Ric}_\M\, g^{-1}_\M\right)
\end{equation}
is
\begin{equation}\begin{array}{rcl}\label{scalocurvo}
   R_s & = & \frac{N_s}{D_s}  \\
   N_s & = &  {\mathcal{K}_0}'(\varpi )^3+\varpi ^3  {\mathcal{K}_0}''(\varpi )^2 \left(2 \varpi
    {\mathcal{K}_0}^{(3)}(\varpi )+5  {\mathcal{K}_0}''(\varpi )\right) \\
   &&+\varpi ^2  {\mathcal{K}_0}'(\varpi
   ) \left(-\varpi ^2  {\mathcal{K}_0}^{(3)}(\varpi )^2+9  {\mathcal{K}_0}''(\varpi )^2+\varpi
   \left(\varpi   {\mathcal{K}_0}^{(4)}(\varpi )+4  {\mathcal{K}_0}^{(3)}(\varpi )\right)
    {\mathcal{K}_0}''(\varpi )\right) \\
   &&+\varpi   {\mathcal{K}_0}'(\varpi )^2 \left(9
    {\mathcal{K}_0}''(\varpi )+\varpi  \left(\varpi   {\mathcal{K}_0}^{(4)}(\varpi )+6
    {\mathcal{K}_0}^{(3)}(\varpi )\right)\right) \\
    D_s & = & \varpi   {\mathcal{K}_0}'(\varpi ) \left(\varpi   {\mathcal{K}_0}''(\varpi )+ {\mathcal{K}_0}'(\varpi
   )\right)^3
\end{array}\end{equation}
Given these base manifold data,  we introduce a K\"ahler potential
for a metric on the canonical bundle
$\operatorname{tot}\left[K\left(\M\right)\right]$ in accordance with
the Calabi Ansatz, namely
\begin{equation}\label{carabina}
    \mathcal{K}\left(\varpi,\lambda\right)={\mathcal{K}_0}(\varpi )\, +
    \, U(\lambda);  \quad \lambda  =
    \underbrace{\exp\left [\mathcal{P}(\varpi) \right]}_{\text{fiber metric $\mathcal{H}(\varpi)$}}  \mid w\mid^2 =
    \parallel w \parallel^2
\end{equation}
where $\lambda$ is  the  square norm of a section of
the canonical bundle and $\exp\left [\mathcal{P}(\varpi) \right]$ is
some   fiber metric. The determinant of the
corresponding K\"ahler metric $g_E$ on the total space of the
canonical bundle  is
\begin{equation}\begin{array}{rcl}\label{detGE}
 \mathrm{det}{g_E} & = &   2 \varpi  \Sigma (\lambda ) e^{\mathcal{P}(\varpi )} \Sigma '(\lambda ) \left(\varpi
   \mathcal{K}_0''(\varpi ) \mathcal{P}'(\varpi )+\mathcal{K}_0'(\varpi ) \left(\varpi
   \mathcal{P}''(\varpi )+2 \mathcal{P}'(\varpi )\right)\right)  \\
   &&+2 \varpi
   e^{\mathcal{P}(\varpi )} \Sigma '(\lambda ) \mathcal{K}_0'(\varpi ) \left(\mathcal{K}_0'(\varpi
   )+\varpi  \mathcal{K}_0''(\varpi )\right)+2 \varpi  \Sigma (\lambda )^2 e^{\mathcal{P}(\varpi
   )} \Sigma '(\lambda ) \mathcal{P}'(\varpi ) \left(\varpi  \mathcal{P}''(\varpi
   )+\mathcal{P}'(\varpi )\right) \\
\end{array}\end{equation}
where  we   set
\begin{equation}\label{sigmus}
    \Sigma(\lambda) =\lambda \, U'(\lambda)
\end{equation}
If we impose the Ricci-flatness condition, namely, that the
determinant of the metric $g_E$ is  a constant which we can always assume to be one since any other
number can be reabsorbed into the normalization of the fiber
coordinate $w$, by integration we get
\begin{equation}\label{lambdusingothic}
   \lambda =  \frac{1}{48} \left(A \, \mathfrak{w}^3+2 B\,
   \mathfrak{w}^2+4\, F \, \mathfrak{w}\right)
\end{equation}
where we have set
\begin{equation}\begin{array}{rcl}  \label{coriandolo}
    \Sigma(\lambda)  &= &  2\, \mathfrak{w} \\
  A &=& 4 \varpi  e^{\mathcal{P}(\varpi )} \mathcal{P}'(\varpi ) \left[\varpi
   \mathcal{P}''(\varpi )+\mathcal{P}'(\varpi )\right]  \\
  B &=& 6 \varpi  e^{\mathcal{P}(\varpi )} \left[\varpi  \mathcal{K}_0''(\varpi ) \mathcal{P}'(\varpi
   )+\mathcal{K}_0'(\varpi ) \left(\varpi  \mathcal{P}''(\varpi )+2 \mathcal{P}'(\varpi
   )\right)\right] \\
  F &=& 12 \varpi  e^{\mathcal{P}(\varpi )} \mathcal{K}_0'(\varpi ) \left[\mathcal{K}_0'(\varpi )+\varpi
   \mathcal{K}_0''(\varpi )\right]
\end{array}\end{equation}
In our complex three dimensional case, setting
\begin{equation}\label{qvar}
    x_u = \log \mid u\mid, \quad x_v = \log \mid v\mid, \quad
    \quad  x_w = \log \mid w \mid,
\end{equation}
the corresponding three moments can be named with the corresponding
gothic letters, and we have
\begin{equation}\label{momentacci}
    \mathfrak{u} =
    \partial_{x_u}\mathcal{K}\left(\varpi,\lambda\right),
    \quad   \vv  =
    \partial_{x_v}\mathcal{K}\left(\varpi,\lambda\right),
        \quad  \mathfrak{w} =
    \partial_{x_w}\mathcal{K}\left(\varpi,\lambda\right).
\end{equation}
As the fiber coordinate $w$ appears only in
the function $U(\lambda)$ via the squared norm $\lambda$,  we have
\begin{equation}\label{vugoticona}
    \mathfrak{w} = 2 \, \lambda \, U'(\lambda) =
    \Sigma(\lambda)
\end{equation}
and this justifies the position \eqref{coriandolo}. At this point
the function $U(\lambda)$ can be easily determined by first
observing that, in view of eqn.~\eqref{lambdusingothic} we can also
set
\begin{equation}\label{toppolina}
    U(\lambda) = \mathrm{U}\left(\mathfrak{w}\right)
\end{equation}
and we can use the chain rule
\begin{equation}\label{catenaria}
   \partial_\mathfrak{w}\,\mathrm{U}\left(\mathfrak{w}\right)=
   \frac{\mathfrak{w} \lambda '(\mathfrak{w})}{2 \lambda
   (\mathfrak{w})} =  \frac{3 A \mathfrak{w}^2+4 B \mathfrak{w}+4 F}{2 A \mathfrak{w}^2+4 B \mathfrak{w}+8
   F}
\end{equation}
which by integration yields the universal function
\begin{equation}\label{Udiwgoth}
 \mathrm{U}(\mathfrak{w})=   -\frac{2 \sqrt{4\, A\, F-B^2} \, \arctan
 \,\left(\frac{A\,
   \mathfrak{w}+B}{\sqrt{4 \,A\, F-B^2}}\right)+B \,\log
   \left(A \, \mathfrak{w}^2\, +\, 2\, B \mathfrak{w}\, +\, 4\,
   F\right)-3 \, A \,\mathfrak{w}}{2 \,A}
\end{equation}
The function $U(\lambda)$ appearing in the K\"ahler potential can
  be obtained by substituting for the argument $\mathfrak{w}$ in
\eqref{Udiwgoth} the unique real root of the cubic equation
\eqref{lambdusingothic}, namely:
\begin{equation}\begin{array}{rcl}\label{fischietto}
    \mathfrak{w} & = &  \displaystyle \frac{\sqrt[3]{8 \sqrt{\left(162 A^2 \lambda +9 A B F-2 B^3\right)^2-4 \left(B^2-3 A
   F\right)^3}+1296 A^2 \lambda +72 A B F-16 B^3}}{3 \sqrt[3]{2} A} \\[12pt]
   && \displaystyle +\frac{4 \sqrt[3]{2}
   \left(B^2-3 A F\right)}{3 A \sqrt[3]{8 \sqrt{\left(162 A^2 \lambda +9 A B F-2
   B^3\right)^2-4 \left(B^2-3 A F\right)^3}+1296 A^2 \lambda +72 A B F-16 B^3}}-\frac{2
   B}{3 A} \\
\end{array}\end{equation}
\paragraph{Consistency conditions for the Calabi Ansatz.}
\label{consistentia} In order for the Calabi Ansatz to yield a
 solution of the Ricci-flatness condition it is
necessary that the universal function $\mathrm{U}(\mathfrak{w})$ in
eqn.~\eqref{Udiwgoth} should depend only on $\mathfrak{w}$, which happens if and only if the coefficients $A,B,F$
are constant. In the case under consideration, where the invariant
combination of the complex coordinate $u,v$ is the one provided by
$\varpi$ as defined in eqn. \eqref{invarpi}, imposing such a
consistency condition would require the solution of three ordinary
differential equations for two functions $\mathcal{P}(\varpi)$ and
$\mathcal{K}_0(\varpi)$, namely:
\begin{equation}\begin{array}{rcl}\label{equinate}
  k_1 &=& 4 \varpi  e^{\mathcal{P}(\varpi )} \mathcal{P}'(\varpi ) \left[\varpi
   \mathcal{P}''(\varpi )+\mathcal{P}'(\varpi )\right]  \\
  k_2 &=& 6 \varpi  e^{\mathcal{P}(\varpi )} \left[\varpi  \mathcal{K}_0''(\varpi ) \mathcal{P}'(\varpi
   )+\mathcal{K}_0'(\varpi ) \left(\varpi  \mathcal{P}''(\varpi )+2 \mathcal{P}'(\varpi
   )\right)\right] \\
  k_3 &=& 12 \varpi  e^{\mathcal{P}(\varpi )} \mathcal{K}_0'(\varpi ) \left[\mathcal{K}_0'(\varpi )+\varpi
   \mathcal{K}_0''(\varpi )\right]
\end{array}\end{equation}
where $k_{1,2,3}$ are three constants. It is clear from their
structure that the crucial differential equation is the first one.
If we could find a solution for it then it would suffice to identify
the original K\"ahler potential $\mathcal{K}_0(\varpi )$ with a
linear function of $\mathcal{P}(\varpi )$ and we could solve the
three of them. So far we were not able to find any analytical
solution of these equations but if we could find one, we still should
verify that the K\"ahler metric following from such $\mathcal{K}_0$
is a good metric on the Hirzebruch surface $\mathbb{F}_2$.
\par
On the contrary for the well known K\"ahler potentials obtained from
the Kronheimer construction that define a one parameter family of
\textit{bona fide} K\"ahler metrics on $\mathbb{F}_2$ and were
discussed in \cite{Bruzzo:2017fwj,noietmarcovaldo}, namely those
presented in eqn.~\eqref{genKron}, equations \eqref{equinate} cannot
be satisfied and no Ricci-flat metric on the canonical bundle can be
obtained by means of the Calabi Ansatz.
\paragraph{The general case with the natural fiber metric  $\mathcal{H}=\frac{1}{\mathrm{det}\left(g_\M\right)}$.}
\label{naturalla} If we consider the general case of a toric
two-dimensional compact manifold $\M$ with a K\"ahler
metric $g_\M$ derived from a K\"ahler potential of the
form:
\begin{equation}\label{taralluccio}
    \mathcal{K}_0 = \mathcal{K}_0\left(|u|^2 ,|v|^2\right)
\end{equation}
choosing as fiber metric the natural one for the canonical bundle,
namely setting:
\begin{equation}\label{carneade0}
    \lambda =\mathcal{H} \, |w|^2 \, =
    \,\frac{1}{\mathrm{det}\left(g_\M\right)} \, |w|^2
\end{equation}
and going through the same steps as in section \ref{F2Calab} we
arrive at an identical result for the function
$\mathrm{U}(\mathfrak{w})$ as in equation \eqref{Udiwgoth} but with
the following coefficients:
\begin{equation}\label{cumulone}
    A = 2 \,
    \frac{\mathrm{det}\left(\mathrm{Ric}_\M\right)}{\mathrm{det}\left(g_\M\right)}, \quad B = 3 \,
    \mathrm{Tr}\left(\mathrm{Ric}_\M\, g^{-1}_\M\right), \quad F =6
\end{equation}
It clearly appears why the Calabi Ansatz works perfectly if the
starting metric on the base manifold is KE. In that
case the Ricci tensor is proportional to the metric tensor:
\begin{equation}\label{KalEinst}
    \mathrm{Ric}_{i\bar\jmath} = \kappa \, g_{i\bar\jmath}
\end{equation}
and we get:
\begin{equation}\label{robilante}
    \mathrm{det}\left(\mathrm{Ric}_\M\right) = \kappa^2
    \, \mathrm{det}\left(g_\M\right) , \quad \mathrm{Tr}\left(\mathrm{Ric}_\M\,
    g^{-1}_\M\right) = 2 \, \kappa
\end{equation}
which implies:
\begin{equation}\label{ABFconst}
    A=2 \, \kappa^2 ,  \quad B = 6   \, \kappa, \quad
    F = 6.
\end{equation}
\subsection{The AMSY symplectic formulation for the Ricci-flat metric
on $\operatorname{tot}\left[K\left(\M_B\right)\right]$}
 According with the discussion
of the AMSY symplectic formalism presented in section \ref{amysone},
given the K\"ahler potential of a toric complex three manifold
$\mathcal{K}(|u|,|v|,|w|)$,   we can define the moments
\begin{equation}\label{momentini}
    \mathfrak{u} =
    \partial_{x_u}\mathcal{K},
    \quad   \vv  =
    \partial_{x_v}\mathcal{K},
    \quad  \mathfrak{w} =
    \partial_{x_w}\mathcal{K}
\end{equation}
and we can obtain the   symplectic potential by means of the
Legendre transform:
\begin{equation}\label{legendretr2}
    {G}\left(\mathfrak{u},\vv ,\mathfrak{w}\right)
    = x_u \,\mathfrak{u} \, + \, x_v \,\vv  + \, x_w
    \,\mathfrak{w}\,  - \, \mathcal{K}(|u|,|v|,|w|)
\end{equation}
The main issue   in the use of eqn.~\eqref{legendretr2} is the
inversion transformation that expresses the coordinates $x_i \, =
\,\{x_u,x_v,x_w\}$ in terms of the three moments $\mu^i\, =
\{\mathfrak{u},\vv ,\mathfrak{w}\}$. Once this is done one
can calculate the metric in moment variables utilizing the Hessian
as explained in section \ref{amysone}. Relying once again on the
results of that section we know that the K\"ahler 2-form has the
following universal structure:
\begin{equation}\label{uniKal3}
  \mathbb{K} \,= \,  \mathrm{d}\mathfrak{u}\wedge \mathrm{d}\phi \, + \, \mathrm{d}\vv \wedge \mathrm{d}\tau \,
  + \, \mathrm{d}\mathfrak{w} \wedge \mathrm{d}\chi
\end{equation}
and the metric is expressed as displayed in eqn.~\eqref{sympametra})
\paragraph{The symplectic potential in the case
with $\mathrm{SU(2)\times U(1) \times U(1)}$ isometries}
\label{varposympoloto} In the case where the K\"ahler potential has
the special structure which guarantees an $\mathrm{SU(2)}\times
\mathrm{U(1)}\times\mathrm{U(1)}$ isometry, namely it depends only on the two
variables $\varpi$ (see eqn.~\eqref{invarpi})) and $ |w|^2$, also the symplectic potential takes a more restricted form.
Indeed we find
\begin{equation}\begin{array}{rcl}\label{specG}
    G\left(\mathfrak{u},\vv ,\mathfrak{w}\right) & = &\underbrace{\left(\vv -\frac{\mathfrak{u}}{2}\right)
   \log (2 \vv -\mathfrak{u})+\frac{1}{2} \mathfrak{u} \log
   (\mathfrak{u})\, -\, \frac{1}{2} \,\vv  \log (\vv )}_{\text{universal part $G_0(\mathfrak{u},\vv )$}}\,
   + \,\underbrace{\mathcal{G}(\vv ,\mathfrak{w})}_{\text{variable
   part}}
\end{array}\end{equation}
where $\mathcal{G}(\vv ,\mathfrak{w})$ is a function of
two variables that encodes the specific structure of the metric.
Note that when we freeze the fiber moment coordinate $\mathfrak{w}$
to some fixed constant value, for instance $0$, the function
$\mathcal{G}(\vv ,0)=\mathcal{D}(\vv )$ can be
identified with the boundary function that appears in
eqns.~\eqref{GBsymplectic},\eqref{lupetto}, namely in the symplectic
potential for the K\"ahler metric of the base manifold.
\par
With the specific structure \eqref{specG} of the symplectic
potential we obtain the following form for the Hessian
\eqref{hessiano}:
\begin{equation}\label{Gijspec}
    \mathbf{G} = \left(
\begin{array}{ccc}
 -\frac{\vv }{\mathfrak{u}^2-2 \mathfrak{u} \vv } &
   \frac{1}{\mathfrak{u}-2 \vv } & 0 \\
 \frac{1}{\mathfrak{u}-2 \vv } & \frac{-2 \vv
   (\mathfrak{u}-2 \vv )
   \mathcal{G}^{(2,0)}(\vv ,\mathfrak{w})+\mathfrak{u}+2
   \vv }{2 \vv  (2 \vv -\mathfrak{u})} &
   \mathcal{G}^{(1,1)}(\vv ,\mathfrak{w}) \\
 0 & \mathcal{G}^{(1,1)}(\vv ,\mathfrak{w}) &
   \mathcal{G}^{(0,2)}(\vv ,\mathfrak{w}) \\
\end{array}
\right)
\end{equation}
\section{The general form of the symplectic potential for the Ricci
flat metric on
{$\operatorname{tot}\left[K\left(\M^{KE}_B\right)\right]$}}
\label{formullageneralla} Having seen that KE
metrics do indeed exist  in  the form
described in eqs.\eqref{trottolina},\eqref{reducedhessian}, it is
natural to inquire how we can utilize the Calabi Ansatz  to
write immediately the symplectic potential for a Ricci-flat metric on the
canonical bundle of $\M_B^{KE}$ without going through the
process of   inverting the Legendre transform.
Namely,  we would
like to make  the  back and forth trip via inverse and direct Legendre transform only once and in full
generality rather than case by case. Our goal  is
not only a simplification of the computatinal steps but it  also involves
 a conceptual issue. Indeed, when we introduce intermediate
steps that rely on the variable $\varpi$ whose range is
$[0,+\infty)$ we necessarily have to choose a branch of a cubic
equation whose coefficients are determined by the root parameters
$\lambda_{1,2}$. On the contrary, if we are able to determine
directly the symplectic potential in terms of the symplectic
coordinates, then we can explore the behavior of the metric and of
its curvature on the full available range of variability of these
latter and we learn more on the algebraic and topological structure
of the underlying manifold.
\par
So let us anticipate the final result of our general procedure. As
we did in the previous section we assume that the Ricci form of
$\M_B$ is proportional to the K\"ahler form via a
coefficient
\begin{equation}\label{ricappo}
    \kappa = \frac{k}{4}
\end{equation}
as in eqs.\eqref{kallopietra},\eqref{KalEinst}. The complete
symplectic potential for the Ricci-flat metric on $\M_T =
\operatorname{tot}\left[K\left(\M_B\right)\right]$ has the following
structure:
\begin{equation}\begin{array}{rcl}\label{pulcherrima}
  G_{\M_T^{KE}}\left(\mathfrak{u},\vv ,\mathfrak{w}\right) &=& G_0\left(\mathfrak{u},\vv \right)\, +\,
  \mathcal{G}^{KE}\left(\vv ,\mathfrak{w}\right)  \\
  G_0\left(\mathfrak{u},\vv \right) &= & \left(\vv -\frac{\mathfrak{u}}{2}\right) \log [2
   \vv -\mathfrak{u}]+\frac{1}{2} \mathfrak{u} \log
   [\mathfrak{u}]-\frac{1}{2} \vv  \log [\vv ]  \\
\mathcal{G}^{KE}\left(\vv ,\mathfrak{w}\right)
&=&\left(\frac{\kappa  \mathfrak{w}}{2}+1\right)
\mathcal{D}^{KE}\left(\frac{2
   \vv }{\kappa  \mathfrak{w}+2}\right)-\frac{1}{2} \vv  \log
   \left(\frac{\kappa  \mathfrak{w}}{2}+1\right)+\frac{1}{2}
   \mathfrak{w} \log (\mathfrak{w}) \\
   &&\displaystyle+\frac{(\kappa  \mathfrak{w}+3) \log
   (\kappa  \mathfrak{w} (\kappa  \mathfrak{w}+6)+12)}{2 \kappa }
   +\frac{\sqrt{3} \arctan\left(\frac{\kappa  \mathfrak{w}+3}{\sqrt{3}}\right)}{\kappa }
  \\
\end{array}\end{equation}
where the second equation is a repetition for the reader's convenience
of  eqn. \eqref{lupetto} and
$\mathcal{D}^{KE}\left(\vv _0\right)$ is the boundary
function 
defined in equation
\eqref{sympaKE}; the relation between the two independent roots
$\lambda_{1,2}$ and the parameter $\kappa$ ia provided by
equations \eqref{rinominopara},\eqref{ricappo}. The reason while we
have used the argument
\begin{equation}\label{v0defi}
   \vv _0= \frac{2
   \vv }{\kappa\mathfrak{w}+2}
\end{equation}
is that the symplectic variable $\vv _0$ associated with the
base-manifold metric and the symplectic variable $\vv $
associated with the metric on the canonical bundle
$\M^{KE}_T$ are not the same; their relation is
precisely that in eqn.~\eqref{v0defi} which is a direct consequence of
the Calabi Ansatz as we explain below.

\paragraph{Derivation of the
formula for
$\mathcal{G}^{KE}\left(\vv ,\mathfrak{w}\right)$. } The
general formula \eqref{pulcherrima} is a direct yield of the direct
Legendre transform after  the Calabi Ansatz:
\begin{equation}\label{GsymMT}
    G_{\M_T^{KE}}\left(\mathfrak{u},\vv ,\mathfrak{w}\right)
   =x_u \,\mathfrak{u} \, + x_v \, \vv  \, + \, x_w \,  \mathfrak{w} \, - \, \mathcal{K}_0(\vv _0) \, -
    \, U(\lambda)
\end{equation}
where
\begin{equation}\begin{array}{rcl}\label{rollandoforte}
    \lambda & = &  \frac{ w\, \bar{w}}{\text{
    det}g_{\M_B}}=  \text{const}\, \times \, w\, \bar{w}\, \, \exp\left[
     \kappa \, \mathcal{K}_0(\vv _0) \right] \,
    = \, \Lambda(\mathfrak{w})  =\frac{1}{24} \mathfrak{w} \left(\kappa^2
    \mathfrak{w}^2+6\kappa
    \mathfrak{w}+12\right) \\ \displaystyle
    \frac{\mathfrak{w}}{2} & = & \lambda \, U'(\lambda)  \\
   U(\lambda)& = & \mathbb{U}(\mathfrak{w})= \displaystyle \frac{-3 \log \left(2 \left(\kappa ^2 \mathfrak{w}^2+6 \kappa
   \mathfrak{w}+12\right)\right)+3 \kappa  \mathfrak{w}-2 \sqrt{3} \arctan\left(\frac{\kappa  \mathfrak{w}+3}{\sqrt{3}}\right)}{2 \kappa }
    \\
\mathcal{K}_0(\vv _0)&=&\vv _0
   \mathcal{D}'\left(\vv _0\right)-\mathcal{D}\left(\vv _0\right)+\frac
   {\vv _0}{2}
\end{array}\end{equation}
The last two lines in eqns.~\eqref{rollandoforte} were derived earlier,
respectively in eqns.~\eqref{Udiwgoth},\eqref{K0inv0}. The explicit form
of $\mathbb{U}(\mathfrak{w})$ follows from eqn.~\eqref{Udiwgoth}
using   the KE condition, namely
eqn.~\eqref{ABFconst}. From the above relations one easily obtains the
relations
\begin{equation}\begin{array}{rcl}
\label{torneodipallacorda} &&\mathfrak{u}_0 = \displaystyle \frac{2
\mathfrak{u}}{\mathit{k}
   \mathfrak{w}+2},  \quad \vv _0 = \frac{2
   \vv }{\mathit{k} \mathfrak{w}+2} \\[8pt]
&& x_w =\frac{1}{2}\left\{
\log\left[\Lambda(\mathfrak{w})\right] - \kappa \,
\mathcal{K}_0(\vv _0) \right\}
\end{array}\end{equation}
The first two relations can be understood as follows. The momenta
$\mathfrak{u}_0,\vv _0$ are, by definition
\begin{equation}\label{transeat}
    \mathfrak{u}_0= \partial_{x_u}\mathcal{K}_0 \quad ; \quad \vv _0= \partial_{x_vu}\mathcal{K}_0
\end{equation}
while we have
\begin{equation}\label{gloriamundi}
    \mathfrak{u}= \partial_{x_u}\mathcal{K} \quad ; \quad \vv  =
    \partial_{x_v}\mathcal{K}
\end{equation}
By the Calabi Ansatz we get:
\begin{equation}\label{giobattabis}
    \mathfrak{u}= \mathfrak{u}_0 +\partial_{x_u} U(\lambda) \,
    = \, \mathfrak{u}_0 + \partial_{xu}\lambda \, \partial_\lambda U(\lambda) =
    \mathfrak{u}_0 + \kappa \partial_{xu}\mathcal{K}_0 \,\lambda \partial_\lambda
    U(\lambda)\,= \,\mathfrak{u}_0\left( 1+ \frac{\kappa}{2} \,
    \mathfrak{w}\right)
\end{equation}
A completely analogous calculation can be done for the case of
$\vv $. Finally let us note that the coordinate $x_u,x_v$
were already resolved in terms of $u_0,v_0$ in
eqns.~\eqref{inversioneB}:
\begin{equation}\label{riducoinveB}
  x_u=  \frac{1}{2} \left(\log
   \left(\mathfrak{u}_0\right)-\log \left(2
   \vv _0-\mathfrak{u}_0\right)\right)\quad ; \quad x_v=
   \mathcal{D}'\left(\vv _0\right)+\log \left(2
   \vv _0-\mathfrak{u}_0\right)-\frac{1}{2} \log
   \left(\vv _0\right)+\frac{1}{2}
\end{equation}
 The information provided in the above equations \eqref{rollandoforte} - \eqref{riducoinveB} is
sufficient to complete the Legendre transform \eqref{GsymMT} and
retrieve the very simple and elegant result encoded in
eqn.~\eqref{pulcherrima}.
\par
To check the correctness of the general formula \eqref{pulcherrima}
we have explicitly calculated, by means of the {\sc mathematica}
package {\sc metricgrav}\footnote{{\sc metricgrav} just as {\sc
vielbgrav23} is a Mathematica package written by one of us (P.F.)
almost thirty years ago and constantly updated. It will be available
from the site of the publisher De Gruyter to the readers of the
forthcoming book \cite{nuovogruppo}.}, the Ricci tensor for a few
cases of $\M_T^{[\lambda_1,\lambda_2]}$, always finding zero.
\paragraph{The example of the metric $[2,1]$.}
Here we present the explicit form in symplectic
coordinates of the Ricci-flat metric on the canonical bundle of the
KE manifold $\M_B^{[1,2]}$, namely that
determined by the choice: $\lambda_1=1$, $\lambda_2=2$. We get:
\begin{equation}\begin{array}{rcl}
  ds^2_{M_T^{[1,2]}} &=& d\phi^2 \left(-\frac{\mathfrak{u}^2 (9 \mathfrak{w}+14)^2}{343
   \vv ^3}-\frac{16464 \mathfrak{u}^2}{(9 \mathfrak{w}+14)^4}+2
   \mathfrak{u}\right) \\[8pt]
   && \displaystyle+\frac{d\vv ^2 \left(\mathfrak{u} \left(2058
   \vv ^3+(9 \mathfrak{w}+14)^3\right)-686 \vv ^3 (9
   \mathfrak{w}+14)\right)}{\vv  (2 \vv -\mathfrak{u}) (7
   \vv -9 \mathfrak{w}-14) (14 \vv -9 \mathfrak{w}-14) (21
   \vv +9 \mathfrak{w}+14)} \\[8pt]
   && \displaystyle+2 d\tau d\phi \left(-\frac{\mathfrak{u} (9
   \mathfrak{w}+14)^2}{343 \vv ^2}-\frac{16464 \mathfrak{u} \vv }{(9
   \mathfrak{w}+14)^4}+\mathfrak{u}\right)+\frac{d\mathfrak{u}
   d\vv }{\mathfrak{u}-2 \vv }+\frac{d\mathfrak{u} (\mathfrak{u}
   d\vv -\vv  d\mathfrak{u})}{\mathfrak{u} (\mathfrak{u}-2
   \vv )} \\[8pt]
   && \displaystyle+\frac{36 \mathfrak{u} \mathfrak{w} \left(27 \mathfrak{w}^2+126
   \mathfrak{w}+196\right) d\chi d\phi}{(9 \mathfrak{w}+14)^3}+d\tau^2
   \left(-\frac{16464 \vv ^2}{(9 \mathfrak{w}+14)^4}-\frac{(9
   \mathfrak{w}+14)^2}{343 \vv }+\vv \right) \\[8pt]
   && \displaystyle+\frac{6174 \vv ^2
   d\vv  d\mathfrak{w}}{(7 \vv -9 \mathfrak{w}-14) (14
   \vv -9 \mathfrak{w}-14) (21 \vv +9
   \mathfrak{w}+14)} \\[8pt]
   && \displaystyle+\frac{d\mathfrak{w}^2 \left(5647152 \vv ^3-343
   \vv ^2 (9 \mathfrak{w}+14)^4+(9 \mathfrak{w}+14)^6\right)}{2 \mathfrak{w} (9
   \mathfrak{w}+14) \left(27 \mathfrak{w}^2+126 \mathfrak{w}+196\right) (7
   \vv -9 \mathfrak{w}-14) (14 \vv -9 \mathfrak{w}-14) (21
   \vv +9 \mathfrak{w}+14)} \\[8pt]
   && \displaystyle+\frac{36 \vv  \mathfrak{w} \left(27
   \mathfrak{w}^2+126 \mathfrak{w}+196\right) d\tau d\chi}{(9
   \mathfrak{w}+14)^3}+\frac{2 \mathfrak{w} \left(27 \mathfrak{w}^2+126
   \mathfrak{w}+196\right) d\chi^2}{(9 \mathfrak{w}+14)^2}
\end{array}\end{equation}
\par
\section{The Ricci-flat metric on $\operatorname{tot}\left[K\left(\M^{KE}_B\right)\right]$ versus the
metric cone on the Sasakian fibrations on
$\M^{KE}_B$}\label{sasacco} This last section  is probably the most relevant one since it clarifies an
unexpected distinction that opens new directions of investigations.
To develop our argument it is appropriate to begin by reformulating
the geometry of the Ricci-flat metric derived from the Calabi ansatz
in terms of vielbeins. In that language the comparison with the
Sasaki-Einstein metrics of \cite{Gauntlett:2004yd} will become much
more transparent.
\par
With some straightforward yet cumbersome algebraic analysis we have
verified that, after the change of variables
\eqref{cambiovariabile}, the Ricci-flat metric in action-angle
coordinates coming  from the symplectic potential $G_{\M^{KE}_T}$,
as displayed in \eqref{pulcherrima}, can be rewritten as a sum of
squares in terms of a set of six vielbein one-forms $V^i$:
\begin{equation}\label{sommaquadra}
    ds^2_{\M^{KE}_T} \, = \, \sum_{i=1}^6 \, \left(V^i\right)^2 \,
    \, = \, \underbrace{\delta_{ij} \, V^i_\mu \, V^j_\nu \,}_{\text{metric }\mathbf{g}_{\mu\nu}} dy^\mu dy^\nu \quad
    ; \quad \underbrace{y^\mu \, = \,\{\theta ,\mathfrak{v},\mathfrak{w},\phi ,\tau ,\chi
    \}}_{\text{coordinates}}
\end{equation}
The explicit general structure of the sechsbein $V^i$, whose matrix
of components $V^i_\mu$ must be invertible and reproduces the metric
\eqref{sommaquadra} is the following one:
\begin{align}
\label{seigambe}
  V^1 & = \sqrt{\mathfrak{v}}\, d\theta \\
  V^2 & = \sqrt{\mathfrak{v}} \, d\phi \, \sin (\theta )  \\
  V^3 & = \frac{d\mathfrak{v}}{\mathfrak{A}(\mathfrak{v},\mathfrak{w})} \\
  V^4 & = \mathfrak{B}(\mathfrak{v},\mathfrak{w}) \left[ d\mathfrak{w}\, + \,
   \mathfrak{C}(\mathfrak{v},\mathfrak{w})\, d\mathfrak{v} \right]  \\
  V^5 & = \mathfrak{D}(\mathfrak{v},\mathfrak{w})\, \left[d\tau \, + \,  \left(1-\cos (\theta )\right)\, d\phi \right] \\
  V^6 & = \mathfrak{E}(\mathfrak{v},\mathfrak{w})\, \left[d\chi \, + \,\mho (\mathfrak{v},\mathfrak{w})
  \left[\,d\tau\, + \,
    \left(1-\cos (\theta )\right)\,d\phi\,
    \right]\right]
\end{align}
For the metric given by the symplectic potential in
\eqref{pulcherrima} the six functions
$\mathfrak{A}(\mathfrak{v},\mathfrak{w})$,
$\mathfrak{B}(\mathfrak{v},\mathfrak{w})$,
$\mathfrak{C}(\mathfrak{v},\mathfrak{w})$,
$\mathfrak{D}(\mathfrak{v},\mathfrak{w})$,
$\mathfrak{E}(\mathfrak{v},\mathfrak{w})$,
$\mho(\mathfrak{v},\mathfrak{w})$ are explicitly given below, where,
in order to make formulae more easily readable, we have renamed
$\lambda_1 =\alpha$, $\lambda_2 = \beta$ : {\fontsize{6.4}{6.5}
\begin{align}
\label{ciamabuti} \mathfrak{A}(\mathfrak{v},\mathfrak{w}) &=-\frac{i
\sqrt{64 \mathfrak{v}^3 (\alpha +\beta ) \left(\alpha ^2+\alpha
\beta
   +\beta ^2\right)^6-4 \mathfrak{v}^2 \left(\alpha ^2+\alpha  \beta +\beta
   ^2\right)^3 \left(2 \left(\alpha ^2+\alpha  \beta +\beta ^2\right)+3
   \mathfrak{w} (\alpha +\beta )\right)^4+\alpha ^2 \beta ^2 \left(2 \left(\alpha
   ^2+\alpha  \beta +\beta ^2\right)+3 \mathfrak{w} (\alpha +\beta )\right)^6}}{2
   \sqrt{\mathfrak{v}} \left(\alpha ^2+\alpha  \beta +\beta ^2\right)^{3/2}
   \left(2 \left(\alpha ^2+\alpha  \beta +\beta ^2\right)+3 \mathfrak{w} (\alpha
   +\beta )\right)^2}\\
\mathfrak{B}(\mathfrak{v},\mathfrak{w}) &=
\frac{\sqrt{N_{\mathfrak{B}}(\mathfrak{v},\mathfrak{w})}}{\sqrt{D_{\mathfrak{B}}(\mathfrak{v},\mathfrak{w})}} \\
N_{\mathfrak{B}}(\mathfrak{v},\mathfrak{w}) & = 64 \mathfrak{v}^3
(\alpha +\beta ) \left(\alpha ^2+\alpha  \beta +\beta
   ^2\right)^6-4 \mathfrak{v}^2 \left(\alpha ^2+\alpha  \beta +\beta ^2\right)^3
   \left(2 \left(\alpha ^2+\alpha  \beta +\beta ^2\right)+3 \mathfrak{w} (\alpha
   +\beta )\right)^4+\alpha ^2 \beta ^2 \left(2 \left(\alpha ^2+\alpha  \beta
   +\beta ^2\right)+3 \mathfrak{w} (\alpha +\beta )\right)^6\\
D_{\mathfrak{B}}(\mathfrak{v},\mathfrak{w}) & =2 \mathfrak{w}
\left(8 \left(\alpha ^2+\alpha  \beta +\beta ^2\right)^3+9
   \mathfrak{w}^3 (\alpha +\beta )^3+24 \mathfrak{w}^2 \left(\alpha ^2+\alpha
   \beta +\beta ^2\right) (\alpha +\beta )^2+24 \mathfrak{w} \left(\alpha
   ^2+\alpha  \beta +\beta ^2\right)^2 (\alpha +\beta )\right)\times \\
   &\quad \times \left(8
   \mathfrak{v}^3 (\alpha +\beta ) \left(\alpha ^2+\alpha  \beta +\beta
   ^2\right)^3-4 \mathfrak{v}^2 \left(\alpha ^2+\alpha  \beta +\beta ^2\right)^3
   \left(2 \left(\alpha ^2+\alpha  \beta +\beta ^2\right)+3 \mathfrak{w} (\alpha
   +\beta )\right)+ \right. \\
   &\left. \quad +\alpha ^2 \beta ^2 \left(2 \left(\alpha ^2+\alpha  \beta
   +\beta ^2\right)+3 \mathfrak{w} (\alpha +\beta
   )\right)^3\right)\\
\mathfrak{C}(\mathfrak{v},\mathfrak{w})   & = \frac{24
\mathfrak{v}^2 \mathfrak{w} (\alpha +\beta ) \left(\alpha ^2+\alpha
   \beta +\beta ^2\right)^3 \left(8 \left(\alpha ^2+\alpha  \beta +\beta
   ^2\right)^3+9 \mathfrak{w}^3 (\alpha +\beta )^3+24 \mathfrak{w}^2 \left(\alpha
   ^2+\alpha  \beta +\beta ^2\right) (\alpha +\beta )^2+24 \mathfrak{w}
   \left(\alpha ^2+\alpha  \beta +\beta ^2\right)^2 (\alpha +\beta )\right)}{64
   \mathfrak{v}^3 (\alpha +\beta ) \left(\alpha ^2+\alpha  \beta +\beta
   ^2\right)^6-4 \mathfrak{v}^2 \left(\alpha ^2+\alpha  \beta +\beta ^2\right)^3
   \left(2 \left(\alpha ^2+\alpha  \beta +\beta ^2\right)+3 \mathfrak{w} (\alpha
   +\beta )\right)^4+\alpha ^2 \beta ^2 \left(2 \left(\alpha ^2+\alpha  \beta
   +\beta ^2\right)+3 \mathfrak{w} (\alpha +\beta )\right)^6}\\
\mathfrak{D}(\mathfrak{v},\mathfrak{w})   & =\frac{\sqrt{-8
\mathfrak{v}^3 (\alpha +\beta ) \left(\alpha ^2+\alpha  \beta
   +\beta ^2\right)^3+4 \mathfrak{v}^2 \left(\alpha ^2+\alpha  \beta +\beta
   ^2\right)^3 \left(2 \left(\alpha ^2+\alpha  \beta +\beta ^2\right)+3
   \mathfrak{w} (\alpha +\beta )\right)-\alpha ^2 \beta ^2 \left(2 \left(\alpha
   ^2+\alpha  \beta +\beta ^2\right)+3 \mathfrak{w} (\alpha +\beta )\right)^3}}{2
   \sqrt{\mathfrak{v} \left(\alpha ^2+\alpha  \beta +\beta ^2\right)^3 \left(2
   \left(\alpha ^2+\alpha  \beta +\beta ^2\right)+3 \mathfrak{w} (\alpha +\beta
   )\right)}}\\
\mathfrak{E}(\mathfrak{v},\mathfrak{w})   & =\sqrt{2}
\sqrt{\frac{\mathfrak{w} \left(4 \left(\alpha ^2+\alpha  \beta
+\beta
   ^2\right)^2+3 \mathfrak{w}^2 (\alpha +\beta )^2+6 \mathfrak{w} \left(\alpha
   ^3+2 \alpha ^2 \beta +2 \alpha  \beta ^2+\beta ^3\right)\right)}{\left(2
   \left(\alpha ^2+\alpha  \beta +\beta ^2\right)+3 \mathfrak{w} (\alpha +\beta
   )\right)^2}}\\
   \mho(\mathfrak{v},\mathfrak{w}) &= \Omega(\mathfrak{v},\mathfrak{w}) \, \equiv \,\frac{3 \mathfrak{v}
    (\alpha +\beta )}{2 \left(\alpha ^2+\alpha  \beta +\beta
   ^2\right)+3 \mathfrak{w} (\alpha +\beta )}
\end{align}
}

\paragraph{Properties of the sechsbein and comparison with Sasakian 5-manifolds.}
The general structure of the \textit{sechsbein} in \eqref{seigambe}
and \eqref{ciamabuti} is very interesting since it  highlights   the double fibration structure of the underlying manifold
$\M_{T}$ which is a line-bundle (the canonical one) on the base
manifold $\M_B$ which, in its turn, is a (singular) $\mathbb{P}^1$
fibration over a base $\mathbb{P}^1$:
\begin{equation}\label{doppiafibbra}
    \M_{T} \, \stackrel{\pi_1}{\longrightarrow} \, \M_{B} \,
    \stackrel{\pi_2}{\longrightarrow} \, \mathbb{P}^1
\end{equation}
The projection onto the base manifold is produced by considering the
limit $\mathfrak{w}\to 0$. Very much informative is the development
in series of the sechsbein for small values of the coordinate
$\mathfrak{w}$. The limit is regular for $\mathfrak{w}=0$; two of the
sechsbein $(V^4,V^6)$ vanish and the remaining four 1-forms
$V^1,V^2,V^3,V^5$ attain the values
$\mathbf{e}^3,\mathbf{e}^4,\mathbf{e}^1,\mathbf{e}^2$ corresponding
to the vierbein of the K\"ahler-Einstein 4-dimensional metrics as
given in eq. \eqref{vierbeine} with the function
$\mathcal{FK}^{KE}(\vv)$ as given in table \ref{casoni}. At order
$\sqrt{\mathfrak{w}}$ there is no deformation of the base manifold
vierbein, but the two vielbein corresponding to the fiber directions
do appear. At order $\mathfrak{w}$ we see the beginning of the
deformation of the base-manifold vierbein. Precisely we find:
\begin{equation}\label{canna}
\begin{array}{rclll}
V^1 &=&\mathbf{e}^3&\null&\null\\
V^2 &=&\mathbf{e}^4&\null&\null\\
V^3 &=&\mathbf{e}^1&\null&+\mathfrak{w}\, \Delta\mathbf{e}^1 \, + \, \mathcal{O}(\mathfrak{w}^2)\\
V^4 &=&0&+\sqrt{\mathfrak{w}}\, \Delta\pmb{\Phi}_{\mathfrak{w}} +\mathcal{O}(\mathfrak{w}^{3/2}) &\null \\
V^5 &=&\mathbf{e}^2 & \null  &+\mathfrak{w}\, \Delta\mathbf{e}^2\,+ \, \mathcal{O}(\mathfrak{w}^2)\\
V^6 &=&0&+ \sqrt{\mathfrak{w}}\, \Delta\pmb{\Phi}_\chi  +\mathcal{O}(\mathfrak{w}^{3/2})\,  & \null\\
\end{array}
\end{equation}
where the deformations of the base manifold vielbein are as it
follows:
\begin{align}\label{cantovello}
\Delta\mathbf{e}^1 &= -\frac{3 (\alpha +\beta )
\sqrt{\frac{\mathfrak{v}}{\alpha
   ^2+\alpha  \beta +\beta ^2}} \left(\alpha ^2 \beta ^2-2 \mathfrak{v}^3 (\alpha
   +\beta )\right)}{2 ((\mathfrak{v}-\alpha ) (\beta -\mathfrak{v}) (\alpha
   \beta +\mathfrak{v} (\alpha +\beta )))^{3/2}}\,  d\mathfrak{v}\\
\Delta\mathbf{e}^2 &=\frac{3 (\alpha +\beta ) \left(\mathfrak{v}^3
(\alpha +\beta )-2 \alpha ^2 \beta
   ^2\right) }{4 \left(\alpha
   ^2+\alpha  \beta +\beta ^2\right)^{3/2} \sqrt{\mathfrak{v}
   (\mathfrak{v}-\alpha ) (\beta -\mathfrak{v}) (\alpha  \beta +\mathfrak{v}
   (\alpha +\beta ))}} \, \left[d\tau+\left(1-  \cos\theta \right)\,d\phi \right]
\end{align}
and the initial fibre-vielbein are instead the following ones:
\begin{align}\label{canberra}
\Delta\pmb{\Phi}_{\mathfrak{w}} &= \frac{d\mathfrak{w}}{\sqrt{2}
\mathfrak{w}}+\frac{3 \mathfrak{v}^2 \,(\alpha +\beta )
   }{(\mathfrak{v}-\alpha ) (\mathfrak{v}-\beta ) (\alpha  \beta
   +\mathfrak{v} (\alpha +\beta ))}\,d\mathfrak{v}\\
\Delta\pmb{\Phi}_\chi  &=\sqrt{2} \left(d\chi
   +\frac{3 \mathfrak{v} (\alpha
+\beta ) }{2 \left(\alpha ^2+\alpha  \beta +\beta
^2\right)}\left[d\tau+\left(1-  \cos\theta \right)\,d\phi
\right]\right)
\end{align}
\paragraph{Comparison with the Sasaki-Einstein metrics.}
It is now the appropriate moment to consider the Sasaki-Einstein
metrics introduced in \cite{Gauntlett:2004yd}. In the coordinates
utilized by those authors we have:
\begin{align}
 ds^2_{SE_5} &= \frac{\text{dy}^2 (1-c y)}{2
\left(a+2 c y^3-3 y^2\right)}+\frac{\left(a+2 c
   y^3-3 y^2\right) (c \text{d$\phi $} \cos (\theta )+\text{d$\psi $})^2}{18
   (1-c y)}+\frac{1}{6} (1-c y) \left(\text{d$\theta $}^2+\text{d$\phi $}^2 \sin
   ^2\theta \right)\\
   &+\frac{\pmb{\Phi} _{\text{SE}}^2}{9}\label{5sasacchi}\\
\pmb{\Phi} _{\text{SE}} & =\left[d\xi\,+\,y \left(d\psi +c d\phi
\cos \theta +d\psi\right)-d\phi  \cos \theta \right]
\end{align}
If in equation \eqref{5sasacchi} we apply the following coordinate
transformation and renaming of the parameters:
\begin{equation}\label{overrule}
   a\,\to \, \frac{1}{4} \left(3 \beta  k^2+4\right),\quad c\, \to \, 1,\quad y\to 1-\frac{k
   \mathfrak{v}}{2},\quad \psi \, \to \, -\tau -\phi
\end{equation}
we find the following interesting result:
\begin{align}
    ds^2_{SE_5} & = \, \frac{k}{12}\,\widehat{ds}^2_{5} \label{omero}\\
    \widehat{ds}^2_{5}&= {ds}^2_{KE_4}+\frac{4}{3 k}\, \pmb{\Phi}_{\text{SE}}^2\label{virgilio}\\
{ds}^2_{KE_4} & =
\frac{d\mathfrak{v}^2}{\mathcal{F}\mathcal{K}_{\text{KE}}(\mathfrak{v})}\,
+ \, \mathcal{F}\mathcal{K}_{\text{KE}}(\mathfrak{v}) \left[d\tau
+\left(1 -\cos \theta \right)\,d\phi\right]^2 \, +\,
\mathfrak{v} \left(d\phi^2 \sin ^2\theta +d\theta^2\right)\label{esiodo}\\[3pt]
\mathcal{K}_{\text{KE}}(\mathfrak{v})&= \frac{3 \beta -k
\mathfrak{v}^3+3 \mathfrak{v}^2}{3 \mathfrak{v}} \label{caramba}
\end{align}
As one sees the line element ${ds}^2_{KE_4}$ in \eqref{esiodo}
exactly coincides with the line element of the K\"ahler Einstein
metrics discussed in previous sections and presented in eq.
\eqref{metrauniversala}. It remains to be seen what is the
appearance  of the 1-form $\pmb{\Phi} _{\text{SE}}$ after the
transformation \eqref{overrule}. If we add also the coordinate
transformation:
\begin{equation}\label{aggiunta}
    \xi \, \to \, p \,\chi \, +\,\tau \, +\, \phi \quad ; \quad p \in \mathbb{R}
\end{equation}
we find:
\begin{equation}\label{crinologo}
    \pmb{\Phi} _{\text{SE}}\, = \, p \, \left[d\chi\, +\,\frac{k}{2 p} \,
    \mathfrak{v}\, \left(d\tau + \left(1-\cos\theta \right)\,d\phi\right)\right]
\end{equation}
Comparing eq.\eqref{crinologo} with \eqref{canberra} we see that we
obtain:
\begin{equation}\label{cagnolotto}
    \pmb{\Phi} _{\text{SE}} \, \ltimes \, \Delta\pmb{\Phi}_\chi
\end{equation}
if we choose the constant $p$ as we do below:
\begin{equation}\label{troglodita}
    p \, = \, k \, \frac{\alpha^{2}+\alpha\,\beta+\beta^{2}}{3 \,
    (\alpha \, + \, \beta)} \, = \, 1
\end{equation}
Hence, using eq.s
\eqref{overrule},\eqref{aggiunta},\eqref{troglodita},\eqref{rinominopara}
we can conclude that the Sasaki-Einstein metric of
\cite{Gauntlett:2004yd} with the choice of the parameters $a,c$
provided in \eqref{overrule} is proportional through the constant
$\frac{k}{12}$ to the following five dimensional metric:
\begin{equation}\label{culatello}
    \widehat{ds}^2_{5} \, = \,\underbrace{{ds}^2_{KE_4}}_{\text{KE metric on $\M_B$}} \, + \,\frac{4}{3\, k}\,
    \,\underbrace{\left(d\chi \, + \,\Omega(\mathfrak{v},0) (\mathfrak{v},\mathfrak{w})
  \left[\,d\tau\, + \,
    \left(1-\cos (\theta )\right)\,d\phi\,
    \right]\right)^2}_{\lim_{\mathfrak{w}\to 0}\frac{V^6}{\sqrt{2\mathfrak{w}}}}
\end{equation}
As we have explicitly checked, the metric is an Einstein metric,
since its Ricci tensor in intrinsic components takes the following
form:
\begin{equation}\label{krumiro}
    \mathcal{R}ic[\widehat{g}_5] \, = \, \frac{k}{6} \, \delta_{ij}
\end{equation}
and on an Einstein space the standard metric of the metric cone is
certainly Ricci-flat. Hence writing a new sechsbein:
\begin{equation}\label{Ebeina}
    \mathbb{E}_{cone} \, = \, \{\mathbf{E}^1\, \dots, \mathbf{E}^6\}
\end{equation}
where:
\begin{align}\label{ruminia}
\mathbf{E}^1 & = \,R\, \mathbf{e}_3 &; \quad &
\mathbf{E}^2 & = \,R\, \mathbf{e}_4 \\
\mathbf{E}^3 & = \,R\, \mathbf{e}_1 &;\quad &
\mathbf{E}^4 & = \,R\, \mathbf{e}_2 \\
\mathbf{E}^5 & = \,R\, \sqrt{\frac{4}{3\, k}}\,
    \,\left(d\chi \, + \,\frac{k}{2} \mathfrak{v}\,
  \left[\,d\tau\, + \,
    \left(1-\cos (\theta )\right)\,d\phi\,
    \right]\right) &;\quad &
\mathbf{E}^6 & =  2\sqrt{\frac{3}{k}} dR
\end{align}
With the Mathematica Code {\sc Vielbgrav23} we have calculated the
the curvature 2-form $\mathfrak{R}^{ab}_{con}$ and the intrinsic
components  of the  Riemann tensor $\mathcal{R}ie_{cone}$  for the
metric provided by the sechsbein \eqref{ruminia}. As the metric is
Ricci-flat, the Riemann tensor coincides with the Weyl tensor
$\mathcal{W}_{cone}(R,\mathfrak{v})$. Similarly we have done for the
Ricci-flat metric constructed with the Calabi ansatz, using the
sechsbein defined in \eqref{seigambe} with the functions displayed
in \eqref{ciamabuti}. In this way we have obtained the curvature
2-form $\mathfrak{R}^{ab}_{CA}$ and the intrinsic components of the
Weyl tensor associated with the Calabi Ansatz metric
$\mathcal{W}_{CA}(\mathfrak{w},\mathfrak{v})$. In order to make the
comparison more precise the sechsbein \eqref{seigambe} has been
reordered in a similar way to the ordering utilized in the metric
cone case:
\begin{equation}\label{CAbeina}
    \mathbb{E}_{CA} \, = \, \{V^1,V^2,V^3,V^5,V^6,V^4\}
\end{equation}
The two metrics exactly coincide on the base-manifold $\M_B$ that is
one of the KE manifolds with two conical singularities discussed at
length in the previous sections of the present paper and are both
Ricci-flat in 6-dimensions, yet from all what we said in the present
section they seem to be intrinsically different, since the Sasakian
Einstein 5-dimensional metric as described in \eqref{culatello}
cannot be obtained by fixing the fibre coordinate $\mathfrak{w}$ to
some appropriate constant value. Yet one might think that there
exists some clever change of coordinates that can map one metric
into the other. To show that this is not the case we resorted to the
calculation of Weyl tensor invariants for the two metrics to compare
them. Furthermore we calculated polynomial 2-form curvature
invariants, in particular the following   6-forms: 
\begin{equation}
\label{tricurvoni}
\begin{array}{rcl}
\mathrm{Ch} & = &\mathfrak{R}^{ab}\wedge \mathfrak{R}^{bc}\wedge
\mathfrak{R}^{ca} \,  = \,Tr\left(\mathfrak{R}\wedge
\mathfrak{R}\wedge \mathfrak{R}\right) \\
 \mathrm{E} & = &\mathfrak{R}^{ab}\wedge \mathfrak{R}^{cd}\wedge
\mathfrak{R}^{fg} \, \epsilon_{abcdfg} \,\\
\end{array}
\end{equation}
The considered Weyl invariants are instead the following ones:
\begin{equation}
\label{invarianti}
\begin{array}{|rcl|rcl|}
\hline Quad_1 & = & \mathcal{W}[abij]\mathcal{W}[ijab]\\Cub_1
&=&\mathcal{W}[abij]\mathcal{W}[ijpq]\mathcal{W}[pqab]\\
Cub_2 &= &\mathcal{W}[ijpq]\mathcal{W}[rpsq]\mathcal{W}[rsij]\\
Cub_3
&=&\mathcal{W}[ijpq]\mathcal{W}[rpsq]\mathcal{W}[risj]\\
Quart_1
&=&\mathcal{W}[abij]\mathcal{W}[ijpq]\mathcal{W}[pqmn]\mathcal{W}[mnab]\\Quart_2
&=&\mathcal{W}[i, j, p, q]\mathcal{W}[p, r, q, s]\mathcal{W}[r, m,
s, n]\mathcal{W}[m, n, i, j]\\
\hline
\end{array}
\end{equation}
The result of the calculations with one special choice of
$\alpha,\beta$ is displayed in table \ref{poltroni}.
\begin{table}[htb!]
\begin{equation}
\begin{array}{|c|l|l|}
\hline
\text{Inv} &\parbox{3cm}{\small \ \hfill\\ metric cone \\ over the Sasaki\\ manifold \\ \ \hfill}&
  \text{Calabi Ansatz for the Ricci-flat metric }\\
\hline
\mathrm{Ch}&0&0\\
\hline \mathrm{E}& 0 & \frac{20736 \left(63780651422208
\mathfrak{v}^6+(9
   \mathfrak{w}+14)^{12}\right)}{343 \mathfrak{v}^6 (9 \mathfrak{w}+14)^{12}} \times \text{Vol}
   \\
\hline Quad_1 & \frac{96}{49 R^4 \mathfrak{v}^6} & \frac{6 (9
\mathfrak{w}+14)^4}{117649 \mathfrak{v}^6}+\frac{8131898880}{(9
   \mathfrak{w}+14)^8}\\
   \hline
   Cub_1 & \frac{384}{343 R^6 \mathfrak{v}^9} & \frac{6 (9 \mathfrak{w}+14)^6}{40353607 \mathfrak{v}^9}+\frac{432}{343
   \mathfrak{v}^6}+\frac{348097316216832}{(9
   \mathfrak{w}+14)^{12}}\\
\hline Cub_2 & \frac{192}{343 R^6 \mathfrak{v}^9}& \frac{3 (9
\mathfrak{w}+14)^6}{40353607 \mathfrak{v}^9}+\frac{216}{343
   \mathfrak{v}^6}+\frac{174048658108416}{(9
   \mathfrak{w}+14)^{12}}\\
\hline Cub_3 &\frac{96}{343 R^6 \mathfrak{v}^9}& \frac{3 (9
\mathfrak{w}+14)^6}{80707214 \mathfrak{v}^9}+\frac{108}{343
   \mathfrak{v}^6}+\frac{87024329054208}{(9 \mathfrak{w}+14)^{12}} \\
   \hline
Quart_1 &\frac{4608}{2401 R^8 \mathfrak{v}^{12}}&\frac{18 (9
\mathfrak{w}+14)^8}{13841287201 \mathfrak{v}^{12}}+\frac{576 (9
   \mathfrak{w}+14)^2}{117649 \mathfrak{v}^9}+\frac{20736}{\mathfrak{v}^6 (9
   \mathfrak{w}+14)^4}+\frac{31631121143724146688}{(9
   \mathfrak{w}+14)^{16}}\\
   Quar_2 &\frac{1152}{2401 R^8 \mathfrak{v}^{12}}&\frac{9 (9 \mathfrak{w}+14)^8}{27682574402 \mathfrak{v}^{12}}+\frac{144 (9
   \mathfrak{w}+14)^2}{117649 \mathfrak{v}^9}+\frac{5184}{\mathfrak{v}^6 (9
   \mathfrak{w}+14)^4}+\frac{7907780285931036672}{(9
   \mathfrak{w}+14)^{16}}\\ \ \hfill  & \ \hfill & \ \hfill \\
   \hline
\end{array}
\end{equation}
\caption{\label{poltroni} Table of the polynomial invariants defined
in \eqref{tricurvoni} and \eqref{invarianti} evaluated in the case of the metric
cone over the Sasaki-Einstein manifold  and in the Calabi Ansatz case of Ricci-flat metrics. Since
the calculations are very long when the parameters are left symbolic
we evaluated the invariants in the reference case
$\alpha=1,\beta=2$.}
\end{table}
Inspecting this  table   one realizes that, as anticipated in
the introduction, the Ricci-flat metric constructed by means of the
Calabi Ansatz translated into action-angle variables is   {different} from that associated with the Sasaki-Einstein metric recalled in \eqref{5sasacchi}. The strongest
evidence is given by vanishing of the invariant $E$ in one case and
the non-zero and coordinate dependent structure of the same
invariant in the second case. However also the other invariants
corroborate the same evidence. One can calculate the value of the
coordinate $R$ in terms of $\mathfrak{v},\mathfrak{w}$ by equating
the quadratic invariant $Quad_1$ of the two metrics. Substituting
the result in the other invariants the two expressions do not agree
for the other invariants. This being clarified one can try to
understand the reason of the disagreement. In the Sasaki Einstein
approach \eqref{virgilio} corresponds to the standard construction
of the metric on a $\mathrm{U(1)}$ bundle. To the metric of the base
manifold, in the present case $\M_B^{KE}$, one adds the square of a
1-form $\Phi$ of the type $\Phi \, = \,d\tau\, +
\,\text{$\mathrm{U(1)}$-connection on $\M_B$}$. The remaining
coordinate is the radial one $R$. In the Calabi ansatz approach,
instead, one directly constructs the  {line bundle} on the
base manifold. A Sasaki-Einstein manifold could be retrieved a
posteriori through the construction of a sphere bundle $R =|w|$,
where the complex coordinate $w$ is the canonical bundle fibre
coordinate. The two procedures do not commute. The investigation of
the Sasaki-Einstein manifold underlying the constructed Ricci-flat
metric via Calabi ansatz is a new research project that we leave for
the future. Similarly the identification of a  gauge 4-dimensional
gauge theory dual to the here presented D3--brane solution of Type
IIB supergravity is one of the goals we plan to pursue in the near
future.

\paragraph{Completeness.} The metrics in the family $\operatorname{Mat}(\mathcal{FV})_{KE}$ are complete \cite{abreu2009toric},
so that, applying Theorem 4.3 in Calabi's paper \cite{Calabi-Metriques}, we would get that the Ricci-flat 6-dimensional metric on the canonical bundle is complete. However one should check that Calabi's theorem also applies in the singular case. 

\appendix
\section{The issue of (2,1)-forms} \label{21forms}
In this appendix we consider the issue of the (2,1)-forms, giving a
proof that no (anti)-self-dual $(2,1)$ forms  exist on the KE
manifolds previously studied. A (2,1)-form  can be written as
\begin{equation}\label{la21}
    \Omega_{\mathit{i}\mathit{j}\bar{\mathit{k}}}\left(z,\bar{z}\right)dz^\mathit{i}\wedge dz^\mathit{j} \wedge
    d\bar{z}^{\bar{\mathit{k}}}
\end{equation}
The dual  (2,1)-form is
\begin{equation}\label{dualus}
    \star_g\Omega \, = \,\tilde{\Omega}_{\ell p\bar{q}}\left(z,\bar{z}\right)dz^\ell\wedge dz^p \wedge
    d\bar{z}^{\bar{q}}
\end{equation}
where:
\begin{equation}\label{bonzo}
    \tilde{\Omega}_{\ell p\bar{q}}\left(z,\bar{z}\right)\, = \,
    \frac{1}{\sqrt{\text{det}g}} \, g_{\ell\bar{m}}\,
    g_{p\bar{n}}\,g_{\bar{q}s}\,
    \epsilon^{\bar{m}\bar{n}\bar{k}}\epsilon^{ijs}\,\Omega_{\mathit{i}\mathit{j}\bar{\mathit{k}}}\left(z,\bar{z}\right)
\end{equation}
Hence the (anti)-selfduality condition is expressed by the equation:
\begin{equation}\label{carolingio}
    \pm \,\mathit{i}\,\tilde{\Omega}_{\ell p\bar{q}}\left(z,\bar{z}\right)\, =
    \, \Omega_{\ell p\bar{q}}\left(z,\bar{z}\right)
\end{equation}
 Given the complex structure tensor
$\mathfrak{J}$ and its eigenvectors one writes the
complex differentials:
\begin{equation}\label{compdiff}
    dz^{i} \,=\,\omega^i_\ell(\mu)d\mu^\ell + \mathit{i} \,
    d\Theta^i \quad ; \quad d\bar{z}^{i} \,=\,\omega^i_\ell(\mu)d\mu^\ell - \mathit{i} \,
    d\Theta^i
\end{equation}
where the real 1-forms $\omega^i \, \equiv
\,\omega^i_\ell(\mu)d\mu^\ell$ depending only on moment variables
are defined by the complex structure tensor and hence by the
explicit form of the metric in terms of the Hessian. Using this
formalism a (2,1)-form is written as
\begin{equation}\label{curello} \begin{array}{rcl}
    \Omega^{(2,1)}=&\Omega_{ij|k}\left(\mu\right)\,(\omega^i+\mathit{i}\,d\Theta^i)\wedge(\omega^j+\mathit{i}\,d\Theta^j)\wedge
    (\omega^k-\mathit{i}\,d\Theta^k)\\
    &= Q_{IJK}\left(\mu\right)\, dy^I\wedge dy^J \wedge dy^K
\end{array} \end{equation}
where
$
    y^I \, = \, \left\{\mu^i, \Theta^j\right\}
$
is the complete set of the $2n$ real coordinates (moments and
angles).The complex functions $Q_{IJK}\left(\mu\right)$ depend on
the real variables $\mu$.  The (anti)self-duality condition is most easily written in the
 symplectic formalism as the determinant of the metric
tensor in symplectic coordinates is just 1. One gets\begin{equation}\label{carneade}
    Q_{IJK} \, = \,\pm \mathit{i} \, \epsilon_{IJKPQR} \, Q_{PQR}
\end{equation}
The original components $\Omega_{ij|k}\left(\mu\right)$ are supposed
to be complex valued functions of their real arguments which means
that we have a total of 9-complex valued functions, namely a total
of 18 real functions:
\begin{equation}\label{carinus}
  r(\mu) \, = \,
  \left\{f_1(\mu),\dots,f_9(\mu),g_1(\mu),\dots,g_9(\mu)\right\}
\end{equation}
Explicitly we obtain
\begin{equation} \label{caroluccia} \begin{array}{rcl}
  \Omega   &  =  & \  2 \left(\left(f_3-f_5-f_7+i g_3-i g_5-i
   g_7\right) d\tau\wedge d\chi\wedge
   \omega^1 -2 \left(f_8+i g_8\right)
   d\tau\wedge d\chi\wedge \omega^2\right.\\
   &&\left. -2 \left(f_9+i g_9\right) d\tau
   \wedge d\chi\wedge \omega^3 +2
   \left(f_1+i g_1\right) d\tau\wedge
   d\phi\wedge \omega^1 +2 \left(f_2+i
   g_2\right) d\tau\wedge d\phi\wedge
   \omega^2 \right.\\
   &&\left.+\left(f_3+f_5-f_7+i g_3+i
   g_5-i g_7\right) d\tau\wedge d\phi
   \wedge \omega^3 +2 \left(g_2-i
   f_2\right) d\tau\wedge \omega^1
   \wedge \omega^2 \right.\\
   &&\left. +\left(-i f_3-i f_5-i
   f_7+g_3+g_5+g_7\right) d\tau\wedge
   \omega^1 \wedge \omega^3 +2 \left(g_8-i
   f_8\right) d\tau\wedge \omega^2
   \wedge \omega^3 \right. \\
   &&\left. +\left(-i f_3-i f_5+i
   f_7+g_3+g_5-g_7\right) d\chi\wedge
   \omega^1 \wedge \omega^2 +2 \left(g_6-i
   f_6\right) d\chi\wedge \omega^1
   \wedge \omega^3 \right.\\
   &&\left. +2 \left(g_9-i
   f_9\right) d\chi\wedge \omega^2
   \wedge \omega^3 -2 \left(f_4+i
   g_4\right) d\phi\wedge d\chi\wedge
   \omega^1 \right.\\
   &&\left. -\left(f_3+f_5+f_7+i g_3+i
   g_5+i g_7\right) d\phi\wedge d\chi\wedge \omega^2 -2 \left(f_6+i
   g_6\right) d\phi\wedge d\chi\wedge
   \omega^3 \right.\\
   &&\left. +2 \left(g_1-i f_1\right)
   d\phi\wedge \omega^1 \wedge \omega^2
   +2 \left(g_4-i f_4\right) d\phi\wedge \omega^1 \wedge \omega^3
   \right. \\
   &&\left. +\left(i f_3-i f_5-i
   f_7-g_3+g_5+g_7\right) d\phi\wedge
   \omega^2 \wedge \omega^3  \right. \\ && \left. +\left(-i
   f_3+i f_5-i f_7+g_3-g_5+g_7\right) d\tau\wedge d\phi\wedge
   d\chi\right.\\
   &&\left. +\omega^1
   \wedge \omega^2 \wedge \omega^3
   \left(f_3-f_5+f_7+i g_3-i g_5+i
   g_7\right)\right)
\end{array}\end{equation}
which is the most general expression for a (2,1)-form expressed in
the real symplectic coordinate basis. Expanding each of the closed
one-forms in the differentials $d\mu^i$ of the moments one obtains
the explicit form of the 20 components $Q_{IJK}(\mu)$ mentioned in
\eqref{curello}.
For instance in our standard example $\lambda_1=1,\lambda_2 =2$ we
have:
\begin{equation}\label{cucumber} \begin{array}{rcl}
 \omega^1 &=&  \frac{d\vv -\vv
   d\mathfrak{u}}{\mathfrak{u}
   (\mathfrak{u}-2 \vv )}\\
\omega^2 &=& \frac{N_2}{D_2} \\
N_2 &= & 3087 \vv ^3 d\mathfrak{w}
   (\mathfrak{u}-2
   \vv )-\mathfrak{u}
   d\vv  \left(2058
   \vv ^3+(9
   \mathfrak{w}+14)^3\right)+\vv
   d\mathfrak{u} \left(2058
   \vv ^3-343 \vv ^2 (9
   \mathfrak{w}+14)\right.\\
   &&\left.+(9
   \mathfrak{w}+14)^3\right)+686
   \vv ^3 (9 \mathfrak{w}+14)
   d\vv \\
  D_2 &=& \vv  (\mathfrak{u}-2 \vv )
   \left(2058 \vv ^3-343
   \vv ^2 (9 \mathfrak{w}+14)+(9
   \mathfrak{w}+14)^3\right)\\
   \omega^3&=&\frac{N_3}{D_3} \\
   N_3 & = & 6174 \vv ^2
   d\vv +\frac{d\mathfrak{w}
   \left(5647152 \vv ^3-343
   \vv ^2 (9 \mathfrak{w}+14)^4+(9
   \mathfrak{w}+14)^6\right)}{\mathfrak{w}
   \left(243 \mathfrak{w}^3+1512
   \mathfrak{w}^2+3528
   \mathfrak{w}+2744\right)}\\
   D_3 &=& 2 \left(2058 \vv ^3-343
   \vv ^2 (9 \mathfrak{w}+14)+(9
   \mathfrak{w}+14)^3\right)
\end{array}\end{equation}
and by substitution one straightforwardly obtains the $Q_{IJK}$
components whose expression is too lengthy to be displayed. In
general a complex valued 3-form has the following structure:
\begin{equation}\label{comico} \begin{array}{rcl}
    \Omega^{[3]}&=& \left(X_{20}+i Y_{20}\right) d\tau\wedge
   d\phi\wedge d\chi+\left(X_{10}+i
   Y_{10}\right) d\mathfrak{u}\wedge
   d\tau\wedge d\chi+\left(X_8+i
   Y_8\right) d\mathfrak{u}\wedge d\tau
   \wedge d\phi\\
   &&+\left(X_3+i Y_3\right)
   d\mathfrak{u}\wedge
   d\vv \wedge d\tau
   +\left(X_1+i Y_1\right)
   d\mathfrak{u}\wedge
   d\vv \wedge
   d\mathfrak{w}+\left(X_4+i Y_4\right)
   d\mathfrak{u}\wedge
   d\vv \wedge d\chi
   \\
   &&+\left(X_2+i Y_2\right)
   d\mathfrak{u}\wedge
   d\vv \wedge d\phi
   +\left(X_6+i Y_6\right)
   d\mathfrak{u}\wedge
   d\mathfrak{w}\wedge d\tau
   +\left(X_7+i Y_7\right)
   d\mathfrak{u}\wedge
   d\mathfrak{w}\wedge d\chi
   \\
   &&+\left(X_5+i Y_5\right)
   d\mathfrak{u}\wedge
   d\mathfrak{w}\wedge d\phi
   +\left(X_9+i Y_9\right)
   d\mathfrak{u}\wedge d\phi\wedge
   d\chi+\left(X_{16}+i Y_{16}\right)
   d\vv \wedge d\tau\wedge
   d\chi\\
   &&+\left(X_{14}+i Y_{14}\right)
   d\vv \wedge d\tau\wedge
   d\phi+\left(X_{12}+i Y_{12}\right)
   d\vv \wedge
   d\mathfrak{w}\wedge d\tau
   +\left(X_{13}+i Y_{13}\right)
   d\vv \wedge
   d\mathfrak{w}\wedge d\chi
   \\
   &&+\left(X_{11}+i Y_{11}\right)
   d\vv \wedge
   d\mathfrak{w}\wedge d\phi
   +\left(X_{15}+i Y_{15}\right)
   d\vv \wedge d\phi\wedge
   d\chi+\left(X_{19}+i Y_{19}\right)
   d\mathfrak{w}\wedge d\tau\wedge
   d\chi\\
   &&+\left(X_{17}+i Y_{17}\right)
   d\mathfrak{w}\wedge d\tau\wedge
   d\phi+\left(X_{18}+i Y_{18}\right)
   d\mathfrak{w}\wedge d\phi\wedge
   d\chi
\end{array} \end{equation}
where the $X_i$ and $Y_i$ are real functions of the momenta $\mu$.
The self-duality condition \eqref{carneade} reduces to an algebraic
relation that expresses all the $Y_i$ in terms of the $X_i$,
precisely:
\begin{equation}\label{carnillo}
   \begin{array}{lccr}
 Y_1& = & \pm & X_{20} \\
 Y_2& = & \pm & X_{19} \\
 Y_3& = & \pm & -X_{18} \\
 Y_4& = & \pm & -X_{17} \\
 Y_5& = & \pm & -X_{16} \\
 Y_6& = & \pm & X_{15} \\
 Y_7& = & \pm & X_{14} \\
 Y_8& = & \pm & X_{13} \\
 Y_9& = & \pm & X_{12} \\
 Y_{10}& = & \pm & -X_{11} \\
 \end{array}\qquad
 \begin{array}{lccr}
 Y_{11}& = & \pm & X_{10} \\
 Y_{12}& = & \pm & -X_9 \\
 Y_{13}& = & \pm & -X_8 \\
 Y_{14}& = & \pm & -X_7 \\
 Y_{15}& = & \pm & -X_6 \\
 Y_{16}& = & \pm & X_5 \\
 Y_{17}& = & \pm & X_4 \\
 Y_{18}& = & \pm & X_3 \\
 Y_{19}& = & \pm & -X_2 \\
 Y_{20}& = & \pm & -X_1 \\
\end{array}
\end{equation}
The choice of the $\pm$ sign corresponding to self/anti-self
duality, respectively. Comparing  eqn.~\eqref{caroluccia} with
eqn.~\eqref{comico} and using  eqn.~\eqref{cucumber} one obtains the
20 $X_i$ and the 20 $Y_i$ of a generic (2,1)-form as   linear
combination of the 18 free parameter functions  \eqref{carinus} with
coefficients that are rational functions of the moment $\mu$. The
self-duality constraint is a set of 20 linear equations on the 18
parameters. Obviously unless the rank of the $20\times 18$ matrix is
less than 18, there are no nontrivial solutions. We have indeed
verified that the 20 equations do not have nontrivial solutions for
the standard case $\lambda_1=1,\lambda_2=2$ and   for some other
choices of the parameters. Hence no harmonic self-dual (2,1) forms
exist on this KE background and we have exact D3-brane solutions
without 3-form fluxes. \cite{abreu2009toric}
\section{Metric on the Lens space $L(k;\,1)\sim S^3/\mathbb{Z}_k$}\label{AA}
 Let us consider the embedding of $S^3$ in $\mathbb{C}^2$ described by the two complex coordinates in \eqref{U0U1}, which satisfy:
$$|U^0|^2+|U^1|^2=r^2\,.$$
The metric can be written in the following form:
\begin{equation}
    ds^2=|dU^0|^2+|dU^1|^2=\frac{r^2}{4}\,(d\theta^2+\sin^2(\theta) d\phi^2)+\frac{r^2}{4}\,(d\gamma+\cos(\theta)\,d\phi)^2\,.
\end{equation}
Recall that  $0\le \theta \le \pi$, $0\le \phi<2\pi$, $0\le \gamma<4\pi$. One can indeed verify that, being
$$\sqrt{|g|}=\frac{r^3}{8}\,\sin(\theta)\,,$$
the above bounds on the angles yield the correct value of the volume
of $S^3$: $${\rm Vol}(S^3)=\int\,\sqrt{|g|}\,d\theta d\phi d\gamma=2\pi^2\,r^3\,.  $$
Consider now the Lens space $L(k;\,1)\sim S^3/\mathbb{Z}_k$
obtained by performing quotient of $S^3$ by the group $\mathbb{Z}_k$ acting as
\begin{equation} \left(\begin{matrix}U^0\cr U^1\end{matrix}\right)\rightarrow
\left(\begin{matrix}e^{\frac{2\pi i}{k}} & 0\cr 0 & e^{\frac{2\pi i}{k}}\end{matrix}\right)
\left(\begin{matrix}U^0\cr U^1\end{matrix}\right)\,.\end{equation}
This amounts to identifying
$$\gamma\sim \gamma+\frac{4\pi}{k}\,,$$
and has the effect of dividing by $k$ the interval of values of $\gamma$, so that, after the identification, $\gamma$   varies
in the range
$$\gamma\in \left(0,\frac{4\pi}{k}\right)\,.$$
Therefore the effect is to make the replacement  $\gamma=\psi/k$ in the metric, where $0\le \psi<4\pi$:
\begin{equation}
      ds^2=\frac{r^2}{4}\,(d\theta^2+\sin^2(\theta) d\phi^2)+\frac{r^2}{4}\,(d\psi/k+\cos(\theta)\,d\phi)^2=\frac{r^2}{4}\,
      (d\theta^2+\sin^2(\theta) d\phi^2)+\frac{r^2}{4 k^2}\,(d\psi+k\,\cos(\theta)\,d\phi)^2\,.
\end{equation}
One can verify that the curvature is just the same (locally it
amounts to a reparameterization), though globally the space becomes
$S^3/\mathbb{Z}_k$. This identifies $L(k;\,1)\sim S^3/\Z_k$ with the
monopole of charge $k$. Comparing the above metric with the one in
\eqref{metrauniversala} at constant $|v|$, we see that, the matching
of the fiber metric requires $k=2$ and $\tau=\psi/2$.

\bigskip\noindent{\bf Conflict of interest statement.} On behalf of all authors, the corresponding author states that there
 is no conflict of interest.

\par\bigskip\frenchspacing \renewcommand{\sc}{\rm}

\end{document}